\documentclass[%
preprint,
superscriptaddress,
preprintnumbers,
 amsmath,amssymb,
]{revtex4-2}
\usepackage{slashed}
\usepackage{cancel}
\usepackage{lipsum}
\usepackage{graphicx}
\usepackage{dcolumn}
\usepackage{bm}
\usepackage[dvipsnames,usenames]{xcolor}
\usepackage{amsmath}
\usepackage{amssymb} 
\usepackage{hyperref}
\usepackage[qm,braket]{qcircuit}
\usepackage{bbold}
\usepackage{algpseudocode}
\usepackage{mathtools}
\usepackage{multirow}

\newcommand{\nc}{\newcommand}
\newcommand{\rc}{\renewcommand}
\nc{\bfx}{{\bf x}}
\nc{\bfy}{{\bf y}}
\nc{\bfz}{{\bf z}}
\nc{\bfxh}{{\bf \hat{x}}}
\nc{\bfyh}{{\bf \hat{y}}}
\nc{\bfzh}{{\bf \hat{z}}}
\nc{\bfj}{{\bf j}}
\nc{\bfr}{{\bf r}}
\nc{\bfR}{{\bf R}}
\nc{\bfk}{{\bf k}}
\nc{\bfq}{{\bf q}}
\nc{\bfp}{{\bf p}}
\nc{\bfv}{{\bf v}}
\nc{\bfvh}{{\bf \hat{v}}}
\nc{\bfqh}{{\bf \hat{q}}}

\nc{\mel}[3]{\langle #1 | #2 | #3 \rangle}
\nc{\rme}[1]{\lange || #1 || \rangle}
\nc{\nuc}[2]{^{#1}{\rm #2}}
\nc{\me}[1]{\mel{\nuc{6}{Li}}{#1}{\nuc{6}{He}}}
\nc{\mee}[1]{\mel{\nuc{6}{Li},10}{#1}{\nuc{6}{He},00}}
\nc{\ord}[1]{\mathcal{O}(#1)}
\rc{\Re}{{\rm Re}}
\rc{\Im}{{\rm Im}}

\begin{document}	

\preprint{LA-UR-21-31938, INT-PUB-22-021}

\title{ \textit{Ab initio} calculation of the \texorpdfstring{$\beta$}{beta} decay spectrum of \texorpdfstring{$^6$He}{6He}}

\author{G.~B. King}

\affiliation{Department of Physics, Washington University in Saint Louis, Saint Louis, MO 63130, USA}

\author{A. Baroni}

\affiliation{Theoretical Division, Los Alamos National Laboratory, Los Alamos, NM 87545, USA }

\author{V. Cirigliano}

\affiliation{Institute for Nuclear Theory, University of Washington, Seattle WA 91195-1550, USA}

\author{S. Gandolfi}

\affiliation{Theoretical Division, Los Alamos National Laboratory, Los Alamos, NM 87545, USA }

\author{L. Hayen}

\affiliation{Department of Physics, North Carolina State University, Raleigh, North Carolina 27695, USA}

\author{E. Mereghetti}

\affiliation{Theoretical Division, Los Alamos National Laboratory, Los Alamos, NM 87545, USA }

\author{S. Pastore}

\affiliation{Department of Physics, Washington University in Saint Louis, Saint Louis, MO 63130, USA}
\affiliation{McDonnell Center for the Space Sciences at Washington University in St. Louis, MO 63130, USA}

\author{M. Piarulli}

\affiliation{Department of Physics, Washington University in Saint Louis, Saint Louis, MO 63130, USA}
\affiliation{McDonnell Center for the Space Sciences at Washington University in St. Louis, MO 63130, USA}

\begin{abstract}
We calculate the $\beta$ spectrum in the decay of $^6$He using 
Quantum Monte Carlo methods with 
nuclear interactions derived from chiral Effective Field Theory and consistent weak vector and axial currents. 
We work at second order in the multipole expansion, retaining terms suppressed by $\mathcal O(q^2/m_\pi^2)$, where $q$ denotes low-energy scales such as the reaction's $\mathcal Q$-value or the electron energy, and $m_\pi$ the pion mass. We go beyond the impulse approximation by including the effects of two-body
vector and axial currents.
We estimate the theoretical error on the spectrum by using four potential models in the Norfolk family of local two- and three-nucleon interactions, which have different cut-off, fit two-nucleon data up to different energies and use different observables to determine the couplings in the three-body force.
We find the theoretical uncertainty on the $\beta$ spectrum, normalized by the total rate, to be well below the permille level, and to receive contributions of comparable size from first and second order corrections in the multipole expansion.
We consider corrections to the $\beta$ decay spectrum induced by beyond-the-Standard Model charged-current interactions in the Standard Model Effective Field Theory, with and without sterile neutrinos, and discuss the sensitivity of the next generation of experiments to these interactions.
\end{abstract}

\maketitle

\tableofcontents

\section{Introduction}
\label{sec:intro}

Nuclear $\beta$ decays have been instrumental in establishing the Standard Model (SM) as the theory of the electroweak interactions \cite{Sudarshan:1958vf,Feynman:1958ty,Weinberg:2004kv}. In the era of the Large Hadron Collider, $\beta$ decays still provide very sensitive probes of physics beyond the Standard Model (BSM), which are highly competitive and complementary to searches at the energy frontier \cite{Bhattacharya:2011qm,Cirigliano:2012ab,Cirigliano:2013xha,Gonzalez-Alonso:2013uqa,Falkowski:2017pss,Alioli:2017ces,Gupta:2018qil,Alioli:2018ljm,Falkowski:2020pma,Torre:2020aiz,Boughezal:2021tih,Cirigliano:2021yto}.
Superallowed $0^+ \rightarrow 0^+$ transitions, combined with theoretical progress in the evaluation of radiative corrections \cite{Seng:2018yzq,Seng:2018qru,Czarnecki:2019mwq,Seng:2020wjq,Hayen:2020cxh,Shiells:2020fqp}, allow for the extraction of the $V_{ud}$ element of the Cabibbo-Kobayashi-Maskawa (CKM) quark mixing matrix  with uncertainity at the level 
 $\delta V_{ud} \sim 3 \cdot 10^{-4}$
\cite{Seng:2018yzq,Seng:2018qru,Czarnecki:2019mwq,Hardy:2020qwl},
probing BSM scales up to 10 TeV. Improved measurements of the neutron lifetime and 
$\beta$ asymmetry \cite{Pattie:2017vsj,Markisch:2018ndu,Brown:2017mhw,UCNt:2021pcg}
and the percent determination of the nucleon axial charge $g_A$ from lattice QCD 
\cite{Chang:2018uxx,Gupta:2018qil,Aoki:2019cca,Cirigliano:2022hob} test right-handed charged currents at subpercent level.
Global analyses of superallowed transitions, neutron decay and mirror $\beta$ decays limit the strength of non-standard vector, axial, scalar and tensor charged-current interactions to be less than a thousandth of the weak interactions  
\cite{Falkowski:2020pma,Cirigliano:2021yto}. At the same time charged- and neutral-current Drell-Yan production at the Large Hadron Collider is starting to directly access scales of a few TeVs \cite{ATLAS:2019lsy,ATLAS:2019erb,ATLAS:2020yat,CMS:2021ctt},
and the good agreement between precise theoretical predictions and Drell-Yan data allow for the exclusion of interactions at the effective scale $ \Lambda = 4-5$ TeV \cite{Cirigliano:2012ab,Alioli:2018ljm,Gupta:2018qil,Torre:2020aiz,Boughezal:2021tih}.

$\beta$ decay spectra provide sensitive probes of charged currents with different chiral structure from the SM. The interference of BSM currents with the $V$-$A$ SM interactions induces a distinctive $m_e/E_e$---where $m_e$ and $E_e$ are the electron's mass and energy, respectively---dependence in the $\beta$ spectrum, the so called   ``Fierz interference term'' \cite{Jackson:1957zz}, usually denoted by $b$.
The first direct neutron measurements constrain the Fierz interference term to be $-0.018 < b < 0.052$ at the 90\% confidence level \cite{Saul:2019qnp,Sun:2019pnb}. 
The Fierz interference term induced by scalar currents is tested in Fermi transitions, with the most recent analysis of superallowed $\beta$ decays yielding $b < 3.3 \cdot 10^{-3}$ ( 90\% confidence level) \cite{Hardy:2020qwl}.
Measurements
of spectra of purely Gamow-Teller transitions, such as the decay of $^6$He to $^6$Li, aim to push  the constraint on the Fierz interference term induced by tensor and pseudoscalar currents to the level  $b < 10^{-3}$ \cite{Cirigliano:2019wao,Garcia}, probing tensor currents at the 10 TeV mass scale. 
In addition, modifications to the shape of the $\beta$ spectrum can reveal the existence of sterile neutrinos, with minimal or non-minimal interactions \cite{Shrock:1980vy,Derbin:2018dbu,Calaprice:1983qn,Deutsch:1990ut,Bolton:2019pcu,Dekens:2021qch}.

With experimental sensitivity approaching the permille  level, it is crucial to provide comparably accurate theoretical predictions of the $\beta$ spectrum in the SM, including small corrections from the momentum dependence of nuclear matrix elements, electromagnetic corrections, and isospin breaking effects. For light to medium mass nuclei, accurate
calculations of low-energy nuclear observables are currently feasible using the microscopic or {\it ab initio} description of nuclei. Within this approach nuclei and their properties emerge from the underlying nucleonic 
dynamics and ensuing many-nucleon correlations and electroweak currents. 
The first \textit{ab initio} calculation of the $^6$He $\rightarrow$ $^6$Li $\beta$ decay spectrum was performed in Ref. \cite{Schiavilla2002},
using the Variational Monte Carlo method with the Argonne $v18$ two-nucleon potential, supplemented by the Urbana-IX three-nucleon interaction.
Ref. \cite{Glick-Magid:2021uwb,Glick-Magid:2021xty} repeated the calculation in the no-core shell model.
In this paper, we compute both SM and BSM nuclear matrix elements for the $^6$He $\rightarrow$ $^6$Li  decay using Quantum Monte Carlo (QMC) methods~\cite{Carlson2015} to solve for the structure and dynamics of the strongly-correlated many-body problem for nuclei. QMC methods allow to retain the complexity of many-nucleon dynamics whose effects are essential to explain electroweak data in a wide range of energy and momentum transfer~\cite{Carlson:1997qn,Pastore:2008ui,Pastore:2009is,Girlanda:2010vm,Pastore:2011ip,Pastore:2012rp,Datar:2013pbd,Piarulli2013,Pastore:2014oda,Carlson:2014vla,Bacca:2014tla,Baroni:2015uza,Baroni:2016xll,Pastore:2017uwc,Baroni:2018fdn,Schiavilla:2018udt,NevoDinur:2018hdo,Gandolfi:2020pbj, King:2020wmp,King:2021jdb,Andreoli:2021cxo}. Here, we base our calculations on the Norfolk two- and three-nucleon chiral effective field theory potentials and consistent electroweak currents~\cite{Pastore:2008ui,Pastore:2009is,Girlanda:2010vm,Piarulli2013,Baroni:2015uza,Baroni:2016xll,Baroni:2018fdn,Schiavilla:2018udt,Piarulli:2016vel}. 

This paper is structured as follows: In Section~\ref{sec:spectrum}, we
introduce the multipole expansion of the SM weak vector and axial currents and express the differential decay rate with respect to the electron energy including terms up to second order in the multipole expansion.
Section~\ref{sec:vmc} reports on the QMC calculations of the multipoles entering the decay rate and in Section~\ref{sec:results} we discuss the uncertainties of the leading and subleading multipoles and the ensuing theoretical uncertainty in the SM decay rate, which limits the sensitivity to beyond the SM physics. 
We then discuss BSM signatures. In Section \ref{sec:SMEFT} we introduce the effective Lagrangians that mediate $\beta$ decays in the presence of BSM interactions and discuss their corrections to the $\beta$ spectrum.
In Section \ref{BSMplots} we examine the implications of controlling the uncertainty on the spectrum at better than the permille level on non-standard charged-current interactions. We conclude in Section~\ref{sec:conclusions}.
The appendices contain some technical details. 
In Appendix~\ref{app:Lagrangians} we provide a list of standard and non-standard Lagrangians which mediate $\beta$ decays in the Standard Model EFT (SMEFT), 
while a sketch of the derivation of the multipole expansion for these currents is carried out in Appendix~\ref{app:multi}.
Formal expressions of the many-body chiral EFT current operators are given in Appendix~\ref{app:Onebody}.  The expression of the fully differential unpolarized decay rate is given in Appendix~\ref{app:rate}. Appendix~\ref{app:radcorr} is devoted to higher-order electroweak and recoil corrections.

\section{Differential decay rate in the Standard Model}
\label{sec:spectrum}

$\beta$ decays are sensitive to a variety of physics scales, namely, the $\mathcal{Q}$-value of the reaction---typically a few MeVs---which determines the momentum of the outgoing electron and neutrino; the nuclear binding momentum $ \gamma = \sqrt{ m_N B} \sim m_\pi$ where $B$ is the binding energy, which is the relevant scale in the nuclear matrix elements; and $\Lambda_\chi$, the scale at which chiral EFT breaks down. We can take advantage of the scale separation $\mathcal{Q} \ll \gamma \ll \Lambda_\chi$ by organizing the nuclear matrix elements in a double expansion in $\mathcal{Q}/\gamma$ and $\gamma/\Lambda_\chi$, combining  chiral EFT with a multipole expansion 
of the weak matrix elements~\cite{Donnelly:1975ze,Walecka:1995mi}. 
After performing the multipole expansion, the 
differential cross section in the SM  
can be expressed in terms of few matrix elements of the axial and vector charge and current densities,
which are generalized to include scalar, pseudoscalar and tensor currents in the SMEFT.

In the SM, $\beta$ decays are mediated by the exchange of a $W$ boson between purely left-handed quarks, electrons and neutrinos. At scales much smaller than the electroweak,
and focusing on the first generation of quarks,
the effective Lagrangian can be expressed in terms of the local four-fermion interaction
\begin{eqnarray}\label{eq:LagSM}
\mathcal L_{\rm SM} = -\frac{4 G_F}{\sqrt{2}} V_{ud} \bar{e}_L \gamma^\mu \nu_L \bar u_L \gamma_\mu d_L + \textrm{h.c.},
\end{eqnarray}
where $G_F = (\sqrt{2} v^2)^{-1} = 1.166 \times 10^{-5}$ GeV$^{-2}$ is the Fermi constant extracted from muon decay,
$v=246$ GeV is the Higgs vacuum expectation value,  and $V_{ud}$ is the $u$-$d$ element of the Cabibbo-Kobayashi-Maskawa mixing matrix, $V_{ud}= 0.97373(31)$ \cite{Zyla:2020zbs,Hardy:2020qwl}.
The Lagrangian \eqref{eq:LagSM} receives weak and electromagnetic corrections, which we will discuss in the following sections.

At the nuclear level, Eq. \eqref{eq:LagSM} leads to the weak Hamiltonian 
\begin{eqnarray}\label{eq:Hweak2}
H_w =  \frac{G_F}{\sqrt{2}} V_{ud}\int d^3 \textbf{x} \,  j_\mu^{\textrm{ lept}}({\bf x}) \mathcal J_{V-A}^\mu({\bf x}),
\end{eqnarray}
where 
\begin{eqnarray}
j_\mu^{\rm lept} = 2 \bar e_L \gamma_\mu \nu_L.
\end{eqnarray}
and $\mathcal J_{V-A}^\mu$ denotes the hadronic realization of the quark current $\bar u \gamma^\mu (1-\gamma_5) d$. The derivation only assumes that such a realization exist, and we give its explicit representation in chiral EFT in  Appendix \ref{app:Onebody}.
The weak Hamiltonian can be expanded in infinite sum of multipole operators with definite total angular momentum $J$ and parity $\pi$. The transition $^6\textrm{He}(0^+)\rightarrow^6\textrm{Li}(1^+)$  only receives contributions from operators with $J^\pi=1^+$.
The general expression for the 
differential decay rate, derived, for example, in Ref.~\cite{Walecka:1995mi} and reported
in Eq. \eqref{eq:dGammageneral}, then 
contains the multipoles $C_1$, $L_1$, $E_1$,  and $M_1$.
These are defined in terms of the coordinate space charge and current densities in Eqs. \eqref{multiSM1}--\eqref{multiSM4}.
In chiral EFT, a momentum space representation is more convenient, and  $C_1$, $L_1$, $E_1$,  and $M_1$ can be expressed in terms of the axial charge, $\rho({\bf q};A)$, and the vector and axial currents,  $\,{\bf j }({\bf q};V)$ and ${\bf j }({\bf q};A)$ \cite{Schiavilla2002}:
\begin{eqnarray}
C_1(q;A)&=&\frac{i}{\sqrt{4\pi}}\langle {}^6{\rm Li},10\rvert \rho_+^\dagger(q\hat{\bf z};A)\rvert {}^6{\rm He},00\rangle,
\,\label{eq:ME1} \\
L_1(q;A)&=&\frac{i}{\sqrt{4\pi}}\langle {}^6{\rm Li},10\rvert \hat{\bf z}\cdot{\bf j}_+^\dagger(q\hat{\bf z};A)\rvert {}^6{\rm He},00\rangle,\, \label{eq:ME2} \\
E_1(q;A)&=&\frac{i}{\sqrt{2\pi}}\langle {}^6{\rm Li},10\rvert \hat{\bf z}\cdot{\bf j}_+^\dagger(q\hat{\bf x};A)\rvert {}^6{\rm He},00\rangle,\,\label{eq:ME3}\\ 
M_1(q;V)&=&-\frac{1}{\sqrt{2\pi}}\langle {}^6{\rm Li},10\rvert \hat{\bf y}\cdot{\bf j}_+^\dagger(q\hat{\bf x};V)\rvert {}^6{\rm He},00\rangle \, ,  \label{eq:ME4}
\end{eqnarray}
where the momentum carried by the current is ${\bf q}={\bf p}_e+{\bf p}_{\nu}$,
with ${\bf p}_e$ and ${\bf p}_\nu$
the electron and antineutrino momenta,
and $q = |{\bf q}|$.  The subscript  $+$ denotes the charge-changing operators $\rho^\dagger_+=\rho^\dagger_x+i\,\rho^\dagger_y$ and ${\bf j}^\dagger_+={\bf j}^\dagger_x+i\,{\bf j}^\dagger_y$, and the Cartesian and spherical components refer to the isospin space~\cite{Shen2017}. In Eqs. \eqref{eq:ME1}--\eqref{eq:ME4}, the states are characterized by the quantum numbers $J$ and $M_J$, denoting the total angular momentum and the projection along the $z$ axis of the initial and final nuclear states.

At zero momentum, the electric and longitudinal multipoles are related by 
\begin{equation}\label{LErelation}
L^{(0)}_{1}(A)  = \frac{1}{\sqrt{2}} E^{(0)}_1(A) = \sqrt{\frac{3}{4\pi}} g_A\, \, {\rm RME} \, ,
\end{equation}
where the reduced matrix element denotes the standard Gamow-Teller matrix element
\begin{equation}
{\rm RME}=\frac{\sqrt{2\, J_f+1}}{g_A} \,
\frac{\langle J_f M |  \hat{\bf z}\cdot{\bf j }_{\pm}(A) | J_i M\rangle}{\langle J_iM, 10 |J_f M\rangle} = -\sqrt{3} \,
\langle J_f M |  \sigma_z \tau^+ | J_i M\rangle \ ,
\end{equation}
where $\langle J_iM, 10 |J_f M\rangle$ is a Clebsch-Gordan coefficient and, in our case, $J_i=0$ and $J_f=1$.
$M_1$ encompasses the contribution from weak magnetism, while $C_1$ receives contributions from the induced tensor and induced pseudoscalar form factors, $d(q^2)$ and $h(q^2)$
in the notation of Ref. \cite{Holstein:1974zf}.

The momentum $q = |\textbf{q}|$ is limited by the reaction's $\mathcal{Q}$-value, and, being $\mathcal{Q} \ll m_\pi$,  the matrix elements can be expanded in powers of $q/m_\pi$.
From the definitions in Eqs. \eqref{multiSM1}--\eqref{multiSM4}, it can be proven that $L_1$
and $E_1$ only have even powers of $q$, while $C_1$ and $M_1$ odd powers
\cite{Walecka:1995mi}, 
so that Eqs. \eqref{eq:ME1} -- \eqref{eq:ME4} can be expanded as:
\begin{eqnarray}
C_1(q;A)&=& - i \frac{q r_\pi }{3} \left(C^{(1)}_1(A)  - \frac{(q r_\pi)^2}{10} C^{(3)}_{1}(A)  +{\cal O}\left[(q r_\pi)^4\right] \right) \, , \label{Fit1}\\
L_1(q;A)&=& -\frac{i}{3} \left( L^{(0)}_1(A) - \frac{(q r_\pi)^2}{10} L^{(2)}_{1}(A)+{\cal O}\left[(q r_\pi)^4\right] \right)\, , \label{Fit2}\\
M_1(q;V)&=& -i \frac{q r_\pi}{3} \left( M_1^{(1)}(V)  - \frac{(q r_\pi)^2}{10} M^{(3)}_{1}(V)  +{\cal O}\left[(q r_\pi)^4\right] \right) \, ,\label{Fit3}\\
E_1(q;A)&=&-\frac{i}{3} \left( E_1^{(0)}(A)-\frac{(q r_\pi)^2}{10} E_1^{(2)}(A)+{\cal O}\left[(q r_\pi)^4\right]\right)\, , \label{Fit4}
 \end{eqnarray}
where, in order to have dimensionless coefficients, we  introduced  $r_\pi = 1/m_{\pi^+} =  1.41382$ fm. In the equations above, the $C^{(i)}_1$, $L^{(i)}_1$, $M^{(i)}_1$, and $E^{(i)}_1$ are simply the coefficients of the expansion in $qr_\pi$, and they will be determined by the interpolation procedure discussed in Section~\ref{sec:results}.
Their operator definitions in the impulse approximation are given, for example, in Ref.
\cite{Holstein:1974zf}.
Since for the decay under consideration $q r_\pi \lesssim 0.03$,
we consider up to second order terms in $qr_\pi$ to reach the uncertainty goal of $10^{-4}$ in $\beta$ decay spectra. 

After summing over the lepton and nuclear spins, integrating over the neutrino energy and the angle between the electron and neutrino momentum,
the SM decay rate, differential with respect to the electron energy, is given by
\begin{eqnarray}
\frac{d \Gamma}{d \varepsilon} &=& (1+\Delta_R^V)(1+\delta_R(Z, \varepsilon))\frac{G_F^2  W_0^5 V_{ud}^2}{2\pi^3} \sqrt{1 - \frac{\mu_e^2}{\varepsilon^2}}\,  \varepsilon^2 (1 - \varepsilon)^2F_0(Z, \varepsilon)L_0(Z, \varepsilon) S(Z, \varepsilon)R_N(\varepsilon) \nonumber \\
& &\frac{4 \pi}{2 J_i + 1} \frac{1}{9} \Bigg\{ 
3 \left| L_1^{(0)} \right|^2  \left[1 + \alpha ZW_0R\left(\frac{2}{35}-\frac{233}{630}\frac{\alpha Z}{W_0R}-\frac{1}{70}\frac{\mu_e^2}{\varepsilon}-\frac{4}{7}\varepsilon\right)\right]  \nonumber \\
& &+ 2 W_0 r_\pi \Bigg[\left(1 - 2 \varepsilon + \frac{\mu_e^2}{\varepsilon }\right)  {\rm Re} (E_1^{(0)} M_1^{(1) *})- 
\left(1 - \frac{\mu_e^2}{ \varepsilon}\right)
  {\rm Re} (L_1^{(0)} C_1^{(1) *})   \Bigg] \nonumber \\
& & + \frac{(W_0 r_\pi)^2}{3} \Bigg[
\left(
3 - 4 \varepsilon (1 - \varepsilon) - \mu^2_e \frac{2 + \varepsilon}{\varepsilon}
\right) |C_1^{(1)}|^2 
- \frac{3}{5}\left( 1 - \frac{\mu_e^2}{\varepsilon} (2 - \varepsilon)\right)  \textrm{Re} \left( L_1^{(0)} L_1^{(2) *} \right) \nonumber \\ & & +  \left(3 - 10 \varepsilon (1 - \varepsilon) + \mu_e^2 \frac{4 - 7 \varepsilon}{\varepsilon}
\right) \left( \left|M_1^{(1)}\right|^2 - \frac{1}{5} \textrm{Re}\left( E^{(0)}_{1} E^{(2)}_{1}\right) \right) \Bigg] \nonumber \\
& & - \frac{4}{7}  \frac{\alpha Z W_0 r_\pi^2}{R}  (1- \varepsilon) \left(\frac{E_1^{(0)}E_1^{(2)}}{2}-L_1^{(0)}L_1^{(2)}\right)\Bigg\}, \label{rate}
\end{eqnarray}
where we introduced the scaled variables $E_e = W_0 \varepsilon$ and $m_e = W_0 \mu_e$, with $W_0 = M_i - M_f = 4.016$ MeV in the case of the 
$^6$He -$^6$Li transition.
$M_i$ and $M_f$ denote the masses of the initial and final state nucleus. Eq. \eqref{rate} is accurate up to corrections of $\mathcal O((W_0 r_\pi)^3)$. In Eq. \eqref{rate}, 
we kept the effects of nuclear recoil at leading order in $W_0/M_f$. The other effect of nuclear recoil is that the electron endpoint energy shifts from $E_e = W_0$ to  $E_e = W_0 - \frac{W_0^2 - m_e^2}{2 M_f}$. 

In addition to the leading terms in the multipole expansion, Eq. \eqref{rate} includes electromagnetic effects, which are not negligible at the precision we are working. The most important contribution is from the Fermi function, given by $F_0(Z,\varepsilon)$ 
\begin{eqnarray}
F_0(Z,\varepsilon) = 4 (2 |{\bf p}_e| R )^{2(\gamma_0 - 1)} \frac{|\Gamma(\gamma_0 + i y)|^2}{|\Gamma(2\gamma_0+1)|^2} e^{\pi y}, \quad \gamma_0 = \sqrt{1- (\alpha Z)^2}, \quad y = \frac{\alpha Z }{|v_e|},
\label{eq:Fermi_function}
\end{eqnarray}
where $R = \sqrt{5/3 \langle r^2_{\rm ch} \rangle}$, 
and $\langle r^2_{\rm ch} \rangle$
is the charge radius of $^6$Li, $\sqrt{\langle r^2_{\rm ch} \rangle} = 2.5890(390)$ fm \cite{Tilley:2002vg}. 
$Z$ is the charge of the daughter nucleus and the electron velocity $v_e = |{\bf p}_e|/E_e$.
Other large corrections arise from the radiative corrections 
$\Delta_R^V $ and $\delta_R(Z,\varepsilon)$, which contribute to the half-life at the level of few percent. These and other higher order corrections, encoded in the functions $L_0(Z,\varepsilon)$, $S(Z,\varepsilon)$ 
$R_N(\varepsilon)$ and in the explicit $\mathcal O(\alpha)$ and $\mathcal O(\alpha^2)$ terms in Eq. \eqref{rate}, are discussed in Appendix \ref{app:radcorr}. In this work, we did not attempt a rederivation of electromagnetic corrections in chiral EFT and instead followed closely the literature, as summarized in Ref.  \cite{Hayen2018}.
To assess the importance of higher-order terms in the multipole expansion, we define the leading contribution to the spectrum as \begin{equation}\label{Gamma0}
\frac{d \Gamma_0}{d\varepsilon}  =
\frac{G_F^2  W_0^5 V_{ud}^2}{2\pi^3} \sqrt{1 - \frac{\mu_e^2}{\varepsilon^2}}\,  \varepsilon^2 (1 - \varepsilon)^2 \, \frac{4 \pi}{3}  
 \left| L_1^{(0)} \right|^2 F_0(Z, \varepsilon)L_0(Z, \varepsilon) S(Z, \varepsilon)R_N(\varepsilon),
\end{equation}
where we included in the definition of 
${d \Gamma_0}/{d\varepsilon}$ 
some radiative corrections, for which we just use results in the literature.

\section{Theoretical Framework}
\label{sec:theory}

\subsection{Quantum Monte Carlo Methods}
\label{sec:vmc}

In this work, we employ Quantum Monte Carlo methods~\cite{Carlson2015}---both the Variational (VMC) and the Green's function Monte Carlo (GFMC) methods---and the Norfolk chiral effective field theory many-body interactions and electroweak currents~\cite{Piarulli:2016vel,Piarulli2013,Baroni:2018fdn,Baroni:2015uza,Baroni:2016xll} to evaluate the required nuclear matrix elements. This computational scheme has been most recently described in Refs.~\cite{King:2020wmp,King:2021jdb} where some of the present authors evaluated Gamow-Teller matrix elements entering $\beta$ decays and electron captures in light nuclei as well as muon capture rates in $A=3$ and $6$ nuclei. Here, we will not provide the details of the computational method nor the interactions. We will instead limit ourselves to briefly summarize the salient points of the calculation and defer the interested reader to Ref.~\cite{King:2020wmp} and references therein for additional technicalities. 

The Norfolk potentials consist of local two- and three-nucleon interactions formulated in configuration space, and derived from a chiral effective field theory that retains, in addition to nucleons and pions, $\Delta$-isobars as explicit degrees of
freedom~\cite{Piarulli2013,Piarulli:2016vel,Baroni:2016xll,Baroni:2018fdn}. They are denoted below as NV2+3, where the two-body interaction (NV2) is constructed up to N$^3$LO in the chiral expansion, and the three-body force (NV3) retains up to N$^2$LO contributions \footnote{
For the nuclear force, we use the nomenclature 
that is standard in the
literature adopting Weinberg's power counting.
Namely, NLO, N$^2$LO and N$^3$LO denote, respectively, corrections scaling as $\mathcal O(Q^2)$, 
$\mathcal O(Q^3)$ and $\mathcal O(Q^4)$, where $Q = m_\pi/\Lambda_\chi$ is the expansion parameter of chiral EFT.}.

In the QMC calculation, theoretical uncertainties arise from deficiencies in the nuclear wave function, i.e. from how well the QMC wave function reproduces the actual ground states for a given nuclear interaction, and 
from the nuclear interactions themselves. 
In chiral EFT, the uncertainties in the nuclear interactions stem from the two- and three-nucleon data used to determine the unknown low-energy constants (LECs) in the nuclear potential and currents,
from the residual dependence of observables on the cutoff used in the calculation and from the truncation error arising from working at a finite order in the chiral expansion.
In order to assign a theoretical error to our estimates, we perform the calculations using four models of Norfolk interactions. These models differ in the cutoff utilized to regularize divergences, in the number of nucleon-nucleon scattering data used to constrain the LECs entering the NV2 interaction, and in the fitting procedure implemented to constrain the NV3 interaction. In particular, NV2+3 models belonging to class I (denoted with NV2+3-I) are fitted up to 125 MeV and use $\sim$2700 data points, while the NV2+3-II models are fitted up to 200 MeV and use $\sim$3700 data points. For each class, two different sets of cutoff are implemented. Specifically, the coordinate space regulators are
\begin{eqnarray}
C_{R_S}(r) = \frac{1}{\pi^{3/2} R_S^3} e^{-\left(r/R_S\right)^2}, \qquad C_{R_L}(r) = 1 - \frac{1}{ (r/R_L)^6 e^{2(r-R_L)/R_L} + 1} \, , 
\end{eqnarray}
where $C_{R_L}$ regulates divergences at $r\sim 0$ in the pion exchange potential, while contact interactions are regulated by $C_{R_S}$.
Models labeled with an `a' use the combination ($R_S$, $R_L$)=(0.8, 1.2) fm (models NV2-Ia and NV2-IIa), while those labeled with a `b' use 
($R_S$, $R_L$)=(0.7, 1.0) fm (models NV2-Ib and NV2-IIb). The NV2 models are supplemented by a three-body force at N$^2$LO. At this order, there are two LECs characterizing the NV3's strength. They are determined  by a simultaneous fit to either the trinucleon binding energy and the triton beta decay reduced matrix element~\cite{Baroni:2018fdn} or the trinucleon binding energy and the $nd$ scattering length~\cite{Piarulli:2017dwd}. Norfolk models based on the former procedure are denoted with a `*', that is NV2+3$^*$.

Nuclear wave functions are constructed in two steps. First, a trial variational Monte Carlo (VMC) wave function $\Psi_T$, which accounts for the effect of the nuclear interaction via the inclusion of correlation operators, is generated by minimizing the energy expectation value with respect to a number of variational parameters. The second step improves on $\Psi_T$  by eliminating excited state contamination. This is accomplished in a Green’s function Monte Carlo
(GFMC) calculation which propagates the Schr\"{o}dinger equation in imaginary time $\tau$.
The propagated wave function $\Psi(\tau) = \exp\left[-(H-E_0)\tau\right]\Psi_T$, for large values of $\tau$, converges to the exact
wave function with eigenvalue $E_0$. Ideally, the matrix elements should be evaluated in between
two propagated wave functions. In practice, we evaluate mixed estimates in which only one wave function is
propagated, while the remaining one is replaced by $\Psi_T$. The calculation of diagonal and off-diagonal matrix elements is discussed at length in Refs.~\cite{Carlson:2014vla} and~\cite{Pervin:2007}.
We will present both VMC and GFMC results. As discussed in Section \ref{UQ}, while the latter are more accurate and are in excellent agreement with the experimental half-life, the two calculations of the spectral shape
show minimal differences, well below the 10$^{-3}$ level, justifying the use of the numerically cheaper VMC in future studies.

\subsection{Power counting and  many-body electroweak currents}\label{pwc}

Accompanying the Norfolk many-body interactions are one- and two-body axial and vector currents derived within the same chiral effective field theory formulation with pions, nucleons and $\Delta$'s~\cite{Baroni:2018fdn,Baroni:2015uza,Baroni:2016xll,Schiavilla:2018udt,Pastore:2008ui,Pastore:2009is}.  
We use the axial and vector charges, $\rho(A)$ and $\rho(V)$, and currents, ${\bf j}(A)$ and ${\bf j}(V)$, to evaluate the SM multipoles 
of  Eqs.~\eqref{eq:ME1}--\eqref{eq:ME4}. The current operators are arranged in powers of a second expansion parameter, namely  
$m_\pi/\Lambda_\chi$ or equivalently $|{\bf p}|/\Lambda_\chi$, where $|{\bf p}|\sim m_\pi$ denotes a typical nuclear physics
scales such as the binding momentum, the inverse of the nuclear radius   $R_A = 1.2\, A^{1/3}$ fm (with $A$ being the mass number), 
or the typical nucleons' momenta inside nuclei. 
As a consequence, the coefficients of the $q$-expansion in Eqs. \eqref{Fit1} - \eqref{Fit4}, {\it i.e.}, the multipoles $C^{(i)}_1$, $L^{(i)}_1$, $M^{(i)}_1$, and $E^{(i)}_1$, 
have a second expansion in  $Q^n$, where $Q \equiv m_\pi/\Lambda_\chi$ is the chiral expansion parameter.
We can then express any of the multipoles---here generically denoted with $\mathcal{M}^{(i)}$---as 
\begin{equation}\label{eq:multichiral}
    \mathcal{M}^{(i)} = \sum_{n=0}^{\infty} \, \mathcal{M}^{(i,\, n)}\, . 
\end{equation}
where the superscripts $i$ and $n$ indicate the orders of  the multipole and chiral expansions, respectively.

The scaling in $Q^n$ of the chiral electroweak currents derived in Refs.~\cite{Baroni:2018fdn,Baroni:2015uza,Baroni:2016xll,Schiavilla:2018udt,Pastore:2008ui,Pastore:2009is} is reported in Table~\ref{tb:scalingQA}. The aforementioned references adopt a different convention where the counting is carried out in powers of $Q^\nu$ with $\nu=n-3$. 
Without going too much into details, the operators consist of one-body contributions obtained from the non-relativistic reduction of the covariant 
axial and vector nucleonic four-vector currents. We denote with `1b(NR)' and `1b(RC)' leading order and first order terms in the non-relativistic expansion.  Two-body currents include 
contributions of one- and two-pion range (OPE and TPE) as well as short-range currents encoded in contact-like operators (CT). OPE currents involving nucleons' virtual excitations into a $\Delta$ are denoted with `OPE-$\Delta$' while those involving sub-leading terms in the pion-nucleon chiral Lagrangian are denoted with `OPE(sub)'. In addition, there are the so called pion-pole contributions where the external field couples with a pion that is then absorbed by a nucleon. These operators are schematically represented in Fig.~\ref{fig:AVcurr} while their formal expressions are listed for convenience in Appendix~\ref{app:Onebody}. Details on the derivation of the currents can be found in Refs.~\cite{Baroni:2018fdn,Baroni:2015uza,Baroni:2016xll,Schiavilla:2018udt,Pastore:2008ui,Pastore:2009is}.

\begin{center}
\begin{table}[t]
\begin{tabular}{c|c|c|c|c|c}
\hline
\hline
 \multirow{2}{*}{{\rm Operator}}     &  {\rm LO}         &  {\rm NLO} & {\rm N$^2$LO}       & {\rm N$^3$LO}   &  {\rm N$^4$LO} \\
     &  $\nu= -3$        & $\nu= -2$  & $\nu= -1$                              &   $\nu= 0$   & $\nu = 1$\\
\hline
 ${\bf j}^\nu(A)$   &  {\rm 1b(NR)} &  --  & {\rm 1b(RC)}      & {\rm OPE(sub)} & $\times$ \\
                    &               &    & {\rm OPE-$\Delta$} & {\rm CT} \\
                    \hline
multipole & $\{ L_1, E_1\}^{(i,0)}$ & $\{ L_1, E_1\}^{(i,1)}$ & $\{ L_1, E_1\}^{(i,2)}$ & $\{ L_1, E_1\}^{(i,3)}$                  & $\{ L_1, E_1\}^{(i,4)}$            \\        
\hline \hline
$\rho^\nu(A)$  & --    &  {\rm 1b(RC)} & {\rm OPE}         & --  & $\times$ \\
\hline                     
multipole  & $C_1^{(i,0)}$ & $C_1^{(i,1)}$ & $C_1^{(i,2)}$ & $C_1^{(i,3)}$ & $C_1^{(i,4)}$\\
\hline
\hline
 ${\bf j}^\nu(V)$    & --  &  {\rm 1b(NR)} & {\rm OPE}   & {\rm 1b(RC)} & {\rm \,\,\,OPE(sub)}      \\
  &              &             &   {\rm OPE-$\Delta$}         & {\rm \,\,\,TPE }  \\
                    &              &             &             & {\rm CT}  \\
 \hline 
multipole  & $M_1^{(i,0)}$ & $M_1^{(i,1)}$ & $M_1^{(i,2)}$ & $M_1^{(i,3)}$ & $M_1^{(i,4)}$   \\
\hline\hline
\end{tabular}
\caption{Scaling in Q$^\nu=Q^{n-3}$ up to $\nu=1$ of the chiral axial~current, ${\bf j}^\nu(A)$, 
and charge, $\rho^\nu(A)$, operators and of the vector current operator, ${\bf j}(V)$.  The acronyms stand for 1b=one-body, OPE = one-pion-exchange,
TPE = two-pion exchange, 
NR = non-relativistic, RC = relativistic correction, OPE-$\Delta$ = one-pion-exchange currents with an intermediate delta excitation, and sub=sub-leading. ``--'' indicates that no contribution exists at that order, while ``$\times$'' that contributions of that order have not been included.
We also indicate to which multipole operator, 
and at which order in the chiral expansion,
each term contributes. See text for explanation. 
}
\label{tb:scalingQA}
\end{table}
\end{center}

\begin{figure}[tbh]
\begin{center}
\includegraphics[width=0.7\textwidth]{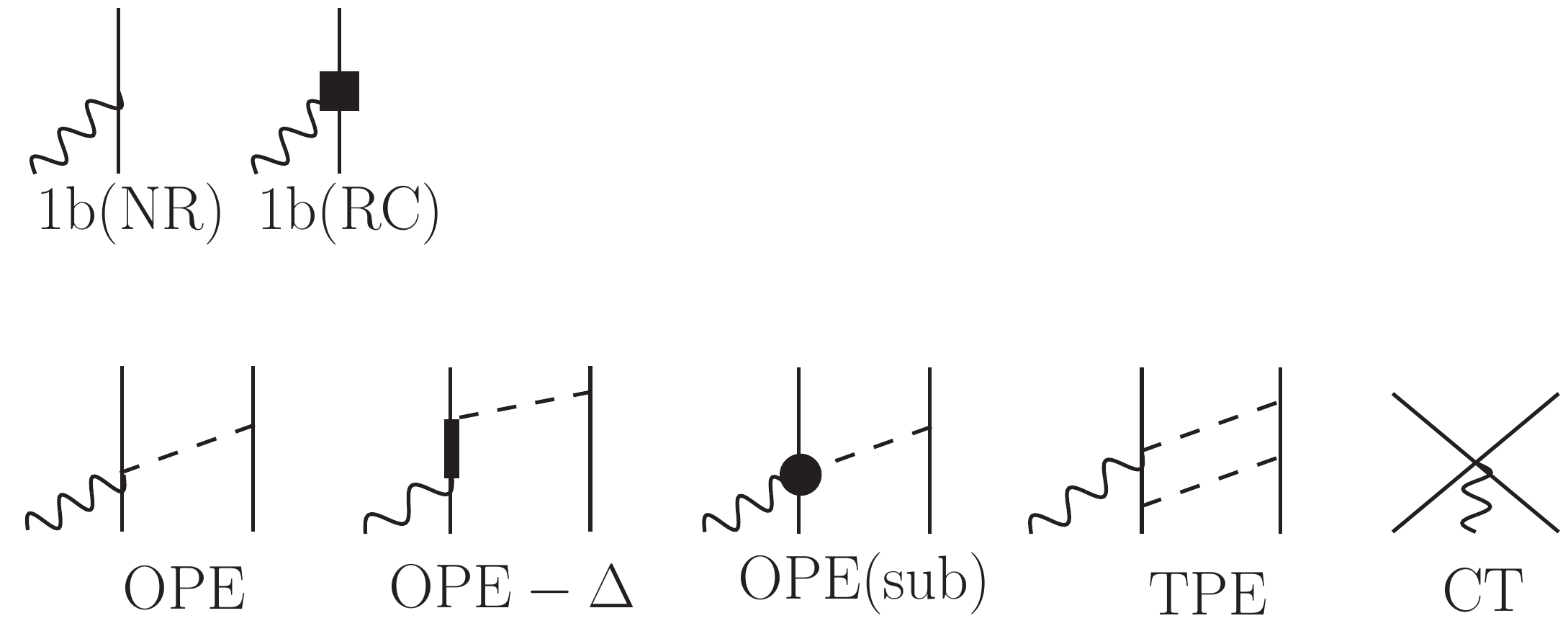}
\end{center}
\caption{Schematic representations of the types of contributions entering  the one- and two-body electroweak currents from Refs.~\cite{Baroni:2018fdn,Baroni:2015uza,Baroni:2016xll,Schiavilla:2018udt,Pastore:2008ui,Pastore:2009is} adopted in this work. Solid, dashed and wavy lines represent nucleons, pions, and axial and vector external fields. The square and the dot represent the relativistic corrections to the leading one-body  operators and sub-leading terms in the pion-nucleon Lagrangian, respectively, while the thick line represents a $\Delta$ intermediate state. See Table~\ref{tb:scalingQA}  and Appendix~\ref{app:Onebody} for the operators' scaling and formal expressions. Not shown in the figure are the pion-pole and  tadpole diagrams (see Refs.~\cite{Baroni:2018fdn,Baroni:2015uza,Baroni:2016xll,Schiavilla:2018udt,Pastore:2008ui,Pastore:2009is} for details.) }
\label{fig:AVcurr}
\end{figure}

For our  discussion, it is sufficient to focus on the leading order terms of both the charge and current operators.  The $E_1^{}$ and $L_1^{}$ multipoles
are proportional to matrix elements of the axial current,
which, at leading order in the chiral expansion is given by the usual Gamow-Teller and pion-pole contributions
(see Eq.~\eqref{eq:axialLO} in  Appendix \ref{app:Onebody}). 
This implies that 
\begin{equation}
E_1^{(i,0)}\,\, {\rm and}\,\, L_1^{(i,0)} \sim \mathcal O(Q^0) = \mathcal O(1)\, ,  \end{equation}
as can be inferred form Table~\ref{tb:scalingQA}. 
In particular, $E_1^{(0,0)}$ and $L_1^{(0,0)}$ are determined by the
zero-momentum Gamow-Teller matrix element in Eq. \eqref{LErelation}.
As illustrated in Table \ref{tb:scalingQA},
two-body axial currents first contribute to $E_1^{(0,2)}$ and $L_1^{(0,2)}$ with the OPE-$\Delta$ term, while subleading OPE diagrams and contact interactions  to 
$E_1^{(0,3)}$ and $L_1^{(0,3)}$.  With the interactions and axial current that we use,  $E_1$ and $L_1$ are accurate up to order $n=3$ $(\nu=0)$ in the chiral expansion.
The $E_1^{(2,0)}$ and $L^{(2,0)}_1$ multipoles are also non-zero, and they are related to the square radius of the Gamow-Teller matrix element, for which the relevant scale is the system's binding momentum.  The one-body induced pseudoscalar form factor, which, as shown in Eq. \eqref{eq:axialLO}, in momentum space scales as $q^2/m_\pi^2$, gives 
additional LO contributions to  $L_1^{(2)}$.
The momentum dependence of the nucleon axial form factor, on the other hand, is suppressed and contributes 
at N$^2$LO in the chiral expansion to both the $E^{(2,2)}_1$ and $L^{(2,2)}_1$ multipoles. The expressions of  the nucleon axial  form factor used in the currents is given in Appendix \ref{app:Onebody}.

The $C_1$ multipole is  induced by the axial charge,
which, as shown in Table~\ref{tb:scalingQA},  at lowest order receives one-body contributions from the 
non-relativistic expansion of the axial form factor,
starting at $\mathcal O(1/m_N)$, and from the induced pseudoscalar form factor. As shown in Appendix \ref{app:Onebody}, the latter contribution is proportional to the electron endpoint energy $W_0$,
which, for power counting purposes, scales as $W_0 \sim O(m_N Q^2)$.
As a consequence,
\begin{eqnarray}
C^{(1,0)}_1(A) \sim  0, \qquad{\rm and}\qquad C^{(1,1)}_1(A) \sim \mathcal O(Q^1)\, .
\end{eqnarray}
We then expect $C^{(1,1)}_1(A)$ to be suppressed with respect to the $L^{(i)}_1$ and $E^{(i)}_1$ multipoles. The two-body axial charge operator scale as $Q^2$ (see Table~\ref{tb:scalingQA})  and thus contributes to  the  $C^{(1,2)}_1$ multipole.   

Finally, the magnetic multipole is induced at the one-body level by the weak magnetic form factor, or equivalently by the one-body vector current at leading order---see Eq.~\eqref{eq:magnetic}  and Table~\ref{tb:scalingQA}---and thus also in this case
\begin{eqnarray}
M^{(1,0)}_1(V) \sim  0, \qquad
M^{(1,1)}_1(V) \sim \mathcal O(Q^1)\, ,
\end{eqnarray}
even though the large isovector anomalous magnetic moment $\kappa_V = \kappa_p - \kappa_n  \sim 3.7$ enhances the formally NLO contribution. 
Two-body currents start to contribute to $M^{(1,2)}_1(V)$ and are found to provide a $6\%~{\rm to}~8\%$ contribution
to the overall matrix element.  

We can use the counting of currents and potentials to give a rough estimate of the truncation error we expect for different multipoles. The coordinate space cut-off
$R_S$ can be converted into a scale $\Lambda_\chi = 2/R_S$, and, for the potentials used in this work $\Lambda_\chi \sim 500 - 550$ MeV. 
For the multipoles $L_1$,  
$E_1$ and $M_1$, the first term that is missed in the calculation
is of order $Q^4 = 0.6\%$, taking $\Lambda_\chi =500$ MeV. This error is, as we will see, smaller than the uncertainties arising from using different models.
For $C_1$, the first missing term in the chiral expansion is $\mathcal O(Q^3) \sim 2\%$. These estimates are merely indicative. This 
type of estimation has been the standard procedure in nuclear physics;
however, for a more robust estimate, one should develop  
Bayesian methods to quantify the uncertainties associated with 
parameters entering the many-body calculations and truncation errors.
Work along this line
is being vigorously pursued by the community \cite{Wesolowski:2018lzj,Melendez:2019izc}, and is beyond the scope of this work.

\section{Standard Model Results and Spectrum}
\label{sec:results}

\begin{figure}[t]
\begin{center}
\includegraphics[width=0.75\textwidth]{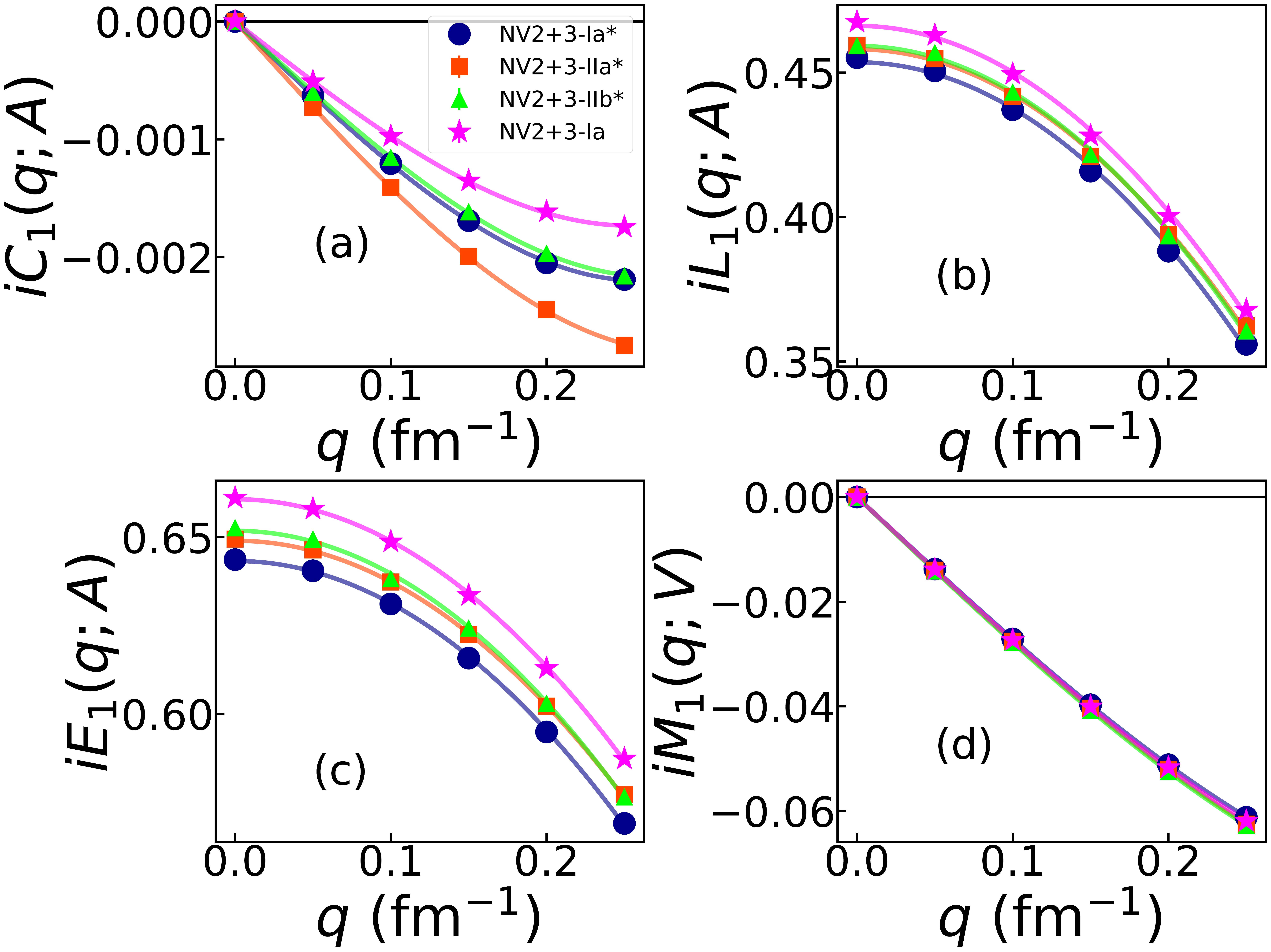}
\end{center}
\caption{VMC multipoles for the NV2+3-Ia$^*$ (blue circles), NV2+3-IIa$^*$ (orange squares), NV2+3-IIb$^*$
(green triangles)
and NV2+3-Ia (magenta stars)
models. The curves of best fit for each case are shown in the same color as the multipoles. 
Statistical errors from the Monte Carlo are included on each point but are too small to be visible in the
figure.
}
\label{fig:all.multipoles}
\end{figure}

In order to determine the matrix elements entering the $\beta$ spectrum in Eq. \eqref{rate}
we calculate  Eqs.~\eqref{eq:ME1}--\eqref{eq:ME4} for six momenta between 0 and $0.25$ fm$^{-1}$ (0 and $\sim$50 MeV),
and we fit them to the functional forms in Eqs.~\eqref{Fit1}-\eqref{Fit4}.
To have a realistic assessment of the theoretical uncertainty, we performed the calculation with four sets of chiral potentials and currents,  NV2-3-Ia$^*$, NV2+3-IIa$^*$,  NV2+3-IIb$^*$, and NV2-3-Ia, using both VMC and GFMC methods.

Figure~\ref{fig:all.multipoles} shows the 
VMC multipoles and the associated curves of best fit for the NV2-3-Ia$^*$, NV2+3-IIa$^*$,  NV2+3-IIb$^*$, and NV2-3-Ia models retaining one- and 
two-body vector and axial current operators. 
The expansion coefficients obtained by fitting VMC multipoles obtained with one-body and 
one- and two-body operators are listed in Table \ref{tab:Multipoles}, where the error denotes the fitting error. Two-body currents have a minor effect on the $L_1$ and $E_{1}$ multipoles, leading to a shift in $L^{(0)}_{1}$, $E^{(0)}_1$, $L^{(2)}_{1}$, $E^{(2)}_1$  of at most $\sim 2\%$, in the case of model Ia. As expected from power counting, two-body currents are more important for $M_1$ and $C_1$. In both cases, two-body currents contribute at $\mathcal O(Q)$ compared to the LO. We see that $M^{(1)}_1$  receives an $\sim 8\%$ correction, for all the models considered here.  In VMC, the corrections to $C^{(1)}_1$ are about 30-40\%.

\begin{table}[t]
    \centering
    \resizebox{\linewidth}{!}{%
    \begin{tabular}{||c c|| r | r || r | r ||r | r || r | r||}
\hline \hline
        &            & \multicolumn{2}{c||}{Model Ia$^*$} & \multicolumn{2}{c||}{Model IIa$^*$}  & \multicolumn{2}{c||}{Model IIb$^*$} & \multicolumn{2}{c||}{Model Ia} \\  
    \hline 
    & & 1b & 1b+2b & 1b & 1b+2b & 1b & 1b+2b &1b &1b+2b\\   
    \hline      \hline 
    $L_1^{(0)}$     &         & $1.3578(3)$    &   $1.3607(4) $  & $1.3662(2)$  & $1.3742(3)$  & $1.3717(2)$ &  $1.3777(1)$ & $1.3641(3)$ & $1.3986(3)$ \\
    $E_1^{(0)}$ &         &  $1.9255(4)$  &   $1.9299(5)$ & $1.9355(2)$  & $1.9471(3)$  & $1.9470(3)$& $1.9555(1)$ &$1.9333(4)$ &$1.9824(4)$\\
    $M_1^{(1)}$  &     &  $-0.5487(1)$    &   $-0.5860(2)$   & $-0.5510(1)$ & $-0.5952(2)$  & $-0.5550(1)$ & $-0.60111(4)$ &$-0.5521(1)$ &$-0.5908(2)$ \\
    $C_1^{(1)}$  &     &  $-0.0182(3)$  &   $-0.0269(3)$ & $-0.0217(3)$  & $-0.0311(3)$ &$-0.0151(3)$ & $-0.0257(3)$ &$-0.0128(3)$ & $-0.0218(3)$ \\
    $L_1^{(2)}$    &    & $23.87(5)$        &  $23.93(6)$        & $23.51(3)$ & $23.51(4)$ & $23.94(3)$& $24.07(1)$ &$23.64(5)$ &$24.12(5)$\\
    $E_1^{(2)}$ &  &  $17.86(7)$       &  $17.99(8)$       & $17.35(5)$ & $17.49(6)$ & $18.07(5)$ & $18.22(2)$ &$17.45(7)$ & $17.80(7)$\\
    \hline \hline
    \end{tabular}}
    \caption{ Expansion coefficients of the  VMC multipole operators, including only one-body currents (1b) or one- and two-body currents (2b).
    The four columns denotes four different NV interactions, as discussed in the text.
    The error denotes the fitting error.}
    \label{tab:Multipoles}
\end{table}

\begin{figure}[tbh]
\begin{center}
\includegraphics[width=0.75\textwidth]{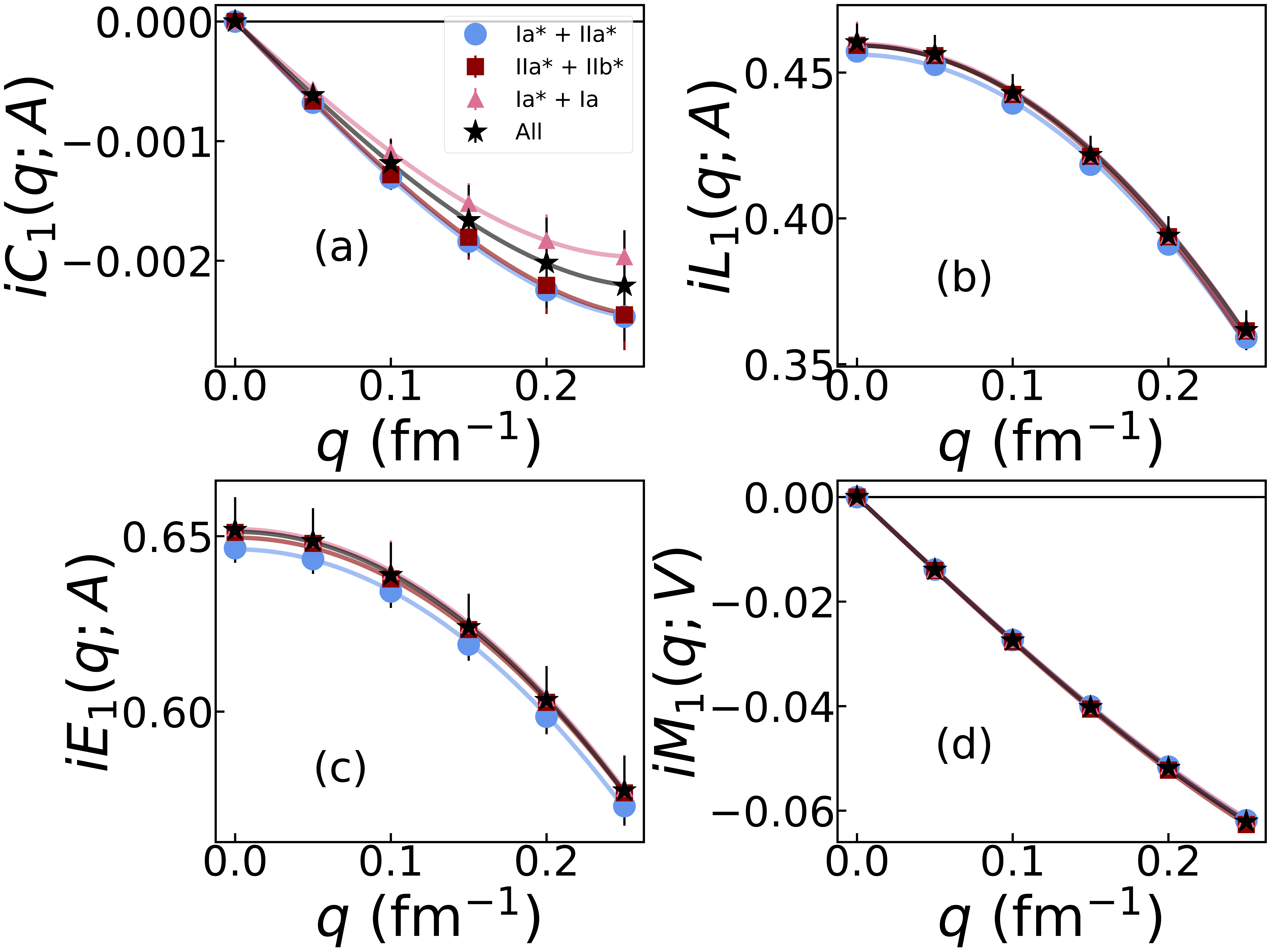}
\end{center}
\caption{Average VMC multipoles for the four NV2+3 models under study. Averages were obtained for NV2+3 potentials with the same cutoff (blue circles), the same energy range of $NN$ scattering data used to fit the
interaction (red squares), the same $NN$ but different $3N$ force (pink triangles), and for all models (black stars). The curves of best fit for each case are shown in the same color as the average multipoles. Details on how the error bars are obtained in each case are
provided in the text.}
\label{fig:avg.multipoles}
\end{figure}

\subsection{Uncertainty Estimation}\label{UQ}

\subsubsection{Variational Monte Carlo}

The expansion coefficients for the NV2+3-Ia$^*$, NV2+3-IIa$^*$, NV2+3-IIb$^*$, and NV2+3-Ia models are presented in
Table \ref{Results}. From Figure \ref{fig:all.multipoles} and Table \ref{tab:Multipoles}, it is clear that 
there will be some degree of uncertainty due to the choice of model. To
account for the model uncertainty, four sets of average multipoles were obtained: An average between
two models with the same cutoff,  same determination of  
the three-body force 
and fit to different range of $NN$ scattering data (NV2+3-Ia$^*$ and NV2+3-IIa$^*$), an average between two models fit to the same 
range of $NN$ scattering data, same determination of the three-body force, and different cut-off (NV2+3-IIa$^*$ and NV2+3-IIb$^*$), an average between two models with consistent NV2 and cut-off but different NV3 interactions (NV2+3-Ia$^*$ and NV2+3-Ia), and an average of all four models. 
These average multipoles were fit to the functional forms of Eqs. \eqref{Fit1} - \eqref{Fit4} with the two
model calculations at each $q$ providing the upper and lower theoretical error bar. The statistical uncertainty
of these averaged fits provides the estimated uncertainty on the expansion coefficients due to the energy range of the fit
($\epsilon_E$), the choice of cutoff ($\epsilon_C$), and the three-body 
force fitting procedure ($\epsilon_{3N}$). 

\begin{table}[t]
\begin{center}
\begin{tabular}{|| l || c | c | c | c | c | c ||}
\hline \hline
Model & $C_1^{(1)}$ & $L_1^{(0)}$ & $L_1^{(2)}$ & $E_1^{(0)}$ & $E_1^{(2)}$ & $M_1^{(1)}$ \\  \hline \hline
 Ia*  & $-0.027$ & $1.361$ & $23.93(6)$ & $1.923$ & $17.99(8)$ & $-0.586$ \\ 
 IIa* & $-0.031$ & $1.374$ & $23.51(4)$ & $1.947$ & $17.49(6)$ & $-0.595$ \\
 IIb* & $-0.026$ & $1.378$ & $24.07(1)$ & $1.955$ & $18.22(2)$ & $-0.601$ \\ 
 Ia   & $-0.022$ & $1.399$ & $24.12(5)$ & $1.982$ & $17.80(7)$ & $-0.591$ \\\hline \hline
 Average &  $-0.026$ & $1.378$ & $23.869$ & $1.954$ & $17.847$ & $-0.593$\\ \hline \hline
 $\epsilon_E(\%)$ & 5.4 & 0.3 & 3.2 & 0.3 & 5.6 & 0.6\\
 $\epsilon_C(\%)$ & 6.9 & 0.01 & 0.3 & 0.07 & 0.8 & 0.3 \\ 
 $\epsilon_{3N}(\%)$ & 7.2 & 0.8 & 7.0 & 0.8 & 13.9 & 0.3 \\
 $\epsilon_{\rm tot}(\%)$ & 11.8 & 0.8 & 7.8 & 0.9 & 15.0 & 0.7 \\
 \hline \hline
\end{tabular}
\end{center}
\caption{Summary of the values of the expansion coefficients 
for the charge ($C_1$), longitudinal ($L_1$), electric ($E_1$), and magnetic ($M_1$) 
VMC multipoles for the NV2+3 models under study. The percent error due to the cutoff $\epsilon_{C}(\%)$, 
energy range of the fit $\epsilon_{E}(\%)$, the three-body force fit  $\epsilon_{3N}(\%)$,
and the total error $\epsilon_{\rm tot}(\%)$ are also presented. Details of 
how the average expansion coefficients are obtained are provided in the text. The uncertainty
on coefficients for individual models is fitting error only. The error is $<0.001$ unless 
otherwise noted.}
\label{Results}
\end{table}

For the fourth case described above, a similar procedure was followed; however, when assigning
the error for the multipoles at each value of $q$, the uncertainty was not taken as simply the spread of the
model calculations. Instead, we summed in quadrature the uncertainties on each point from the other
three average fits to combine the cutoff, energy range, and three-body force uncertainties. The expansion coefficients
obtained from this fit are the ``Average'' results in Table \ref{Results} and their statistical 
uncertainties are cited as $\epsilon_{\rm tot}$. Figure \ref{fig:avg.multipoles} shows the averaged multipoles 
and curves of best fit for these three cases.

As evidenced by the results in Table \ref{Results}, the coefficient $E_1^{(2)}$ has the 
largest model uncertainty at $15\%$. 
The next largest uncertainties are on the coefficients $C_1^{(1)}$ at $11.8\%$ and
$L_1^{(2)}$ at $7.8\%$. The remaining uncertainties are $\lesssim 1\%$ for $L_1^{(0)}$, $E_1^{(0)}$, and $M_1^{(1)}$.
The main driver of the uncertainty in most of the coefficients is the choice of three-body force, though for $C^{(1)}_1$ and $M^{(1)}_1$ the uncertainties from this source and the others are comparable. In fact,
it is the energy range of scattering data used to fit the 
interaction that provides the largest uncertainty on $M^{(1)}_1$. 

Another way that one might obtain the average expansion parameters with an uncertainty would be to instead 
use the results of Table \ref{tab:Multipoles}. The average value of the expansion coefficients for the
four models differ from the values obtained in our procedure by less than one percent.
Then, one could estimate the uncertainty due 
to the energy range used to fit $NN$ interaction from the difference between the NV2+3-Ia$^*$ and NV2+3-IIa$^*$
results, the uncertainty due to the cutoff from the difference between the NV2+3-IIa$^*$ and NV2+3-IIb$^*$ results, and the uncertainty due to the three-body force from the difference between the NV2+3-Ia$^*$ and NV2+3-Ia results. 
Following this procedure, the error on the leading order expansion coefficients increase by factors of $1.5$ to $1.8$. The error on the dominant contributions would be $\lesssim 1.4\%$ in this approach.
For the coefficients $L_1^{(2)}$ and $E_1^{(2)}$, this scheme reduces the uncertainty by a factors of $2.6$ and $3.0$, respectively. However, the approach that we took to obtain the uncertainty is more reasonable as each matrix
element should have its own model uncertainty that then propagates to the expansion coefficients in the fitting procedure.

\subsubsection{Green's Function Monte Carlo}

\begin{table}[t]
\begin{center}
\begin{tabular}{|| l || l || c || c || c || c || c || c ||}\hline \hline
Model & Method & $C_1^{(1)}$ & $L_1^{(0)}$ & $L_1^{(2)}$ & $E_1^{(0)}$ & $E_1^{(2)}$ & $M_1^{(1)}$ \\  \hline \hline
Ia*  &VMC  & $-0.027$ & $1.361$ & $23.93(6)$ & $1.930$ & $17.99(8)$ & $-0.586$ \\
     &GFMC & $-0.047$ & $1.308$ & $23.98(6)$ & $1.856$ & $19.71(9)$ & $-0.559$ \\ \hline
IIa* &VMC  & $-0.031$ & $1.374$ & $23.51(4)$ & $1.947$ & $17.49(6)$ & $-0.595$ \\
     &GFMC & $-0.045(2)$ & $1.341(7)$ & $25(1)$ & $1.901(9)$ & $19(2)$ & $-0.573(4)$ \\ \hline
Ia   &VMC  & $-0.022$ & $1.399$ & $24.12(5)$ & $1.982$ & $17.80(7)$ & $-0.591$ \\
     &GFMC & $-0.045$ & $1.360$ & $23.64(6)$ & $1.929$ & $17.77(9)$ & $-0.575$ \\ \hline
Average   &VMC & $-0.026(3)$ & $1.38(1)$ & $24(2)$ & $1.95(2)$ & $18(3)$ & $-0.593(4)$ \\
          &GFMC & $-0.046(2)$ & $1.34(2)$ & $24(3)$ & $1.90(3)$ & $19(5)$ & $-0.568(9)$ \\ \hline\hline
\end{tabular}
\end{center}
\caption{Expansion coefficients of the GFMC multipoles and comparison with the VMC, using three different Norfolk models. The error denotes fitting error only}
\label{GFMC}
\end{table}

In addition to a calculation of the multipoles using VMC, we also 
performed calculations at the Green's Function Monte Carlo
level, to remove residual excited state contamination in the nuclear wave functions.
Following the procedure of Ref.~\cite{Pervin:2007}, we
perform a mixed estimate extrapolation of the GFMC multipoles. We then
fit the expansion coefficients of Eqs.~\eqref{Fit1} - \eqref{Fit4} as
was done for the VMC calculations. The results of these fits are presented in Table \ref{GFMC} for models Ia*, IIa*, and Ia. 
The GFMC evolution reduces the leading coefficients, $L_1^{(0)}$ and $E_1^{(0)}$, by $\sim 3$-$4\%$, which, as we will see, results in better agreement with the experimental half-life. 
Propagating models Ia* and Ia changes the remaining expansion coefficients at the level of a few percent while for model IIa*  terms that were higher order in $qr_{\pi}$ experienced a more significant percent change after the propagation. This can be understood by looking at the system size 
as a function of $\tau$ during the GFMC propagation.
To further understand this, we calculate the point proton radius 
in GFMC and we observe that for model IIa*, the system size grows much more rapidly in 
$\tau$ than models Ia and Ia*. 
This behavior is due to the proximity of the model IIa* $^6{\rm He}(0^+;1)$ ($^6{\rm Li}(1^+;0)$)
ground state energy in GFMC to the $\alpha + 2n$ ($\alpha+d$) breakup threshold. 
Because of terms going like $e^{-i\bfq\cdot\bfr_i}$ in the current operators, 
the monotonic increase of the system size will impact the convergence of the IIa* matrix elements needed to determine the multipoles.
To account for this in the GFMC extrapolation with model IIa*, we adopt a procedure used 
for systems near threshold and broad resonances~\cite{Pastore:2014oda}. 
We note that while the system sizes grow with $\tau$, the ground state energies of $^6{\rm Li}$ and $^6{\rm He}$
drop rapidly and stabilize near $\tau\approx 0.1 ~{\rm MeV}^{-1}$. This indicates that
spurious contamination has been removed from the wave functions at that point. Under this assumption, we extract the values of the matrix
elements by performing a linear fit to the matrix element in the 
interval $\tau = [0.1~{\rm MeV}^{-1}, 0.3~{\rm MeV^{-1}}]$ and 
extrapolating back to $0.1~{\rm MeV}^{-1}$. We determine the systematic error of this procedure by averaging in the intervals 
$\tau = [0.08~{\rm MeV}^{-1}, 0.3~{\rm MeV^{-1}}]$ and 
$\tau = [0.12~{\rm MeV}^{-1}, 0.3~{\rm MeV^{-1}}]$ to get a
conservative estimate. 

For the GFMC extrapolations, we assigned errors to the average matrix element arising from the
energy range and three-body force following the same 
procedure as was done for the VMC. We also include
the systematic uncertainty from the model IIa* extrapolation by summing it in quadrature with the energy 
range and three-body force uncertainties. For the GFMC average, the coefficients $L_1^{(0)}$, $E_1^{(0)}$, $C_1^{(1)}$, and 
$M_1^{(1)}$ have uncertainties ranging from $\sim1.4\%$ to $\sim5.2\%$. The coefficients $L_1^{(2)}$ and $E_1^{(2)}$ 
have uncertainties of $\sim12\%$ and $\sim26\%$, respectively. In the expression for the rate,
the large error coefficients appear suppressed by powers of $qr_{\pi}$ and do not contribute as strongly as the
coefficients with errors of order $1\%$.

Because of the small cutoff uncertainty, we 
estimate the impact of including model IIb* by re-weighting the matrix elements in the 
average so that model IIa* is counted twice. Averaging under this assumption provides 
coefficients and errors that are consistent with the average results obtained when omitting model IIb*.
Thus, we conclude that we can safely neglect model IIb* in the GFMC average.

\subsection{Experimental comparison and remaining spectral uncertainty}

We can first of all check the $L^{(0)}_1$ 
and $E^{(0)}_{1}$ multipoles by comparing our calculation with experimental half-life
$\tau_{1/2}= 807.25 \pm 0.16 \pm 0.11$ ms \cite{Kanafani:2022tbr,Knecht:2011ir}.
Using the VMC matrix elements in Table \ref{Results}, we obtain
\begin{eqnarray}\label{halflife}
\left. \tau_{1/2}\right|_{\rm VMC}  = \frac{\log{2}}{\Gamma} =  
\frac{1}{V_{ud}^2 g_A^2}\left(1175 \pm  17\right)  \, {\rm ms} = \left(762 
\pm  11 \pm 2  \right)  \, {\rm ms} 
\end{eqnarray}
where we have used $V_{ud}= 0.97370 \pm 0.00031$ and $g_A = 1.2754 \pm 0.0013$. The first error is due to the nuclear matrix elements, while the second to $g_A$. The error from $V_{ud}$ is negligible.
Eq. \eqref{halflife} deviates by about 5\% from the observed value. While small, the discrepancy between the VMC calculation and the experiment is not covered by the uncertainty range in Eq. \eqref{halflife}, indicating that the errors due to either the chiral EFT truncation or the Monte Carlo method are underestimated. For this reason, we evolved the VMC wave functions in the GFMC.
Using the GFMC matrix elements, we obtain
\begin{eqnarray}\label{halflifeG}
\left. \tau_{1/2}\right|_{\rm GFMC}  = \frac{1}{V_{ud}^2 g_A^2}\left(1246 \pm  37\right)  \, {\rm ms} = \left(808 
\pm  24 \pm 2  \right)  \, {\rm ms}, 
\end{eqnarray}
which has a 3\% error and is in perfect agreement with the observed half-life.
The half-life is dominated by $L_1^{(0)}$, $E_1^{(0)}$ and by the Fermi function. The next most important correction comes from inner and outer radiative corrections, $\Delta^V_R$ and $\delta_R(Z,\varepsilon)$, which together shorten the half-life by about 4\%. For $\Delta^V_R$
we use the dispersive evaluation of Refs. \cite{Seng:2018qru,Seng:2018yzq,Seng:2021syx}, $\Delta^V_R = 2.467(22) \cdot 10^{-2}$.
Higher multipoles impact the half-life at the 0.1\% level.

\begin{figure}[th]
\includegraphics[width=0.55\textwidth]{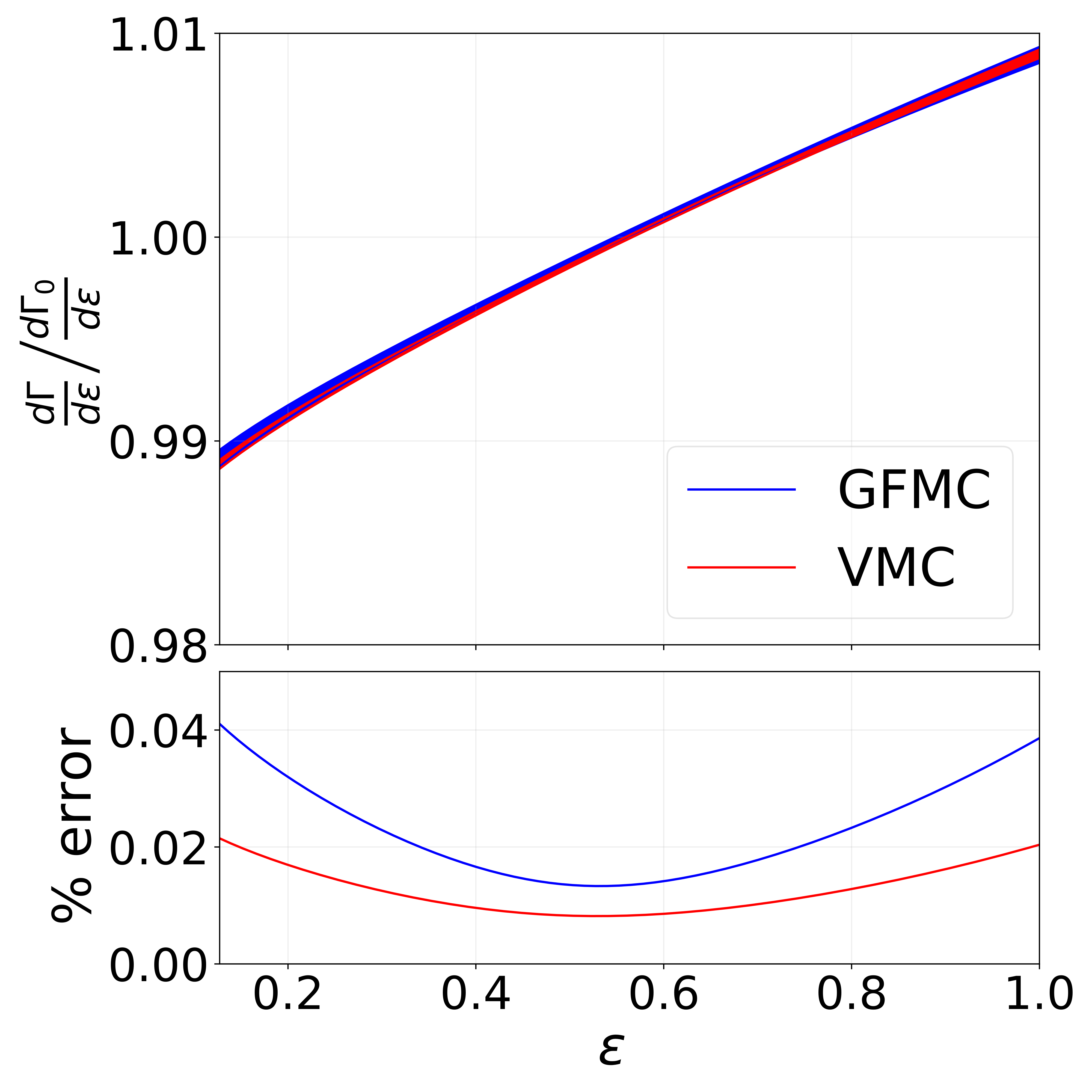}
\caption{ Deviation of the $^6$He $\beta$ spectrum from the expression truncated at leading order in the multipole expansion, given in Eq. \eqref{Gamma0}, and percentage error on the ratio. The blue curves use GFMC matrix elements in Table \ref{GFMC}, while the red line the VMC averages in Table \ref{Results}.
In the top panel, the width of the curves denotes the theoretical error.
}\label{fig:SMspectrum}
\end{figure}

Moving on to the differential decay rate, using the VMC multipoles given in Table \ref{Results}, we find
\begin{eqnarray}\label{numerics}
& &\left. \frac{d \Gamma}{d \varepsilon}\right|_{\rm VMC}= \frac{d \Gamma_0}{d \varepsilon} \Bigg\{ 
1  + \Bigg[ \left(1 - 2 \varepsilon + \frac{\mu_e^2}{\varepsilon }\right) (-1.16 \pm 0.01)  + 
\left(1 - \frac{\mu_e^2}{ \varepsilon}\right)
(3.6 \pm 0.4) \cdot 10^{-2}   \Bigg] \cdot 10^{-2} \nonumber \\
& & 
- \Bigg[ \left( 1 - \frac{\mu_e^2}{\varepsilon} (2 - \varepsilon)\right)  (0.96 \pm 0.08)   + \left(3 - 10 \varepsilon (1 - \varepsilon) + \mu_e^2 \frac{4 - 7 \varepsilon}{\varepsilon}
\right) (0.32 \pm 0.06)  \Bigg] \cdot 10^{-3} \nonumber
\\ & & 
+ (4.1 \pm 1.1) \cdot 10^{-4}  \left(1 - \varepsilon \right)
 \Bigg\}, 
\end{eqnarray}
where the terms in the first line appear at NLO in the multipole expansion, and are proportional to
$M_{1}^{(1)}$ and
$C_{1}^{(1)}$, respectively.
The terms in the second line
appear at N$^2$LO,
the first proportional to $L_1^{(2)}$ and the second to a combination of $|M^{(1)}_1|^2$
and $E_1^{(0)} E_1^{(2)}$. Finally, the term in the third line is an electromagnetic correction proportional to subleading multipoles.
Using the GFMC matrix elements in Table \ref{GFMC}, the differential rate is
\begin{eqnarray}
& &\left. \frac{d \Gamma}{d \varepsilon}\right|_{\rm GFMC}= \frac{d \Gamma_0}{d \varepsilon} \Bigg\{ 
1  + \Bigg[ \left(1 - 2 \varepsilon + \frac{\mu_e^2}{\varepsilon }\right) (-1.15 \pm 0.02)  + 
\left(1 - \frac{\mu_e^2}{ \varepsilon}\right)
(6.6 \pm 0.7) \cdot 10^{-2}   \Bigg]\cdot 10^{-2} \nonumber \\
& & 
- \Bigg[ \left( 1 - \frac{\mu_e^2}{\varepsilon} (2 - \varepsilon)\right)  (0.99 \pm 0.12)   + \left(3 - 10 \varepsilon (1 - \varepsilon) + \mu_e^2 \frac{4 - 7 \varepsilon}{\varepsilon}
\right) (0.35 \pm 0.10)  \Bigg] \cdot 10^{-3}
\nonumber
\\ & & 
+ (4.0 \pm 1.8) \cdot 10^{-4}  \left(1 - \varepsilon \right)
 \Bigg\}. \label{rateGFMC} 
\end{eqnarray}
From Eqs. \eqref{numerics} and \eqref{rateGFMC} we see that the dominant correction is, as expected \cite{Holstein:1974zf,Calaprice1975}, given by the magnetic multipole $M_1$, which contributes to the spectrum at the percent level.  The uncertainty on the ratio $M^{(1)}_1/L_1^{(0)}$, which dominates the error budget, is about $2\%$, and, in our calculation, it receives the main contribution from changing the energy range of the fits to $NN$ scattering data.
The next contributions come from 
$L_1^{(2)}$, $E_1^{(2)}$ 
and $C_1^{(1)}$, which affect
 the energy distribution at the 10$^{-3}$--$10^{-4}$ level. 
$L_1^{(2)}$, $C_1^{(1)}$ and $E_1^{(2)}$ have uncertainties of about $10\%$, $12\%$ and $20\%$, respectively, so that these terms contribute to the theory error at the $10^{-4}$ level.

\begin{figure}[t]
\includegraphics[width=0.49\textwidth]{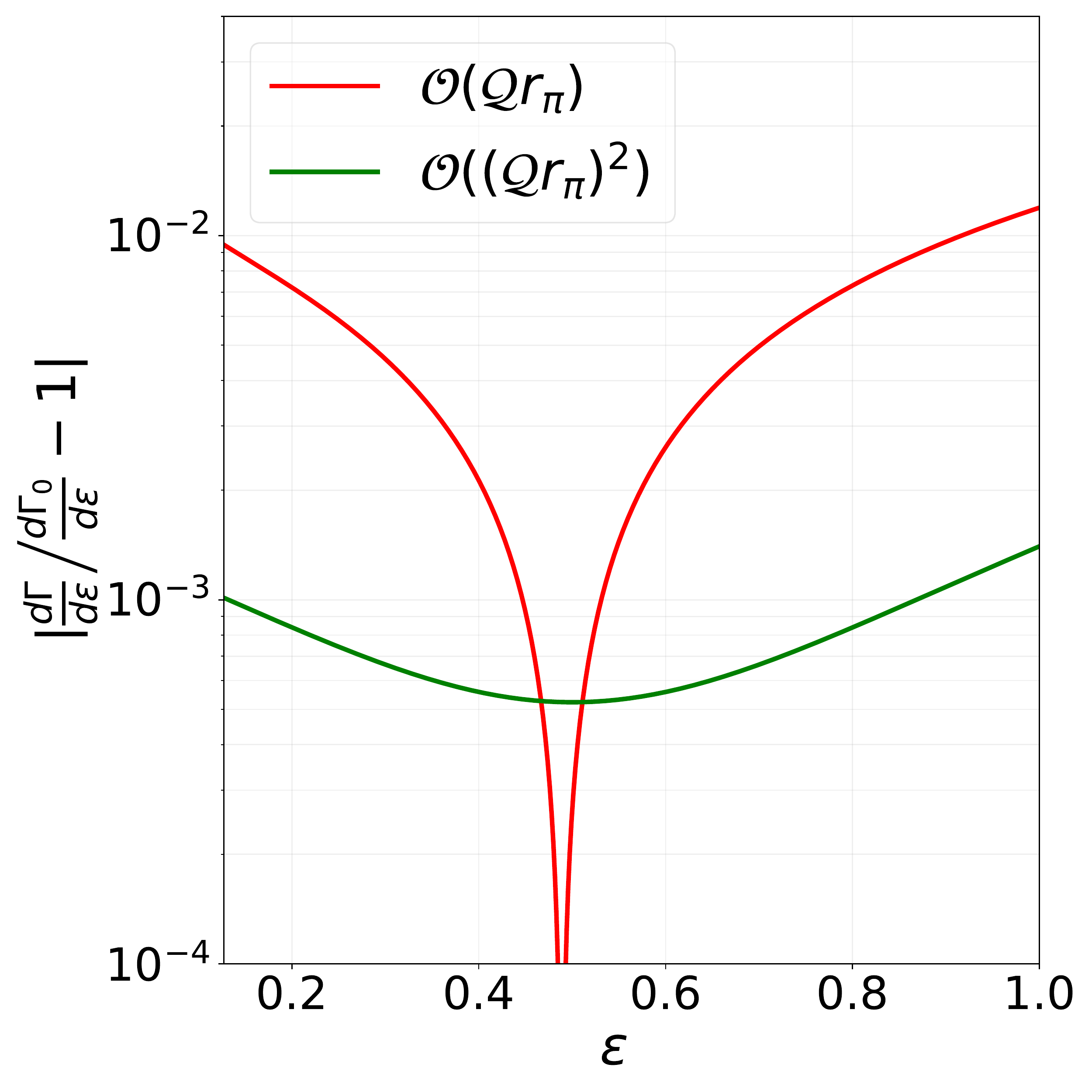}
\includegraphics[width=0.49\textwidth]{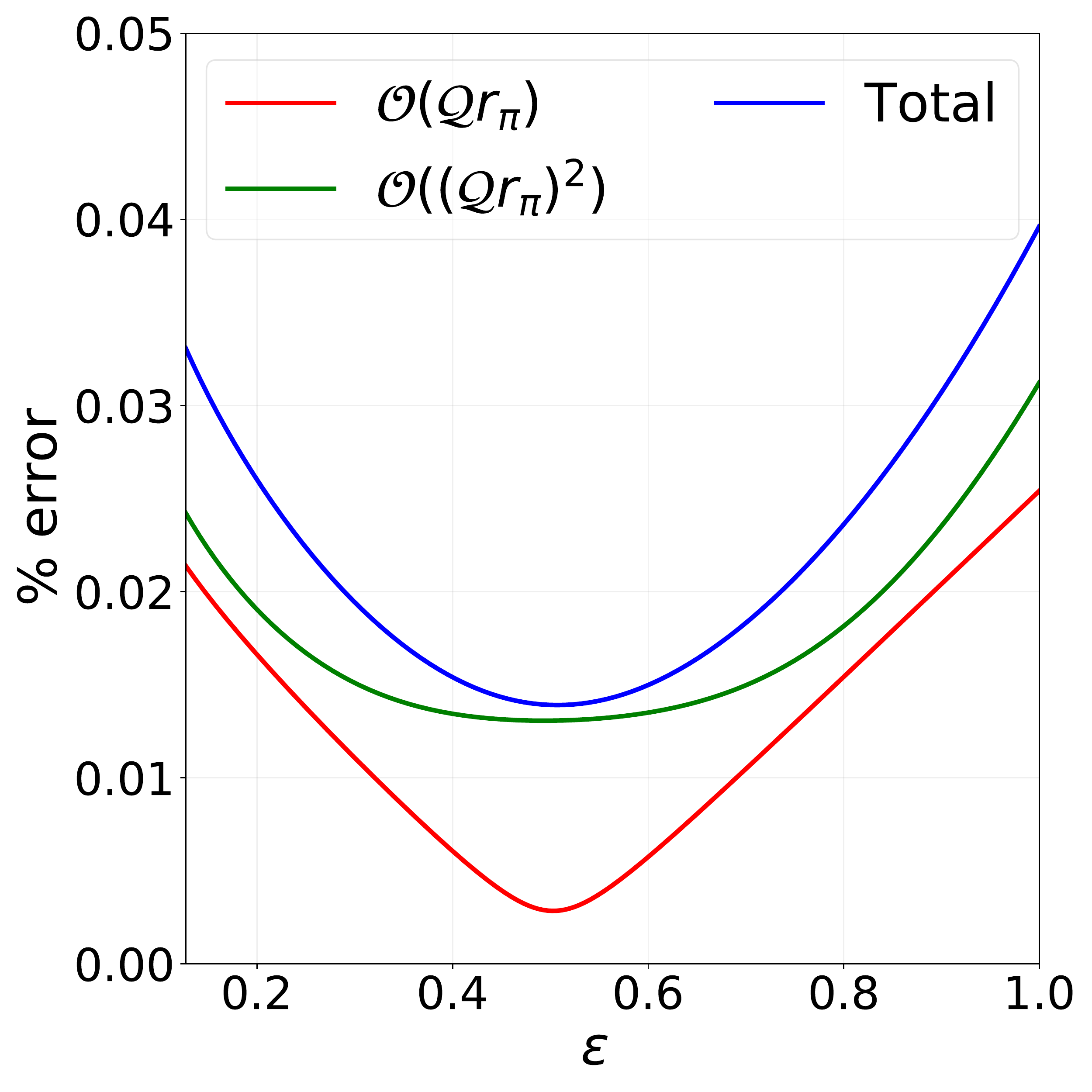}
\caption{ 
 Corrections to the $\beta$ spectrum ($left$)
 and  contributions to the error  on the ratio between $d\Gamma/d\varepsilon$ and 
$d\Gamma_0/d\varepsilon$ ($right$)  
at next-to-leading order ($\mathcal O(\mathcal Q r_\pi)$)
and next-to-next-to leading order
($\mathcal O(\mathcal Q^2 r^2_\pi)$)
in the multipole expansion. The figure uses GFMC matrix elements.
}\label{fig:budget}
\end{figure}

In Figure \ref{fig:SMspectrum}
we show the deviation of the $\beta$ spectrum from the leading term in the multipole expansion, $d\Gamma_0/d\varepsilon$, defined in Eq. \eqref{Gamma0}, using both VMC and GFMC matrix elements. We see that, while GFMC and VMC differ by 6\% on the total rate, the differences largely cancel in the ratio, and the corrections to the spectral shape are very similar in both cases. The bottom panel of Fig. \ref{fig:SMspectrum} shows the error on the ratio. This is somewhat larger in GFMC, but well below $10^{-3}$.
Fig. \ref{fig:budget}
shows the contributions of 
the leading $\mathcal O(\mathcal Q r_\pi)$ correction, arising from $M_1$, and of the 
second order terms
$\mathcal O((\mathcal Q r_\pi)^2)$ (including the formally $\mathcal O(\mathcal Q r_\pi)$ but numerically small contribution from $C_1^{(1)}$), to the differential rate (left) and to the uncertainty (right). 
We see that, while $M_1$ dominates the correction to the spectrum, the second order terms contribute at the $10^{-3}$ level and the first and second order terms give contributions of similar size to the uncertainty. 

\subsubsection{Validation of \texorpdfstring{$M_1$}{M1}}

We can further validate the calculation in several ways.
The magnetic multipole $M_1$ can be cross-checked against data.
Using the conserved vector current hypothesis,  $M_1^{(1)}$ can be expressed in terms of the transition $\Gamma( ^6{\rm Li}(0^+) \rightarrow ^6 {\rm Li} (1^+) \gamma)$
\begin{equation}
    \left|M_1^{(1)}\right| = \frac{3m_\pi}{\sqrt{4\pi}}\sqrt{\frac{\Gamma_{M_1}}{\alpha E_\gamma^3}} = 0.582(6)
    \label{eq:M_1_CVC}
\end{equation}
where the uncertainty is determined by the width, $\Gamma_{M1} = 8.19(17)$ eV \cite{Bergstrom:1975ax,Tilley:2002vg}. Using the average VMC value listed in Table \ref{Results}, our result agrees within 1\% with the experimental value, while individual models are consistent within 3\% when including two body currents.
The GFMC average is a bit lower, but still only 2\% away from Eq. \eqref{eq:M_1_CVC}, and compatible within $\sim 1\sigma$. 
Our results also agree with the \textit{ab initio} calculation of Ref. \cite{Glick-Magid:2021uwb}, which, including only the one-body piece of the current, finds $|M_1^{(1)}| = 0.565$.

Eq. \eqref{eq:M_1_CVC} is valid up to isospin breaking corrections.
To check the level of isospin breaking that we can expect in $M^{(1)}_1$, we computed both the 
$^6{\rm He}(0^+;1) \to ^6{\rm Li}(1^+;0)$ transition 
and the electromagnetic 
transition $^6{\rm Li}(0^+;1) \to ^6{\rm Li}(1^+;0)$ with NV2+3-IIb wave functions in the impulse approximation, that is retaining only one-body current operators at LO in the chiral expansion. Using VMC wave functions, the electromagnetic transition had a value 0.553 and the weak
transition a value of 0.554. Agreement between the two $M_1$ calculations is thus achieved at the $\sim0.1\%$ level. Propagating this 
calculation in GFMC, we find, in  the electromagnetic case, $| M^{(1)}_1|_{\rm EM} = 0.532$ and, for the weak transition, $|M_1^{(1)}| = 0.538$, showing a 0.9\% difference. This analysis is performed with one 
nuclear interaction model and thus does not account for any model uncertainty. 

The change in the level of agreement between VMC and GFMC can be understood as due to how 
the wave functions are generated in each method. For the VMC case, the variational parameters in the $^6{\rm He}(0^+;1)$ and $^6{\rm Li}(0^+;1)$ wave functions are minimized 
separately. Because of explicit isospin symmetry breaking terms in the potential, the 
parameters of the two wave functions differ; however, ispospin symmetry breaking correlations are not turned on in the VMC wave functions. When the trial states are acted on with the imaginary time propagator,   isospin breaking correlations are introduced. This, in turn, increases the disagreement for the electromagnetic and weak $M_1$ transitions. Because the systems differ by changing the isospin of one nucleon, the effect
of this symmetry breaking is small and at the level of the experimental uncertainty in Eq. \eqref{eq:M_1_CVC}.
Since isospin breaking corrections are smaller than the experimental error, we can use $M_1^{(1)}$ extracted from $^6{\rm Li}(0^+;1) \to ^6{\rm Li}(1^+;0)$ to further reduce the error on $M_1$ to the 1\% level. Controlling isospin-breaking effects will become more important with improved measurements of the 
$^6{\rm Li}(0^+) \rightarrow ^6 {\rm Li} (1^+) \gamma$ transition, as those suggested in Ref. \cite{Romig:2014dbe}.

\begin{table}[t]
\begin{center}
\begin{tabular}{||c || c || c || c || c || c  || c ||}\hline \hline
& VMC  &  GFMC  & Calaprice \cite{Calaprice1975}   & Glick-Magid et al. \cite{Glick-Magid:2021uwb} \\ \hline
recoil &
 $ 0.020(3)$ &$ -0.001(3)$ & $-0.0144$ &  $-0.006$\\\hline
 pseudo & $-0.040$ & $-0.038$ & -- & $-0.039$\\ \hline
 2 body & $-0.006$ & $-0.007$ & -- & -- \\\hline 
 total    & $-0.026(3)$  & $-0.046(3)$ & $-0.0144$ & $-0.045$ \\
 \hline \hline
\end{tabular}
\end{center}
\caption{VMC and GFMC one-body and two-body averages for $C^{(1)}_1$ compared with Ref.~\cite{Calaprice1975}
and Ref. \cite{Glick-Magid:2021uwb}. Notice that to obtain $C^{(1)}_{\rm pseudo}$ we rescale the results of Ref. \cite{Glick-Magid:2021uwb} by 
$W_0/(W_0 + \Delta E_c)$, as we are not including Coulomb corrections to $C_1$.}
\label{tab:c1.summary}
\end{table}

\subsubsection{Validation of \texorpdfstring{$C_1$}{C1}}

The axial charge correction $C_1^{(1)}$, unlike the weak magnetism correction,  is not
constrained by an experimental datum. As such, its evaluation is critical for current experimental efforts aiming for the sub-0.1\% level. The prefactor to $C_1^{(1)}$ from Eq. (\ref{rate}) is $2W_0r_\pi \approx 0.05$ and with $C_1^{(1)} \sim 1-2\%$ a relative $\sim 10-20\%$ precision is needed for a theory uncertainty to be smaller than one part in $10^4$. 
As can be seen in Eq. \eqref{axial_charge1},
at lowest order in chiral EFT, the axial charge
receives a $\mathcal O(1/m_N)$
 recoil contribution from the coupling of the axial current to the nucleon, and a
contribution from the induced pseudoscalar form factor, proportional to the energy of the electron and neutrino $E_e + E_\nu = W_0$.
Naively, these would translate in an
 $\mathcal O(m_\pi/m_N)$
and $\mathcal O(W_0/m_\pi)$ correction to $C_1$,
respectively. 
The matrix element of the recoil component, however, vanishes for transitions between the dominant $S$ state components of the wave function \cite{Calaprice1975}, making this component particularly sensitive to other wave function admixtures and two-body currents. 
In our calculation, indeed, $C_1$ is dominated by the induced pseudoscalar contribution,  which, being proportional to  $L_1^{(0)}$, is  fairly stable.
We can separate $C_1^{(1)}$ in three pieces
\begin{equation}
C_1^{(1)} = C^{(1)}_{\rm recoil} + C^{(1)}_{\rm pseudo} + C^{(1)}_{\rm 2-body},
\end{equation}
where $C^{(1)}_{\rm recoil}$ and $C^{(1)}_{\rm pseudo}$ are given by the matrix element of $\rho^{-2}_{\rm recoil}(\mathbf{q},A)$
and $\rho^{-2}_{\rm pseudo}(\mathbf{q},A)$
in Eq. \eqref{axial_charge1}, while
$ C^{(1)}_{\rm 2-body}$ by the matrix element of $\rho^{-1}(\mathbf{q},A)$ in Eq. \eqref{axialcharge2b}.
Results for each component, after averaging over different models as discussed in Section \ref{UQ}, are given in Table \ref{tab:c1.summary}. 

The calculation of $C^{(1)}_{\rm recoil}$ in the decay of $^6$He,
together with other triplet decays in the mass $A=8, 12, 20$ systems,  received a significant amount of attention over 40 years ago but ultimately remained unresolved \cite{Calaprice1975, Kleppinger1977, Holstein1989, Blin-Stoyle1978}. Until recently, the only theoretical determination of the axial charge contribution to $^6$He was performed by Calaprice \cite{Calaprice1975} in Holstein's formalism. A direct comparison can be made by observing \cite{Hayen:2020nej}
\begin{equation}
    C_{\rm recoil}^{(1)} = {\color{blue} -} \sqrt{\frac{3(2J_i+1)}{4\pi}}\frac{d(0)}{2r_\pi M}
\end{equation}
where $d(q^2)$ is the so-called induced tensor form factor.  
Using wave functions tuned to reproduce the experimental energy levels, Calaprice  obtained $d(0)=2.4$ \cite{Calaprice1975}, which can be converted into $C_{\rm recoil}^{(1)} = -0.0144$. This value is in the same ballpark of the VMC calculation, which however has an opposite sign. After GFMC evolution, the recoil contribution is reduced and qualitatively agrees with the result of Ref. \cite{Glick-Magid:2021uwb}.

To further track down the origin of the discrepancy, we notice that one can write
\begin{equation}
    C_{\rm recoil}^{(1)} = - \sqrt{\frac{3}{4\pi}}\frac{g_A }{2r_\pi m_N} \left(   \mathcal M_{\sigma L} + m_N \mathcal M_{\sigma r p} \right),
\end{equation}
where $\mathcal M_{\sigma L}$ and $\mathcal M_{\sigma r p}$ are one-body matrix elements defined in Ref. \cite{Holstein:1974zf}. Calaprice assumed the first matrix element to be dominant and neglected $\mathcal M_{\sigma r p}$. 
To check the assumption, we calculated 
$\mathcal M_{\sigma L}$ in model NV2+3-Ia$^*$. 
In the VMC, the $^6$He and $^6$Li wave functions are expressed in terms of
the action of correlation operators on single particle wave functions with an $\alpha$ core and two nucleons whose wave functions are $p$-wave solutions of an effective $\alpha$-N potential \cite{Pudliner:1997ck,Schiavilla2002}. The two nucleons can be coupled in different $LS$ channels, with strength determined by the parameter $\beta_{LS}$.
In the NV2+3-Ia$^*$ model, $\beta_{00}$ and $\beta_{11}$, parameterizing the $^1S_0$ and $^3 P_0$ components of the $^6$He wave function, are given by $\beta_{00} = 0.931$ and 
$\beta_{11} = -0.364$. In the case of $^6$Li, the $^3S_1$, $^1P_1$
and $^3D_2$ components are given 
by $\beta_{01}= 0.967$,
$\beta_{10}= 0.182$ and
$\beta_{21}= 0.176$, respectively. These  agree fairly well with Ref. \cite{Calaprice1975}, which found a smaller $D$ wave component, $\beta_{21}=-0.03$.
In this model, we obtain
\begin{equation}
    \left. C_{\rm recoil}^{(1)}\right|_{\sigma L}  = 0.013,
\end{equation}
with statistical uncertainties $<0.001$.
We observe a small, approximately linear dependence on $\beta_{21}$, which would shift the value to $0.012$ for $\beta_{21}=-0.03$. To better mimic the shell model calculation of 
Ref. \cite{Calaprice1975}, we turned off 
the “one-pion-exchange-like” correlation operators in the VMC wave functions in a similar fashion as to what was done in
Ref.~\cite{Pastore:2017uwc}, observing a 10\% increase of the matrix element, from 0.013 to 0.014. While the magnitude of the matrix element
agrees very well with Ref. \cite{Calaprice1975},
we were not able to resolve the disagreement on the sign. 
Note the correlation and $D$ wave analyses above have been done only at the VMC level. 

In GFMC the value of $\mathcal M_{\sigma L}$ is further decreased to 
\begin{equation}
    \left. C_{\rm recoil}^{(1)}\right|_{\sigma L}  = 0.009.
\end{equation}
The above 
values have small statistical uncertainties but are not accounting for any possible model dependencies. 
We conclude that: $a)$ the magnitude of $\mathcal M_{\sigma L}$ agrees well with shell model calculations, but is reduced by nuclear correlations and by the GFMC evolution,
$b)$ the contribution of $\mathcal M_{\sigma r p}$ is non-negligible  
in the case of $^{6}$He, varying from
\begin{equation}
\left. C_{\rm recoil}^{(1)}\right|_{\sigma r p}  = 0.007
\end{equation}
in VMC
to 
\begin{equation}
\left. C_{\rm recoil}^{(1)}\right|_{\sigma r p}  = -0.012
\end{equation}
Finally, as anticipated in Refs. \cite{Calaprice1975, Kleppinger1977, Holstein1989, Blin-Stoyle1978}, we find a substantial contributions from two-body currents for all three models.

The induced pseudoscalar contribution, captured by the $h(q^2)$ function in Holstein's formalism, was not considered in Ref. \cite{Calaprice1975}. In agreement with Ref. 
\cite{Glick-Magid:2021uwb}, we find this contribution to dominate $C_1^{(1)}$. We however stress that the induced pseudoscalar contribution to $C_1^{(1)}$ is partially cancelled by the one to $L_1^{(2)}$, so that the single-nucleon induced pseudoscalar form factor gives corrections to the spectrum that are proportional to 
$m_e^2$, as expected.

\section{Charged currents in the SM Effective Field Theory }
\label{sec:SMEFT}

With the theoretical accuracy of the SM spectrum well below  $0.1\%$, we can then study the sensitivity to physics beyond the SM.
If BSM physics arises at a scale $\Lambda \gg v$, its correction to $\beta$ decays can be described in the framework of the Standard Model Effective Field Theory  \cite{Buchmuller:1985jz,Grzadkowski:2010es}, an effective field theory  that complements the SM with the most general set of gauge-invariant effective operators, expressed in terms of SM fields and organized according to their canonical dimension. SMEFT contains a single dimension-five operator  \cite{Weinberg:1979sa}, which, when the Higgs gets its vacuum expectation value, gives rise to a Majorana mass term for the three left-handed neutrinos.
At dimension-six, the SMEFT contains several classes of operators, which, at low-energy, induce new axial, vector, scalar, pseudoscalar and tensor semileptonic interactions between quarks, charged leptons and left-handed neutrino fields \cite{Buchmuller:1985jz,Grzadkowski:2010es}. Since the mechanism behind the origin of neutrino masses is unknown, for generality we extend the SMEFT with a multiplet of $n$ sterile neutrino fields $\nu_R$ ($\nu$SMEFT) \cite{delAguila:2008ir,Cirigliano:2012ab,Liao:2016qyd}. The sterile neutrino is a singlet under the SM group. At dimension-three, $\nu_R$ has a Majorana mass term, while at dimension-four it interacts with active neutrinos via  Yukawa interactions. 
If one considers only these renormalizable interactions, 
after diagonalizing the neutrino mass matrix, the neutrino sector is characterized by $3+n$ mass eigenstates with masses $m_{1}, \ldots m_{3+n}$,
and a $(3+n)\times (3+n)$ unitary mixing matrix $U$,
which generalizes the Pontecorvo-Maki-Nakagawa-Sakata (PMNS) mixing matrix.
For simplicity, we will consider $n=1$, 
so that one has to consider the parameters $m_4$ and $U_{\alpha 4}$, with $\alpha \in \{e, \mu,\tau\}$,
in addition to the SM.
The inclusion of multiple light neutrino states is straightforward.
From oscillation experiments and from  the upper limit from the  KATRIN experiment 
\cite{KATRIN:2019yun,KATRIN:2021uub},
$m_{1,2,3} \lesssim 0.8$ eV, so that they can be neglected in our analysis. We will consider $m_4$ as a free parameter, and mostly focus on the region in which $m_4$ is smaller than the $\mathcal Q$-value, so that $\nu_4$ can be produced in the decay. 
In addition to renormalizable interactions, $\nu_R$  can have lepton-number-conserving non-standard interactions with SM fields at dimension-six \cite{delAguila:2008ir,Cirigliano:2012ab,Liao:2016qyd}, which, as we will see, induce new axial, vector, scalar, pseudoscalar and tensor charged-current interactions involving sterile neutrinos.

After integrating out heavy gauge and quark fields and rotating to the neutrino mass basis, the most general low-energy Lagrangian for $\beta$ decays is given in Refs.  \cite{Lee:1956qn,Cirigliano:2012ab,Dekens:2020ttz}.
If the masses of all active and sterile neutrino states are much smaller than the nuclear scale or the electron mass, 
we can neglect sterile neutrino operators, whose interference with the SM is suppressed by powers of the neutrino masses. In this case, only scalar, pseudoscalar and tensor interactions can give rise to a Fierz interference term, and,
making connections with the notation of Ref. \cite{Cirigliano:2012ab}, 
we can write the relevant interactions as
\begin{eqnarray}
\mathcal L^{(6)} &=& - \frac{4 G_F}{\sqrt{2}} V_{ud}\Bigg\{ \frac{1}{2} \bar e_R \nu_L \, \left( \epsilon_S \bar u   d + \epsilon_P \bar u \gamma_5  d  \right) + \epsilon_T \bar e_R \sigma^{\mu \nu} \nu_L \,   \bar u_R   \sigma_{\mu \nu} d_L  \Bigg\} + \textrm{h.c.}.
\end{eqnarray}
If the masses of sterile neutrinos are non-negligible
compared to the electron mass, there are additional interference terms. Using again the conventions of Ref. \cite{Cirigliano:2012ab},
we write the Lagrangian for $\nu_4$ as
\begin{eqnarray}\label{Lag:nu4}
\mathcal L^{(6)}& =& -\frac{4 G_F}{\sqrt{2}} V_{ud} U_{e4} \Bigg\{ 
\bar e_{L}  \gamma_\mu   \,  \nu_4 \bigg( 
(1 +\epsilon_L)  \bar u_L \gamma^\mu d_L\,        + \epsilon_R\, \bar u_R \gamma^\mu d_R \bigg) +
\frac{1}{2}\bar e_{R}    \nu_4  \left(   \epsilon_S \bar u  d  + \epsilon_P  \bar u \gamma_5  d \right)\nonumber \\ & &  +  \epsilon_T\bar u_R \sigma^{\mu\nu} d_L\,  \bar e_{R}  \sigma_{\mu\nu} \, \nu_4
\Bigg\} 
-\frac{4 G_F}{\sqrt{2}} V_{ud} 
\Bigg\{ 
\bar e_{R}  \gamma_\mu   \nu_4
\bigg(\tilde{\epsilon}_L  \bar u_L \gamma^\mu d_L      +
\tilde{\epsilon}_R   \bar u_R \gamma^\mu d_R \bigg) \nonumber\\
& & +
\frac{1}{2}\bar e_{L}  \nu_4 \left(
  \tilde \epsilon_S \bar u  d + 
  \tilde \epsilon_P  \bar u \gamma_5 d  \right) + \tilde \epsilon_T \bar u_L \sigma^{\mu\nu} d_R\,  \bar e_{L}  \sigma_{\mu\nu} \, \nu_4 
\Bigg\} 
+{\rm h.c.}.
\end{eqnarray}
The conversion between the $\epsilon$, the low-energy EFT (LEFT) couplings 
defined in Ref. \cite{Dekens:2020ttz}
and $\nu$SMEFT is discussed in Appendix \ref{app:Lagrangians}.
The terms in the first bracket of Eq. \eqref{Lag:nu4}
are induced by SMEFT operators involving active neutrinos
and are proportional to the mixing $U_{e4}$. Since $U_{e4}$ is small, we can usually neglect terms proportional to $U_{e4} \times \epsilon$. The terms in the second bracket, on the other hand, are induced by $\nu$SMEFT operators with sterile neutrinos.

\subsection{Beyond the SM corrections to the \texorpdfstring{$\beta$}{beta} spectrum}
\label{sec:MultiBSM}

The multipole expansion can be generalized to non-standard currents induced by SMEFT
operators (see also Ref. \cite{Glick-Magid:2022erc}). At dimension-six in the  
$\nu$SMEFT, for both scalar/tensor and vector/axial operators, the leptonic and hadronic currents have at most spin one. 
The derivation of the multipole expansion therefore proceeds essentially as in the SM.
The additional vector and axial operators in Eq. \eqref{Lag:nu4} generate exactly the same multipoles as in Eqs. \eqref{multiSM1} - \eqref{multiSM4}.
For scalar and pseudoscalar currents, only the $\mathcal C_{J0}$ multipole is present, while for tensor currents only the electric, magnetic and longitudinal multipoles.

Restricting again to the case of $J_i=0$, $J_f=1$, 
we can easily adapt the formulas in Eqs.~\eqref{eq:ME1}--\eqref{eq:ME4} to the case of non-standard currents and define
\begin{eqnarray}
C_1(q,P)&=&\frac{i}{\sqrt{4\pi}}\langle {}^6{\rm Li},10\rvert \rho_{P+}^\dagger(q\hat{\bf z})\rvert {}^6{\rm He},00\rangle \label{eq:SME1}
\, \\
L_1(q,T)&=&\frac{i}{\sqrt{4\pi}}\langle {}^6{\rm Li},10\rvert \hat{\bf z}\cdot {\bf{j}}_{T+}^\dagger(q\hat{\bf z})\rvert {}^6{\rm He},00\rangle\, \label{eq:SME2}\\
E_1(q,T)&=&\frac{i}{\sqrt{2\pi}}\langle {}^6{\rm Li},10\rvert \hat{\bf z}\cdot {\bf{j}}_{T+}^\dagger(q\hat{\bf x})\rvert {}^6{\rm He},00\rangle\, \label{eq:SME3} 
\\
M_1(q,T^\prime)&=&-\frac{1}{\sqrt{2\pi}}\langle {}^6{\rm Li},10\rvert \hat{\bf y}\cdot{\bf{j}}^{\prime \dagger}_{T+}(q\hat{\bf x})\rvert {}^6{\rm He},00\rangle.\label{eq:SME4}
\end{eqnarray}
We now used 
\begin{eqnarray}
\rho_P({\bf q})  = \int d^3 {\bf x} \, e^{i \bf q \cdot x} \mathcal J_P({\bf x}), \qquad {\bf j}^{(\prime)}_{T}({\bf q})  =  \int d^3 {\bf x} \, e^{i \bf q \cdot x} \boldsymbol {\mathcal J}^{(\prime)}_{T}({\bf x}).
\label{eq:rho-jt}
\end{eqnarray}
The pseudoscalar density $\mathcal J_P$ and tensor currents $\mathcal J_T$ and $\mathcal J_T^{(\prime)}$ are defined in Eq.~\eqref{intro5},
and the subscript $+$ again refers to the isospin components.
The multipoles $C_1(q,P)$, $L_{1}(q,T)$, $E_1(q,T)$ and $M_1(q,T^\prime)$ have a $q$ expansion
completely analogous to Eqs. \eqref{Fit1}--\eqref{Fit4}.
All corrections to the $^6$He decay spectrum arising at dimension-six in the $\nu$SMEFT can be expressed in terms on Eqs. \eqref{eq:ME1} -- 
\eqref{eq:ME4} and \eqref{eq:SME1} -- 
\eqref{eq:SME4}.

The same power counting considerations in Section \ref{pwc}  apply to multipoles induced by SMEFT charged-current operators.
For purely GT transitions,  tensor interactions are the most important, as they induce both $E^{(m,0)}_1(T)$ and $L^{(m,0)}_1(T)$
of order 1. 
From Eq. \eqref{T2b}, $M^{(1,0)}_1(T^\prime) = 0$, and thus we will neglect this contribution. Pseudoscalar interactions 
induce the multipole $C^{}_1(q,P)$, which starts at $\mathcal O(q r_\pi)$. This suppression is partially overcome by pion pole dominance of the pseudoscalar
form factor, which implies
\begin{eqnarray}
C^{(1)}_1(P) = \mathcal O\left( \epsilon_\chi^{-1}\right).
\end{eqnarray}
We thus include this contribution in the analysis.
In the case of multipoles induced by non-SM currents, we only consider contributions at LO in both the multipole and chiral expansions, implying, in particular, that we only consider one-body currents, as given in Appendix \ref{app:Onebody}.
To this order, from Eqs.
\eqref{eq:ME1}--\eqref{eq:ME4}, \eqref{eq:SME1}--\eqref{eq:SME4}, 
and \eqref{S}--\eqref{T2b},
we can see that 
\begin{eqnarray}
    L^{(0)}_1(T) &=& \frac{1}{\sqrt{2}} E^{(0)}_1(T)  =  -\frac{2 g_T}{g_A} L^{(0)}_{1}(A) + \mathcal O\left(\epsilon_\chi\right), \\
    C^{(1)}_1(P) &=& -\frac{B}{m_\pi} L^{(0)}_{1}(A)+ \mathcal O\left(\epsilon_\chi\right),\label{pseudo}
\end{eqnarray}
where $g_T$ is the isovector tensor charge, $g_T= 0.989 \pm 0.033$ \cite{Aoki:2021kgd}, and $B = m^2_\pi/(m_u + m_d) \sim 2.8$ GeV. $B$ and $g_T$ are scale dependent and given at the $\overline{\rm MS}$ scale $\mu =2$ GeV. Eq. \eqref{pseudo} confirms the relative enhancement of the pseudoscalar contribution.

We first consider the case in which active and sterile neutrinos have masses much smaller than the electron mass. In this particular case, non-standard axial and vector interactions simply shift the overall normalization of Eq. \eqref{rate}. If BSM interactions induce not only $V-A$ but also $V+A$ right-handed currents, the only effect on the spectrum will be a shift in the relative coefficient between the $(L^{(0)}_1)^2$ and $E^{(0)}_1 M^{(1)}_1$ terms in Eq. \eqref{rate}, as the second originates from the interference of the axial and vector currents. Since this effect arises at recoil order, it will not provide strong constraints on new physics.

Tensor and scalar currents interfere with the standard model via terms proportional to the electron mass. Tensor interactions give rise to a term at $\mathcal O(q^0)$.
The pseudoscalar contribution  is formally $\mathcal O(q)$, but it is enhanced because of the pion-pole dominance of the pseudoscalar form factor. The differential cross section with respect to the electron energy is given by 
\begin{eqnarray}
\frac{d \Gamma_{T}}{d \varepsilon} &=& \frac{d\Gamma_0}{d \varepsilon}  \frac{4 m_e}{3 E_e} 
\frac{1}{\left|L^{(0)}_1(A)\right|^2}\Bigg\{ 
\epsilon_T \textrm{Re}\left( E^{(0)}_1(T) E^{(0)*}_1(A) +  L^{(0)}_1(T) L^{(0)*}_1(A) \right)
\nonumber \\ & & -  \frac{\epsilon_P}{2} (1-\varepsilon) W_0 r_\pi \, \textrm{Re} \left(C_1^{(1)}(P) L^{(0)*}_{1}(A) \right)  \Bigg\}, \label{rateBSM}
\end{eqnarray}
where $d\Gamma_0/d\varepsilon$ is the  SM decay rate at LO in the multipole expansion, defined in Eq. \eqref{Gamma0} \footnote{
Since Eq. \eqref{rateBSM} is induced by BSM physics and thus small, we neglect $\mathcal O(\alpha)$ corrections}.
Eq. \eqref{rateBSM} shows the characteristic $m_e/E_e$ behavior. 
Using Eq. \eqref{pseudo}, Eq. \eqref{rateBSM} becomes
\begin{eqnarray}
\frac{d \Gamma_{T}}{d \varepsilon} &=& \frac{d\Gamma_0}{d \varepsilon}  \frac{m_e}{E_e} \Bigg\{ 
- 8 \frac{g_T \epsilon_T}{g_A}   +  \frac{2}{3}\epsilon_P  (1-\varepsilon) \frac{ W_0}{m_u + m_d}   \Bigg\}. \label{rateBSM2}
\end{eqnarray}
The tensor contributions agrees with the result of Ref. \cite{Jackson:1957zz},
while the enhancement of the pseudoscalar contribution was noted, for example, in Ref. \cite{Gonzalez-Alonso:2013ura}. Eq. \eqref{rateBSM2} is only valid at LO in chiral EFT.

Next, we consider the case of sterile neutrinos with non-negligible mass. Here we 
work at LO in the multipole expansion and consider $m_{\nu_4} < \mathcal Q$, so that $\nu_4$ can be produced.
Considering vector and axial currents, we obtain
\begin{eqnarray}
 \frac{d \Gamma_{\nu_4}}{d \varepsilon} &=& \frac{d \Gamma_0}{d\varepsilon}\Bigg\{ 
 \left(\sqrt{1 - \frac{m_{\nu_4}^2}{E_\nu^2} } - 1 \right)  \left( 1 +\epsilon_L - \epsilon_R  \right) |U_{e4}|^2 + \sqrt{1 - \frac{m_{\nu_4}^2}{E_\nu^2} }  \frac{m_e m_{\nu_4}}{E_e E_\nu} U_{e4}^* (\tilde{\epsilon}_L - \tilde{\epsilon}_R)
\Bigg\},  \nonumber \\ \label{sterilespectrum}
\end{eqnarray}
where $E_\nu = W_0 (1 -\varepsilon)$.
If we turn off all dimension-six operators, $\epsilon_{L,R}=0$ and
$\tilde \epsilon_{L,R}=0$, 
Eq. \eqref{sterilespectrum} is only proportional to the mixing $U_{e4}$. In this case the presence of sterile neutrinos has two effects, one on the normalization and one energy dependent.
In the presence of non-standard interactions of a sterile neutrino with a right-handed electron,  $\tilde \epsilon_L -\tilde \epsilon_R$, we also get new Fierz-like terms.
For tensor interactions
\begin{eqnarray}\label{sterileT}
\frac{d \Gamma_{\nu_4, T}}{d \varepsilon} &=& \frac{d\Gamma_0}{d\varepsilon} \frac{4}{3 \left|L_1^{(0)}(A)\right|^2}\textrm{Re}
\Bigg[ 
   E^{(0)}_1(T) E^{(0)*}_1(A)+L^{(0)}_1(T) L^{(0)*}_1(A)  \Bigg] \nonumber \\ &\times &
\Bigg\{  \frac{m_e}{E_e}  \left(\sqrt{1 - \frac{m_{\nu_4}^2}{E_\nu^2} } - 1 \right) \epsilon_T |U_{e4}|^2
+  \frac{m_{\nu_4}}{E_\nu}  \sqrt{1 - \frac{m_{\nu_4}^2}{E_\nu^2} }   \tilde{\epsilon}_T U^*_{e4}
 \Bigg\}, 
\end{eqnarray}
so that a fourth massive neutrino  would affect the standard Fierz interference term, and, more importantly, generate a new interference term, proportional to the neutrino mass. 
Similarly, the pseudoscalar interactions of sterile and active neutrinos give 
\begin{eqnarray}
\frac{d \Gamma_{\nu_4, P}}{d \varepsilon} &=& -\frac{d \Gamma_0}{d\varepsilon} \frac{2}{3 \left|L_1^{(0)}(A)\right|^2} \Bigg[ (1-\varepsilon) W_0 r_\pi \, \textrm{Re} \left(C_1^{(1)}(P) L^{(0)*}_{1}(A) \right) \Bigg]\nonumber \\
&\times& \Bigg\{
\frac{m_e}{E_e} \left(\sqrt{1 - \frac{m_{\nu_4}^2}{E_\nu^2} } - 1 \right)   \epsilon_P |U_{e4}|^2 
+\frac{m_{\nu_4}}{E_\nu} \sqrt{1 - \frac{m_{\nu_4}^2}{E_\nu^2} }  
   \tilde\epsilon_P U^*_{e4}
 \Bigg\}.
    \label{sterileP}
\end{eqnarray}

\section{Sensitivity to BSM signatures}\label{BSMplots}

\begin{figure}[t]
\includegraphics[width=0.55\textwidth]{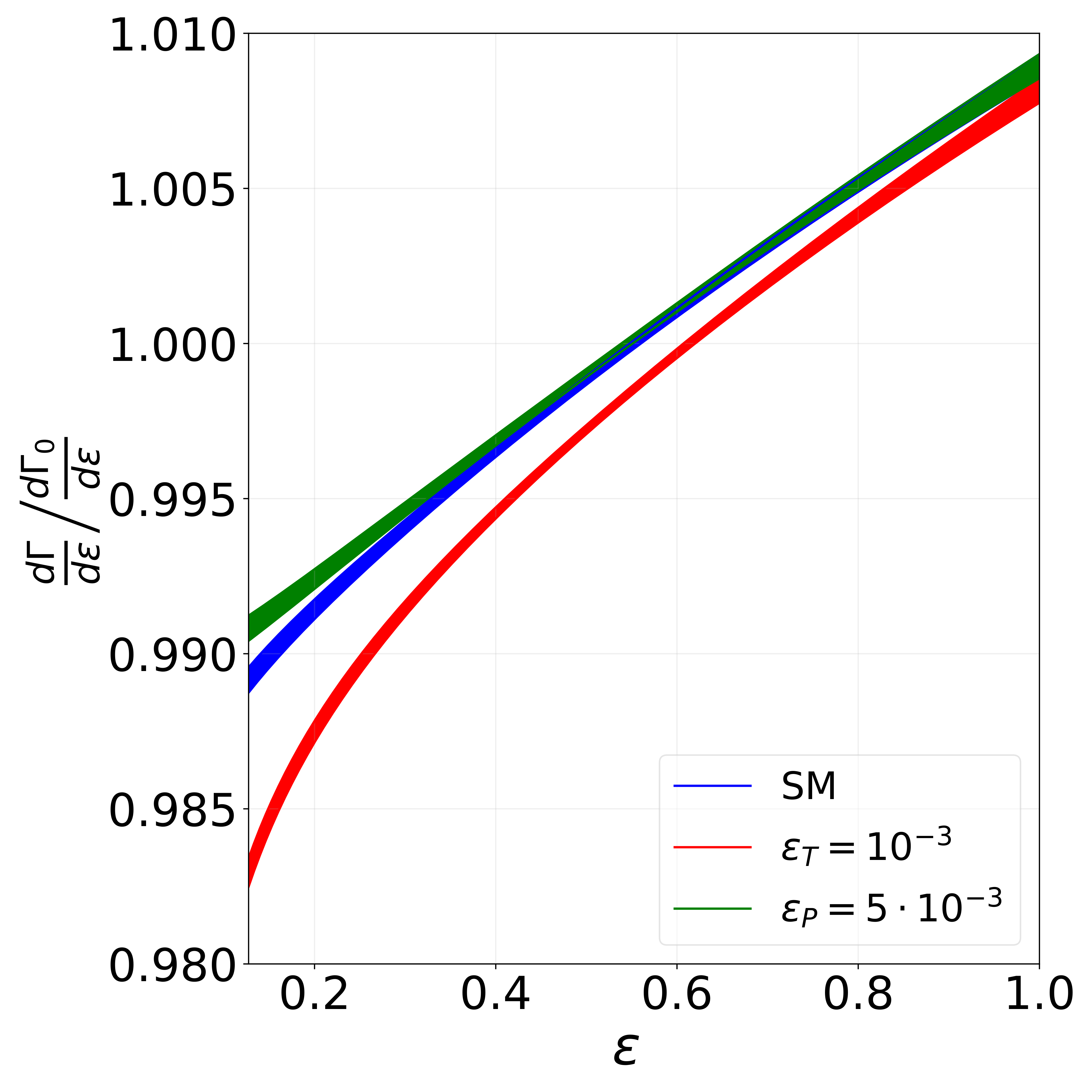}
\caption{
Deviation of the $^6$He $\beta$ spectrum from the expression truncated at leading order in the multipole expansion, given in Eq. \eqref{Gamma0}. The blue curve denotes the SM results, while the red and green lines include the contributions of a tensor and pseudoscalar current, respectively. The width of the bands denotes the theoretical error.}
\label{fig:Tensor}
\end{figure}

We now consider the impact of non-standard charged-current interactions on the energy spectrum. In Fig. \ref{fig:Tensor}, we assume  
all neutrinos to be massless and we set the tensor and pseudoscalar interaction to $\epsilon_T = 10^{-3}$
and $\epsilon_P = 5 \cdot 10^{-3}$, corresponding to new physics scales $\Lambda \sim 8$ and $4$ TeV, respectively. 
Interactions of these size lead to $\sim 10^{-3}$ corrections, which should be resolved in the next generation of experiments.
For both pseudoscalar and tensor interactions, the uncertainty band includes uncertainties on the one-body parameters, $g_T$
and $B$, and the nuclear uncertainties on the multipoles, 
but does not include the truncation to the one-body level, and it is thus slightly underestimated. 
High-invariant mass Drell-Yan production at the LHC  currently probes $\epsilon_T$ at a very similar level \cite{Boughezal:2021tih,Alioli:2018ljm}, while a global analysis of $\beta$ decays
found $\epsilon_T \in [-0.8,1.2] \cdot 10^{-3}$, at the 1$\sigma$ level \cite{Falkowski:2020pma}.
Pseudoscalar interactions are very well constrained by the ratio $\textrm{BR}(\pi \rightarrow e \nu)/\textrm{BR}(\pi \rightarrow \mu \nu)$, which yields $-1.4 \cdot 10^{-7} < \epsilon_P < 5.5 \cdot 10^{-4}$ \cite{Cirigliano:2013xha}. Such values are not in reach of upcoming $^6$He experiments.

We next consider the case of a massive sterile neutrino, which mixes with the electron neutrino with strength $U_{e4}$,
and has non-standard axial, vector, scalar and tensor interactions.
Since the corrections scale in general as $m_{\nu_4}/W_0$, $^6$He decays can probe $m_{\nu_4}$ in the MeV range.
Currently the best limit on the 
a sterile neutrino with mass $m_{\nu_4} = 1$ MeV come from the $\beta$ spectra of $^{20}$F and $^{144}$Pr, and, in the assumption that the neutrino interacts with the SM only via mixing, these constrain a mixing angle $U_{e4} \sim 2\cdot 10^{-3}$  \cite{Calaprice:1983qn,Isakov:1986ez,Deutsch:1990ut,Derbin:2018dbu,Bolton:2019pcu,Dasgupta:2021ies}.
From this we see that we can always neglect terms in Eqs. \eqref{sterilespectrum}, \eqref{sterileT} and \eqref{sterileP} that are proportional to non-standard interactions of active neutrinos, $\epsilon_{L,R,P,T}$, since they are doubly suppressed by $|U_{e4}|^2$ and $v^2/\Lambda^2$. 

\begin{figure}[t]
\includegraphics[width=0.95\textwidth]{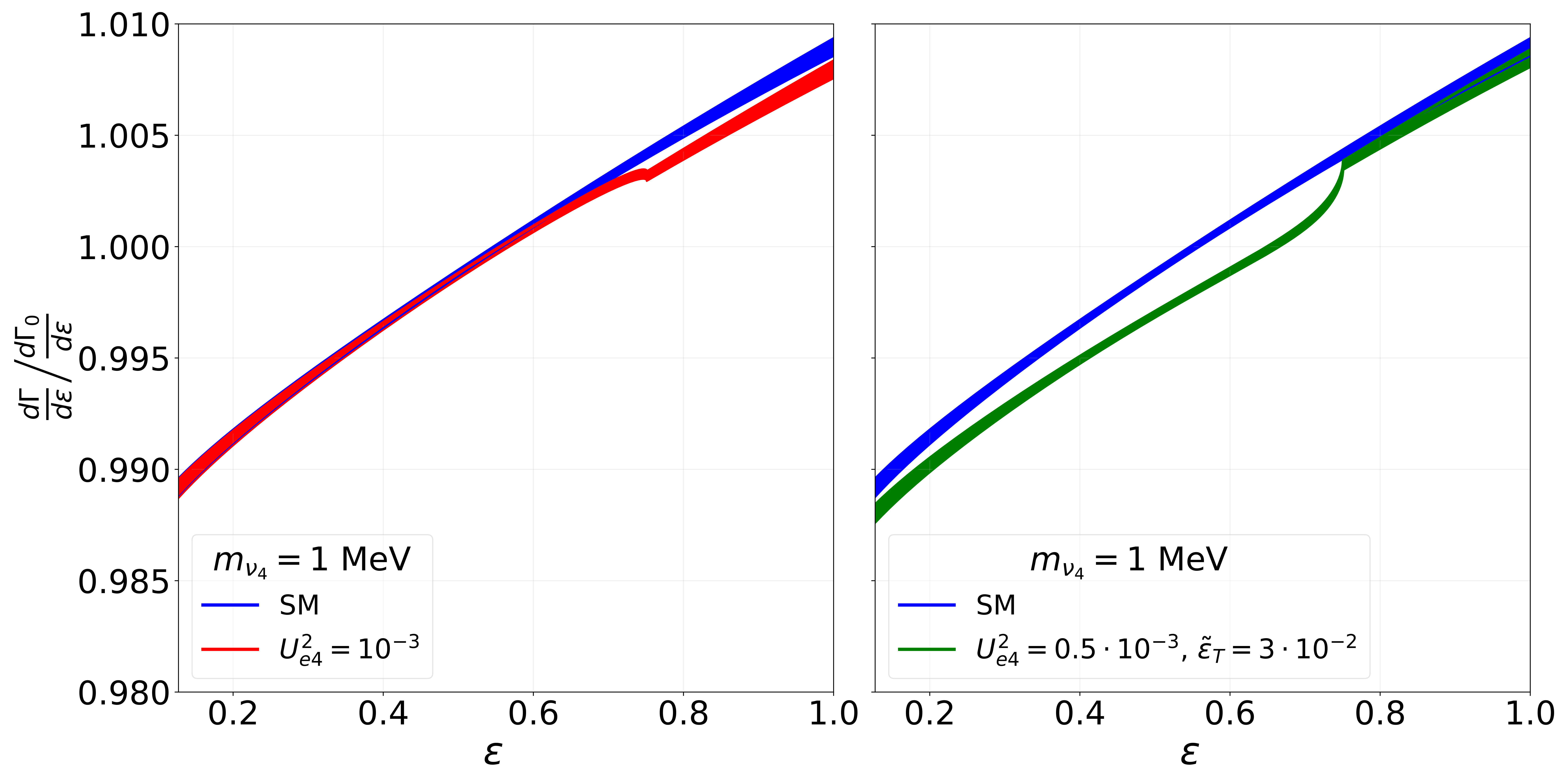}
\caption{Corrections to the $\beta$ spectrum from sterile neutrinos with minimal $(left)$ and tensor $(right)$ interactions, $\tilde{\epsilon}_T$.}
\label{fig:SterileMin}
\end{figure}

In the left panel of Fig. \ref{fig:SterileMin} we show the corrections to the spectrum in the case sterile neutrinos interact with SM particles only via Yukawa interactions. In this case, the spectrum would show a characteristic ``kink'' at $\varepsilon =  1 - m_{\nu_4}/W_0$, due to the emission of a massive neutrino. With mixing $|U_{e4}|^2 = 10^{-3}$, the spectrum receives permille level corrections.

Non-standard interactions of sterile neutrinos cause corrections to the spectrum of order $U_{e4}\, \tilde{\epsilon}_J$, which could thus be relevant for $ \tilde{\epsilon}_J \sim U_{e4} \sim 3\cdot 10^{-2}$, corresponding to new physics scales of 1 TeV.
Sterile neutrinos with an axial coupling to quarks
induce corrections proportional to 
$m_e m_{\nu_4}/(E_e E_\nu)$, which are however fairly small.
More promising is the case of sterile neutrinos with non-standard tensor interactions, $\tilde\epsilon_T$, showed in the right panel of Figure \ref{fig:SterileMin}.
These interactions arise, for example, in leptoquark models \cite{Dekens:2020ttz}.
In this case, an interference term of the form $m_{\nu_4}/E_\nu$
is induced, which has a very different shape compared to tensor interactions of active neutrinos. This is particularly interesting, since the analysis of Ref. \cite{Falkowski:2020pma}
found some preference for a tensor interaction involving sterile neutrinos in $\beta$ decay data.

\section{Conclusions}
\label{sec:conclusions}

We performed  an \textit{ab initio} calculation of the electron energy spectrum in the $\beta$ decay of $^6$He. We used potentials derived from chiral EFT,  with consistent weak vector and axial currents, and adopted Quantum Monte Carlo methods to solve the many-body nuclear problem. We included terms up to second order in the multipole expansion \cite{Walecka:1995mi}, and state-of-the-art electromagnetic corrections, following the treatment of Ref. \cite{Hayen2018}. 
In particular, we included two-body currents,
for the first time in an \textit{ab initio} calculation of the spectrum.
To estimate the theoretical error on the spectrum, we evaluated the matrix elements $L_1$, $E_1$, $M_1$ and $C_1$ with four potential models in the Norfolk 
family of local two- and three-nucleon interactions, derived in chiral EFT  with explicit $\Delta$s and including terms up to N$^3$LO in the chiral expansion. The four interactions have different cut-off, they fit nucleon-nucleon scattering data up to different energies, and use different observables to determine the low-energy constants in the three-body force, and thus they provide a good estimate of the systematic errors in the calculation. 
We find the $^6$He half-life to be in good agreement with experiment. The theoretical uncertainty of about 3\% is  dominated by the determination of the three-body force.  We find the error on the spectral shape to be well below the permille level, and to receive contributions of approximately the same size from $M_1^{(1)}$, $C_1^{(1)}$, $L_1^{(2)}$ and $E_1^{(2)}$.
In the case of $M_1^{(1)}$, which encodes the contribution of weak magnetism, our results agree within theoretical error with the extraction from the electromagnetic transition $^{6}{\rm Li}(0^+,1) \rightarrow ^{6}{\rm Li}(1^+,0) \gamma$, which is exact in the isospin limit. We checked that isospin-breaking terms in the  nuclear potential induce a 1\% difference between $M_1$ and its electromagnetic analog, of the same size as the experimental error. 

$C_1$ is determined by the matrix element of the axial charge density. We find this matrix element to be dominated by the induced pseudoscalar form factor,
in agreement with Ref. \cite{Glick-Magid:2021uwb}.
Finally, we find that $E_1^{(2)}$ and $L_1^{(2)}$, which contribute at N$^2$LO in the multipole expansion, give permille level corrections to the spectrum, and thus need to be included for an accuracy goal of few parts in $10^{-4}$. 
In GFMC, $E_1^{(2)}$ and $L_1^{(2)}$ have relatively large uncertainties,  $26\%$ and $12\%$, respectively. Also in this case, the dominant systematic uncertainty arises from the determination of the three-body force 
and the linear extrapolation of model IIa*. 

Two-body currents play a particularly important role for $M_1$ and $C_1$, which receive a $8\%$ and $\sim 20-30\%$  correction respectively. The effects on $E_1$
and $L_1$ are smaller.

Combining the uncertainties on different matrix elements, we reach a total error on the differential decay rate, normalized by the rate at leading order in the multipole expansion, of at most $4 \cdot 10^{-4}$. We discussed the consequences of such accuracy on non-standard charged-current interactions involving active and sterile neutrinos, showing that the next generation of experiments will be sensitive to tensor interactions and, to a lesser extent, to pseudoscalar interactions of active neutrinos. Future experiments will also constrain sterile neutrinos with mass in the $\sim 1$ MeV region, with both minimal and non-minimal interactions.

\acknowledgments 
We acknowledge stimulating conversations
with A. Garcia, D. Gazit, A. Glick-Magid, M. Hoferichter, J. Men\'endez 
and R. Schiavilla.
A.~B., S.~G. and E.~M. are supported  by the US Department of Energy through  
the Office of Nuclear Physics under contracts DE-AC52-06NA25396,
and  the LDRD program at Los Alamos National Laboratory. Los Alamos National Laboratory is operated by Triad National Security, LLC, for the National Nuclear Security Administration of U.S.\ Department of Energy (Contract No. 89233218CNA000001). The work of S.G. has also been supported by the DOE Early Career research Program. 
L.~H. is supported through the U.S.\ Department of Energy, Low Energy Physics grant DE-FG02-ER41042 and NSF grant PHY-1914133. This work is also supported by the U.S.~Department of Energy under contract DE-SC0021027 (G.~K. and S.~P.), a 2021 Early Career Award number DE-SC0022002 (M.~P.), and the FRIB Theory Alliance award DE-SC0013617 (S.~P. and M.~P.), and  the U.S.\ DOE NNSA Stewardship Science Graduate Fellowship under Cooperative Agreement DE-NA0003960 (G.~K.).
V.~C. is supported by the U.S. Department of Energy under contract  DE-FG02-00ER4113.

 The many-body 
calculations were performed on the parallel computers of the Laboratory Computing Resource Center, Argonne National Laboratory, 
and the computers of 
the Argonne Leadership Computing Facility via the INCITE grant ``Ab-initio nuclear structure and nuclear reactions'', the 2019/2020 ALCC grant ``Low Energy Neutrino-Nucleus interactions'' for the project NNInteractions,
the 2020/2021 ALCC grant ``Chiral Nuclear Interactions from Nuclei to Nucleonic Matter'' for the project ChiralNuc, and by the 2021/2022 ALCC
grant ``Quantum Monte Carlo Calculations of Nuclei up to $^{16}{\rm O}$ and Neutron Matter" for the project QMCNuc. 
This research also used resources provided by the Los Alamos National Laboratory Institutional Computing Program, which is supported by the U.S. Department of Energy National Nuclear Security Administration under Contract No. 89233218CNA000001.

\newpage

\appendix

\section{Effective Lagrangians for charged-current processes}
\label{app:Lagrangians}

\subsection{Charged currents in SMEFT and \texorpdfstring{$\nu$SMEFT}{nuSMEFT}}

The SMEFT Lagrangian includes all operators that are invariant under the SM $SU(3)_c\times SU(2)_L \times U(1)_Y$ gauge group and are built out of SM fields, the left-handed quark and lepton doublets, $Q= (u_L, d_L)^T$ and $L = (\nu_L, e_L)^T$, the right-handed $SU(2)$ singlets $u$,
$d$ and $e$, and the scalar doublet 
\begin{equation}
H = \frac{v}{\sqrt{2}} U(x) \left(\begin{array}{c}
0 \\
1 + \frac{h(x)}{v}
\end{array} \right)\,,
\end{equation}
where  $h(x)$ is the Higgs field, and $U(x)$ is a $SU(2)$ matrix encoding the Goldstone modes.  We will also use $\tilde H = i \tau_2 H^*$.

At dimension-five, the only operator that can be constructed is the LNV Weinberg operator \cite{Weinberg:1979sa}
\begin{equation}
 \mathcal L^{(5)}_{\rm SMEFT} = \epsilon_{kl}\epsilon_{mn}(L_k^T\, C^{( 5)}\,CL_m )H_l H_n\, \label{dim5}
\end{equation}
where $C$ is the charge conjugation matrix. After electroweak symmetry breaking, Eq. \eqref{dim5} induces a Majorana mass term for active neutrinos.
The full, non-redundant, dimension-six Lagrangian is given in Ref. \cite{Grzadkowski:2010es}. For $\beta$ decays, the most important terms are quark and lepton bilinears, which modify the couplings of the $W$ boson to left-handed quarks and leptons and induce new right-handed couplings of the $W$ to quarks, and semileptonic four-fermion operators. 
\begin{eqnarray}\label{SMEFT6}
\mathcal L^{(6)}_{\rm SMEFT}&=&
C^{(6)}_{HL\,3}
(H^\dagger i\overleftrightarrow{D}^I_\mu H)(\bar L \tau^I \gamma^\mu L)
+C^{(6)}_{HQ\, 3}(H^\dag i\overleftrightarrow{D}^I_\mu H)(\bar Q \tau^I \gamma^\mu Q) \nonumber \\
& &
+ C^{(6)}_{Hud}
i(\widetilde H ^\dag D_\mu H)(\bar u \gamma^\mu d) +  C^{(6)}_{LQ\, 3} (\bar L \gamma^\mu \tau^I L)(\bar Q \gamma_\mu \tau^I Q) +
\nonumber \\ & & \Bigg[  C^{(6)}_{LedQ} (\bar L^j e)(\bar d Q^{j}) 
+ C^{(6)}_{Le Qu\, 1}(\bar L ^j e) \epsilon_{jk} (\bar Q^k u)
 + C^{(6)}_{Le Qu\, 3}(\bar L^j \sigma_{\mu\nu} e) \epsilon_{jk} (\bar Q^k \sigma^{\mu\nu} u) + \textrm{h.c.}\Bigg]. \nonumber \\
\end{eqnarray}
All dimension-six operators are lepton-number-conserving (LNC). 
LNV operators arise at dimension-seven and were constructed in Ref. \cite{Lehman:2014jma}. Their contribution to low-energy charged-current operators were considered in Ref. \cite{Dekens:2020ttz}. 

In addition to the SM fields, we introduce a multiplet of sterile neutrinos $\nu_R$, which is a singlet under the SM group.
At dimension-three, this allows to write down a Majorana mass term and a Yukawa interaction, so that the renormalizable $\nu$SMEFT Lagrangian is 
\begin{equation}
\mathcal L_{\nu\rm{SMEFT}} = \mathcal L_{\rm SM} +\bar \nu_R i \gamma^\mu \partial_\mu \nu_R - \left[ \frac{1}{2} \bar \nu^c_{R} \,\bar M_R \nu_{R} +\bar L \tilde H Y_\nu \nu_R + \rm{h.c.}\right],
\end{equation}
where $\bar M_R$ is a symmetric $n\times n$ complex matrix, and $Y_\nu$ is a $3\times n$ matrix of Yukawa couplings. 
The next interactions relevant to $\beta$ decay appear at dimension-six,
\begin{eqnarray}\label{nSMEFT6}
\mathcal L^{(6)}_{\nu\rm{SMEFT}} &=& \mathcal L^{(6)}_{\rm SMEFT}
+C^{(6)}_{H\nu e} (\bar{\nu }_R\gamma^\mu e)({\tilde{H}}^\dagger i D_\mu H)
+ C^{(6)}_{du\nu e} (\bar{d}\gamma^\mu u)(\bar{\nu}_R \gamma_\mu e) \nonumber \\
&+&  C^{(6)}_{Qu\nu L}  (\bar{Q}u)(\bar{\nu}_RL)  + 
C^{(6)}_{L\nu Qd}(\bar{L}_i\nu_R )\epsilon_{ij}(\bar{Q}_j d))
+ C^{(6)}_{LdQ\nu } (\bar{L}_i d)\epsilon_{ij}(\bar{Q}_j \nu_R ) + \textrm{h.c.}\,.
\end{eqnarray}
The first operator induces a coupling of the $W$ boson to $\nu_R$ and a right-handed electron. $C^{(6)}_{du\nu e}$ is a purely right-handed semileptonic charged-current interaction, while the operators on the second line are scalar and tensor interactions of a right-handed neutrino with quarks and left-handed electrons.
Finally, there are LNV operators at dimension-7 \cite{Liao:2016qyd}, which we do not consider here.

\subsection{Charged currents in LEFT}

After electroweak symmetry breaking and integrating out the $W$ boson, the operators in Eq. \eqref{SMEFT6} and \eqref{nSMEFT6} match onto a 
LNC $\beta$ decay Lagrangian in a low-energy, $SU(3) \times U_{\rm em}(1)$ invariant theory (LEFT). 
In the flavor basis, this is given by 
\begin{eqnarray}
\mathcal L^{(6)}_{}& =& -\frac{4 G_F}{\sqrt{2}} \Bigg\{ 
  \bar u_L \gamma^\mu d_L \left[  \bar e_{L}  \gamma_\mu c^{(6)}_{\textrm{VL}} \,  \nu_{L}+ \bar e_{R}  \gamma_\mu \bar c^{(6)}_{\textrm{VL}} \,  \nu_{R} \right]+
  \bar u_R \gamma^\mu d_R \left[\bar e_{L}\,  \gamma_\mu  c^{(6)}_{\textrm{VR}} \,\nu_{L}+\bar e_{R}\,  \gamma_\mu  \bar c^{(6)}_{\textrm{VR}} \,\nu_{R}  \right]\nonumber \\
& & +
  \bar u_L  d_R \left[ \bar e_{R}\, c^{(6)}_{ \textrm{SR}}  \nu_{L} +\bar e_{L}\, \bar c^{(6)}_{ \textrm{SR}}  \nu_{R} \right]+ 
  \bar u_R  d_L \left[ \bar e_{R} \, c^{(6)}_{ \textrm{SL}}    \nu_{L} + \bar e_{L} \, \bar c^{(6)}_{ \textrm{SL}}    \nu_{R} \right]\nonumber \\
&&+  \bar u_R \sigma^{\mu\nu} d_L\,  \bar e_{R}  \sigma_{\mu\nu} c^{(6)}_{ \textrm{T}} \, \nu_{L}+  \bar u_L \sigma^{\mu\nu} d_R\,  \bar e_{L}  \sigma_{\mu\nu} \bar c^{(6)}_{ \textrm{T}} \, \nu_{R}
\Bigg\}  +{\rm h.c.} \,. \label{lowenergy6_l0}
\end{eqnarray}
Here we follow the conventions of Ref. \cite{Dekens:2020ttz}
and denote with 
unbarred and barred lower case coefficients,
such as $c^{(6)}_{\rm VL}$ and $\bar{c}^{(6)}_{\rm VL}$, lepton-number-conserving operators that, in the flavor basis,
involve active and sterile neutrinos, and thus receive matching contributions from dimension-six SMEFT and $\nu$SMEFT operators, respectively. 
In the neutrino mass basis, and assuming one additional light state, $\nu_4$, the Lagrangian becomes
\begin{eqnarray}\label{6final}
\mathcal L^{(6)}& =& -\frac{4 G_F}{\sqrt{2}}  \sum_{i=1}^4 \Bigg\{ 
  \bar u_L \gamma^\mu d_L \left(  \bar e_{R}  \gamma_\mu \left[C^{(6)}_{\textrm{VLR}}\right]_{ei} \,  \nu_i + \bar e_{L}  \gamma_\mu  \left[C^{(6)}_{\textrm{VLL}}\right]_{ei} \,  \nu_i \right) \nonumber  \\ & &  +
  \bar u_R \gamma^\mu d_R \left(\bar e_{R}\,  \gamma_\mu  \left[C^{(6)}_{\textrm{VRR}}\right]_{ei} \,\nu_i+\bar e_{L}\,  \gamma_\mu   \left[C^{(6)}_{\textrm{VRL}}\right]_{ei} \,\nu_i  \right)\nonumber\\
& & +
  \bar u_L  d_R \left( \bar e_{L}\, \left[ C^{(6)}_{ \textrm{SRR}} \right]_{ei} \nu_i +\bar e_{R}\, \left[ C^{(6)}_{ \textrm{SRL}}\right]_{ei}  \nu_i \right)+ 
  \bar u_R  d_L \left( \bar e_{L} \,\left[ C^{(6)}_{ \textrm{SLR}}  \right]_{ei}  \nu + \bar e_{R} \, \left[  C^{(6)}_{ \textrm{SLL}} \right]_{ei}   \nu_i \right)\nonumber\\
&&+  \bar u_L \sigma^{\mu\nu} d_R\,  \bar e_{L}  \sigma_{\mu\nu} \left[C^{(6)}_{ \textrm{TRR}}\right]_{ei} \, \nu^i+  \bar u_R \sigma^{\mu\nu} d_L\,  \bar e_{R}  \sigma_{\mu\nu} \left[ C^{(6)}_{ \textrm{TLL}}\right]_{ei} \, \nu_i
\Bigg\}  +{\rm h.c.}.
\end{eqnarray}
Neglecting lepton-number-violating (LNV) operators, which only arise at dimension-seven in the $\nu$SMEFT,  $C^{(6)}_{\rm VLL}$,
$C^{(6)}_{\rm VRL}$,
$C^{(6)}_{\rm SLL}$,
$C^{(6)}_{\rm SRL}$
and $C^{(6)}_{\rm TLL}$
receive contributions from vector axial, scalar, pseudoscalar and tensor  interactions with left-handed neutrinos,
\begin{align}\label{redefC6}
\left[C_{\rm VLL}^{(6)}\right]_{ei} &=      \left[c_{\rm VL}^{(6)}\right]_{e\alpha} U_{\alpha i}\,,	\qquad 	&\left[C_{\rm VRL}^{(6)}\right]_{ei} &=  \left[c_{\rm VR}^{(6)}\right]_{e \alpha} U_{\alpha i}\,,\nonumber\\
\left[C_{\rm SLL}^{(6)}\right]_{ei} &=   \left[c_{\rm SL}^{(6)}\right]_{e\alpha} U_{\alpha i}\,,		\qquad  &\left[C_{\rm SRL}^{(6)} \right]_{ei}&= \left[ c_{\rm SR}^{(6)}\right]_{e \alpha} U_{\alpha i}\,,\nonumber\\
\left[C_{\rm TLL}^{(6)}\right]_{ei} &=   \left[c_{\rm T}^{(6)}\right]_{e \alpha} U_{\alpha i},
\end{align}
with $i=1,\ldots,4$ and $\alpha$ denotes a charged lepton flavor index, $\alpha \in \{e, \mu,\tau \}$. In practice, we will assume SMEFT operators to be diagonal in lepton flavor, and restrict our attention to the $e e$ components. We will thus drop the flavor subscripts on the $c^{(6)}$ coefficients.
The operators
$C^{(6)}_{\rm VLR}$,
$C^{(6)}_{\rm VRR}$,
$C^{(6)}_{\rm SLR}$
$C^{(6)}_{\rm SRR}$
and $C^{(6)}_{\rm TRR}$
involve sterile neutrinos
\begin{align}\label{redefC6bis}
\left[C_{\rm VLR}^{(6)}\right]_{ei} &= \left[\bar c_{\rm VL}^{(6)}\right]_{e S} U^*_{Si}\,,\qquad 	&
\left[C_{\rm VRR}^{(6)}\right]_{ei} &=\left[ \bar  c_{\rm VR}^{(6)} \right]_{e S} U^*_{Si}\,,\nonumber\\
\left[C_{\rm SLR}^{(6)}\right]_{ei} &= 	\left[ \bar  c_{\rm SL}^{(6)}\right]_{e S} U^*_{Si}\,,\qquad  &\left[C_{\rm SRR}^{(6)}\right]_{ei} &= \left[\bar c_{\rm SR}^{(6)}\right]_{eS} U^*_{Si}\,,\nonumber\\
\left[C_{\rm TRR}^{(6)}\right]_{ei} &=  \left[\bar c_{\rm T}^{(6)}\right]_{e S} U^*_{Si}\,,
\end{align}
where $S$ in the subscript of the PMNS matrix denote the sterile flavor state. In this case, we will absorb the factor of $U_{Si}$ in the coefficient of the effective operators. 

In the body of the paper, we adopt the $\epsilon$ notation defined in Ref. \cite{Cirigliano:2012ab}. For  interactions involving active neutrinos, the relation between the $\epsilon$ couplings and the couplings in Eqs. \eqref{lowenergy6_l0} and \eqref{6final} is given by
\begin{eqnarray}
V_{ud} \left(1 + \epsilon_L\right) &=& c^{(6)}_{\rm VL}, \qquad \qquad \; \; V_{ud}\,\epsilon_R  = c^{(6)}_{\rm VR}, \nonumber \\
V_{ud} \,\epsilon_S &=& c_{\rm SR}^{(6)} + c_{\rm SL}^{(6)}, \qquad 
V_{ud} \,\epsilon_P = c_{\rm SR}^{(6)} - c_{\rm SL}^{(6)}, \qquad
V_{ud} \, \epsilon_T = c_{\rm T}^{(6)},
\end{eqnarray}
while, in the case of sterile neutrinos, we have 
\begin{eqnarray}
V_{ud}\, \tilde\epsilon_L &=& \left[\bar{ c}^{(6)}_{\rm VL}\right]_{e S} U_{S4}^*, \qquad \qquad \;\; V_{ud}\, \epsilon_R  = \left[ \bar{c}^{(6)}_{\rm VR}\right]_{eS}  U_{S4}^*,  \\
V_{ud}\, \tilde \epsilon_S &=& \left[\bar{c}_{\rm SR}^{(6)} + \bar{c}_{\rm SL}^{(6)}\right]_{e S} U^*_{S4}, \qquad 
V_{ud}\, \tilde \epsilon_P = \left[c_{\rm SR}^{(6)} - c_{\rm SL}^{(6)}\right]_{eS} U^*_{S4}, \qquad
V_{ud}\, \tilde\epsilon_T = \left[\bar c_{\rm T}^{(6)}\right]_{e S} U^*_{S4}.\nonumber 
\end{eqnarray}

The matching between SMEFT, $\nu$SMEFT 
and Eq. \eqref{lowenergy6_l0}
was carried out in Ref. \cite{Dekens:2020ttz},
and here we report the results. For non-standard interactions involving active neutrinos, one finds
\begin{eqnarray}\label{match6LNC}
\left[c_{\rm VL}^{(6)}\right]_{\alpha \beta} &=&  V_{ud} \, \delta_{\alpha \beta} - v^2\left[C_{LQ\,3}^{(6)}-C_{HL\,3}^{(6)}\right]_{\alpha\beta}
+ v^2 \left[C_{HQ\,3}^{(6)}\, \delta_{\alpha\beta}\right],\nonumber\\
\left[c_{\rm VR}^{(6)}\right]_{\alpha\beta} &=& \frac{v^2}{2}C_{Hud}^{(6)}\, \delta_{\alpha\beta},\nonumber\\
\left[c_{\rm SR}^{(6)}\right]_{\alpha\beta} &=& -\frac{v^2}{2} \left[C_{LedQ}^{(6)}\right]^*_{\beta \alpha}\,,\nonumber\\
\left[c_{\rm SL}^{(6)}\right]_{\alpha\beta} &=& -\frac{v^2}{2}\left[C_{LeQu\,1}^{(6)}\right]^*_{\beta\alpha}\,,\nonumber\\
\left[c_{\rm T}^{(6)}\right]_{\alpha\beta} &=& -\frac{v^2}{2}\left[C_{LeQu\,3}^{(6)}\right]^*_{\beta\alpha}\,.
\end{eqnarray}
Here $\alpha$ and $\beta$ denote charge lepton flavor indices, $\alpha, \beta \in \{e, \mu, \tau\}$, while we are always assuming the quark flavor indices to be $u$ and $d$.
The matching coefficients of LNC sterile neutrino operators are
\begin{eqnarray}\label{match6LNCsterile}
\left[\bar c_{\rm VL}^{(6)}\right]_{\alpha S} &=&  \frac{v^2}{2}\left[C_{H\nu e}^{(6)} \right]^*_{S \alpha}\,,\nonumber\\
\left[\bar c_{\rm VR}^{(6)}\right]_{\alpha S} &=& -\frac{v^2}{2}\left[C_{du\nu e}^{(6)}\right]^*_{S \alpha}\,,\nonumber\\
\left[\bar c_{\rm SR}^{(6)}\right]_{\alpha S}&=&  \frac{v^2}{2}\left[C_{L\nu Qd}^{(6)}\right]_{\alpha S}-\frac{v^2}{4} \left[C_{LdQ\nu }^{(6)}\right]_{\alpha S}\,,\nonumber\\
\left[\bar c_{\rm SL}^{(6)}\right]_{\alpha S}&=& -\frac{v^2}{2 }\left[C_{Qu\nu L}^{(6)}\right]^*_{S \alpha}\,,\nonumber\\
\left[\bar c_{\rm T}^{(6)}\right]_{\alpha S} &=& -\frac{v^2}{16} \left[ C_{LdQ\nu }^{(6)}\right]_{\alpha S}\,,
\end{eqnarray}
where $S$ denotes a sterile flavor index.
Ref. \cite{Dekens:2020ttz} also reports the contribution of LNV SMEFT and $\nu$SMEFT operators to Eq. \eqref{6final}.

\subsection{Charged currents in the Chiral Lagrangian}
The quark-level SM and SMEFT Lagrangians
lead to interactions between pions and nucleons, which can be organized in chiral perturbation theory \cite{Gasser:1983yg,Gasser:1984gg,Bernard:1995dp}
and are the building blocks for the derivation of the one- and two-body nuclear currents used in this paper. 
Axial and pseudoscalar interactions induce couplings to a single pion, 
\begin{eqnarray}
\mathcal L_\pi &=& 2 G_F F_\pi \Bigg\{  \partial^\mu \pi^-
\left(  \bar e_{R}  \gamma_\mu \left[C^{(6)}_{\textrm{VLR}}-C^{(6)}_{\textrm{VRR}}\right]_{ei} \,  \nu_i + \bar e_{L}  \gamma_\mu  \left[C^{(6)}_{\textrm{VLL}}-C^{(6)}_{\textrm{VRL}}\right]_{ei} \,  \nu_i \right) \nonumber \\
& & + i B \pi^-\,
\left( \bar e_{L} \,\left[ C^{(6)}_{ \textrm{SLR}} - C^{(6)}_{ \textrm{SRR}} \right]_{ei}  \nu_i + \bar e_{R} \, \left[  C^{(6)}_{ \textrm{SLL}} - C^{(6)}_{ \textrm{SRL}} \right]_{ei}   \nu_i \right) \Bigg\} + \ldots, \label{eq:Lpi}
\end{eqnarray}
where $\ldots$ denotes terms with multiple pions and affect the nuclear currents beyond leading order. Axial and pseudoscalar interactions induce terms with an odd number of pions, while scalar and vector terms with an even number of pions.
$F_\pi = 92$ MeV is the pion decay constant, while $B = m_\pi^2/(m_u + m_d)\sim 2.8$ GeV, at the renormalizations scale $\mu=2$ GeV. The strong coupling of the pion to pseudoscalar   operators implies that this interaction dominates the nucleon pseudoscalar density.
In the heavy-baryon formalism, the nucleon Lagrangian at leading order is
\begin{eqnarray}
    \mathcal L_N &=&- \frac{2 G_F}{\sqrt{2}} \bar N \tau^+
    \Bigg\{ v^\mu \left(  \bar e_{R}  \gamma_\mu \left[C^{(6)}_{\textrm{VLR}}+C^{(6)}_{\textrm{VRR}}\right]_{ei} \,  \nu_i + \bar e_{L}  \gamma_\mu  \left[C^{(6)}_{\textrm{VLL}}+C^{(6)}_{\textrm{VRL}}\right]_{ei} \,  \nu_i \right) \nonumber \\
    & & - 2 g_A S^\mu \left(  \bar e_{R}  \gamma_\mu \left[C^{(6)}_{\textrm{VLR}}-C^{(6)}_{\textrm{VRR}}\right]_{ei} \,  \nu_i + \bar e_{L}  \gamma_\mu  \left[C^{(6)}_{\textrm{VLL}}-C^{(6)}_{\textrm{VRL}}\right]_{ei} \,  \nu_i \right)\nonumber 
    \\ & & 
    + g_S \left( \bar e_{L} \,\left[ C^{(6)}_{ \textrm{SLR}} + C^{(6)}_{ \textrm{SRR}} \right]_{ei}  \nu + \bar e_{R} \, \left[  C^{(6)}_{ \textrm{SLL}} + C^{(6)}_{ \textrm{SRL}} \right]_{ei}   \nu_i \right) \nonumber \\
 & &   - 4 g_T \varepsilon^{\mu \nu \alpha \beta}v_\alpha S_\beta \left(\bar e_{L}  \sigma_{\mu\nu} \left[C^{(6)}_{ \textrm{TRR}}\right]_{ei} \, \nu^i +  \bar e_{R}  \sigma_{\mu\nu} \left[ C^{(6)}_{ \textrm{TLL}}\right]_{ei} \, \nu_i \right)
\Bigg\}N,  \label{eq:Lnuc}
\end{eqnarray}
where  $N$ denotes a non-relativistic nucleon field, with velocity $v^\mu = (1,{\bf 0})$
and spin $S^\mu = (1, \boldsymbol{ \sigma}/2)$, in the nucleon rest frame.
In the absence of non-standard currents, the axial charge can be extracted from neutron decay.
In this work, we adopt for the value of the axial charge \cite{Zyla:2020zbs}
\begin{eqnarray}\label{gA}
\frac{g_A}{g_V} = 1.2754 \pm 0.0013.
\end{eqnarray}
This value is slightly larger and with roughly half the uncertainty of the one in the 2018 version of the PDG, 
$\left. g_A\right|_{18} = 1.2723(23)$ \cite{ParticleDataGroup:2018ovx}, which is used in the code. 
In obtaining the half-life and $\beta$ spectrum, 
we rescale the leading multipoles 
$E_1$ and $L_1$, given in Tables \ref{Results} and \ref{GFMC},  by $g_A/\left. g_A\right|_{18}$. For the subleading multipoles, $C_1$ in particular, the difference is well within the theoretical error. In the future, one can envision using 
lattice QCD extraction of the axial charge \cite{Bhattacharya:2016zcn,Berkowitz:2017gql,Chang:2018uxx,Gupta:2018qil,Aoki:2019cca,Aoki:2021kgd}, which is approaching percent level accuracy.

The scalar and tensor isovector charges have been computed in lattice QCD \cite{Bhattacharya:2016zcn,Gupta:2018qil,Aoki:2019cca,Aoki:2021kgd}. We will use for the scalar and tensor isovector charges the averages
of the lattice results with $N_f = 2 +1 +1$ flavors of dynamical quarks, given in Ref. \cite{Aoki:2019cca,Aoki:2021kgd} 
\begin{eqnarray}\label{charges}
\qquad g_S = 1.02 \pm 0.10, \qquad g_T = 0.989 \pm 0.033 .
\end{eqnarray}
Eqs. \eqref{eq:Lpi} and \eqref{eq:Lnuc} are sufficient for the construction of one-body currents at LO. The construction of vector and axial currents to subleading orders is 
reviewed in Refs. \cite{Pastore:2009is,Krebs:2019aka,Baroni:2015uza,Krebs:2016rqz}. For the BSM scalar, pseudoscalar and tensor currents, it is sufficient to work at LO.

\section{Multipole expansion for SM and BSM currents}
\label{app:multi}

The derivation of the multipole expansion for SM currents is reviewed in Ref. \cite{Walecka:1995mi}.
The starting point is the weak Hamiltonian
\begin{eqnarray}\label{eq:Hweak}
H_w &=&  \frac{G_F}{\sqrt{2}} V_{ud} \int d^3 \textbf{x} \,  j_\mu^{\textrm{ lept}}({\bf x}) \mathcal J_{V-A}^\mu({\bf x})  \nonumber \\ &=& - \frac{G_F}{\sqrt{2}} V_{ud}\int d^3 \textbf{x} \, \left( \textbf{j}^{\textrm{lept}}({\bf x}) \cdot \boldsymbol{ \mathcal{J}}_{V-A}({\bf x}) - j_0^{\textrm{lept}}({\bf x}) \mathcal J_{V-A}^0({\bf x}) \right),
\end{eqnarray}
where 
\begin{eqnarray}
j_\mu^{\rm lept} = 2 \bar e_L \gamma_\mu \nu_L.
\end{eqnarray}
Here $\mathcal J_{V-A}^\mu$ denotes the hadronic realization of the quark current $\bar u \gamma^\mu (1-\gamma_5) d$, and the derivation only assumes that such a realization exist. The first few orders of the explicit representation of   $\mathcal J_{V-A}^\mu$ in chiral EFT will be given in Appendix \ref{app:Onebody}.
Introducing the scalar and vector under rotations
\begin{eqnarray}
\ell_0 = 2\langle e \bar\nu | \bar e_L \gamma_0 \nu_L | 0 \rangle,  \qquad {\rm and} \qquad \boldsymbol{\ell} = 2 \langle e \bar\nu | \bar e_L \boldsymbol{\gamma} \nu_L | 0 \rangle,
\end{eqnarray}
we can write the matrix element
\begin{eqnarray}
\langle f e \bar \nu| H_{w} | i \rangle  =  - \frac{G_F V_{ud}}{\sqrt{2}} \int d^3 \textbf{x} \, \left( \boldsymbol{\ell} \cdot \left(\boldsymbol{\mathcal J}_{V-A}\right)_{fi} - \ell_0 \, \left(\mathcal J^0_{V-A}\right)_{fi}\right)\, , \label{intro3}
\end{eqnarray}
where $| i \rangle$ and $| f \rangle$ denote the initial and final nuclear states. We recall here that the leptonic tensor can be written as
\begin{eqnarray}
\ell^\mu&=&e^{-i({\bf p}_e+{\bf p}_\nu)\cdot{\bf x}}\tilde{\ell}^\mu\, ,
\end{eqnarray}
with
\begin{equation}
    \tilde{\ell}^\mu=\overline{u}_e\gamma^\mu(1-\gamma_5)v_\nu\, ,
\end{equation}
with $\overline{u}_e$ and $v_\nu$ spinors of the electron and electronic antineutrino respectively.
The matrix element becomes
\begin{eqnarray}
\langle f e \bar \nu| H_{w} | i \rangle  =  - \frac{G_F V_{ud}}{\sqrt{2}} \int d^3 \textbf{x} e^{-i{\bf q}\cdot{\bf x}}\, \left( \tilde{\boldsymbol{\ell}} \cdot \left(\boldsymbol{\mathcal J}_{V-A}\right)_{fi} - \tilde{\ell_0} \, \left(\mathcal J^0_{V-A}\right)_{fi}\right)\, ,
\end{eqnarray}
we can define now
\begin{eqnarray}
\left({\mathcal J}^\mu_{V-A}\right)_{fi}(-{\bf q})&=&\int d{\bf x}\,\, e^{-i{\bf q}\cdot{\bf x}}\left({\mathcal J}^\mu_{V-A}\right)_{fi}({\bf x}).
\end{eqnarray}
We define $ \left({\mathcal J}^\mu_{V-A}\right)_{fi}(-{\bf q})=\left({\mathcal J}^\mu_{V-A}\right)^\dagger_{fi}({\bf q})$ and, for ease of notation, we write
\begin{eqnarray}
h^\mu_a=\bra{f} {\mathcal J}^{\mu\dagger}_{V-A}({\bf q})\ket{i}
\end{eqnarray}
that leads to the following matrix element for the interaction Hamiltonian
\begin{eqnarray}
\langle f e \bar \nu| H_{w} | i \rangle  =  \frac{G_F V_{ud}}{\sqrt{2}}  \tilde{\ell}_\mu \,\,  h^\mu_a
\end{eqnarray}
Similarly to what has been done in Ref.~\cite{Walecka:1995mi} we can decompose the space part of the leptonic tensor $\tilde{\ell}^\mu$ in terms of spherical coordinates
\begin{eqnarray}
\tilde{\boldsymbol{\ell}}=\tilde{\ell}_3\hat{\bf e}_{q0}^\dagger+\sum_{\lambda=\pm 1} \tilde{\ell}_\lambda \hat{\bf e}^\dagger_{q\lambda}\, 
\end{eqnarray}
where the $\hat{\boldsymbol{e}}_{q, \lambda}$ are defined as \cite{Walecka:1995mi,Schiavilla2002}
\begin{eqnarray}
\hat{\bf e}_{q,\pm 1}=\mp\frac{1}{\sqrt{2}}(\hat{\bf e}_{q1}\pm\hat{\bf e}_{q2})\, ,\qquad \hat{\bf e}_{q0}=\hat{\bf e}_{q3}\, 
\end{eqnarray}
where $\hat{\bf e}_{q3}=\hat{\bf q}$, $\hat{\bf e}_{q2}=\hat{\bf z}\times {\bf q}/\lvert \hat{\bf z}\times {\bf q} \rvert$ and $\hat{\bf e}_{q1}=\hat{\bf e}_{q2}\times\hat{\bf e}_{q3}$.
The matrix element can now be expressed as
\begin{eqnarray}
\langle f e \bar \nu| H_{w} | i \rangle  =  \frac{G_F V_{ud}}{\sqrt{2}} \left(\Tilde{\ell}_0h^0_a-\tilde{\ell}_3\hat{\bf e}_{q0}^\dagger\cdot{\bf h}_a-\sum_{\lambda=\pm 1}\tilde{\ell}_\lambda\hat{
\bf e}_{q\lambda}^\dagger\cdot{\bf h}_a\right)
\end{eqnarray}

We can expand Eq. \eqref{intro3} in a sum of terms with well defined total angular momentum. For the SM current one finds \cite{Walecka:1995mi}
\begin{eqnarray}
\label{eq:fhi}
\langle f  e \bar \nu| H_w | i \rangle &=& -\frac{G_F}{\sqrt{2}} V_{ud} \langle f | \Bigg\{ 
- \sum_{J\ge 1}\sqrt{2\pi (2J+1 )} (-i)^J \sum_{\lambda = \pm 1}\ell_\lambda \left[\lambda \mathcal M_{J -\lambda}(q)+ \mathcal E_{J\, -\lambda}(q)\right] \nonumber \\
& &
+ \sum_{J \ge 0 } \sqrt{4\pi (2 J +1)} (-i)^J \left(\ell_3 \mathcal L_{J0}(q)
- \ell_0 \mathcal C_{J0}(q)
\right) 
\Bigg\} | i \rangle,
\end{eqnarray}
The multipole operators, defined in Ref.~\cite{Walecka:1995mi}, are
\begin{eqnarray}
\mathcal C_{JM}(q) &=& \int d^3 x \,  j_J (q x) Y_{JM}(\Omega_x) \left( \mathcal J^0_V(x) + \mathcal J^0_A(x)\right), \label{multiSM1}\\
\mathcal L_{JM}(q) &=& \frac{i}{q}\int d^3 x \, \nabla [  j_J (q x) Y_{JM}(\Omega_x)] \cdot \left( \boldsymbol{\mathcal J}_V(x) + \boldsymbol{\mathcal J}_A(x)\right), \label{multiSM2}\\
\mathcal E_{JM}(q) &=& \int d^3 x \, [\nabla \times  j_J (q x) \boldsymbol{\mathcal Y}^{M}_{JJ1}(\Omega_x)  ] \cdot \left( \boldsymbol{\mathcal J}_V(x) + \boldsymbol{\mathcal J}_A(x)\right), \label{multiSM3}\\
\mathcal M_{JM}(q) &=& \frac{i}{q}\int d^3 x [  j_J (q x) \boldsymbol{\mathcal Y}^{M}_{JJ1}(\Omega_x)  ] \cdot \left( \boldsymbol{\mathcal J}_V(x) + \boldsymbol{\mathcal J}_A(x)\right), \label{multiSM4}
\end{eqnarray}
in terms of spherical Bessel function $j_J$, spherical harmonics $Y_{JM}$
and vector spherical harmonics $\boldsymbol{\mathcal Y}^{M}_{JJ1}(\Omega_x)$ (see Ref. \cite{Walecka:1995mi} for the relevant definitions). 
The matrix elements $L_1$, $E_1$, $M_1$ and $C_1$
that enter the decay rate in Eqs. \eqref{rate} and 
\eqref{eq:dGammageneral} are the reduced matrix elements of the operators \eqref{multiSM1}-\eqref{multiSM4}.
The matrix elements between a generic multipole ${\cal T}_{JM}$ between initial and final nuclear states can be written as
\begin{eqnarray}
\bra{J_f,M_f}{\cal T}^\dagger_{JM}\ket{J_i,M_i}&=&\alpha (-)^M (-1)^{J_i-M_i}\frac{1}{\sqrt{2L+1}}\bra{J_f,M_f;J_i,-M_i}J,-M\rangle T_J(q)\nonumber\\
\end{eqnarray}
with $\alpha=1$ for the multipole operator ${\cal C}$ and $\alpha=-1$ for the multipole operators ${\mathcal L}$, ${\mathcal E}$ and ${\mathcal M}$. We denote with $T_J(q)$ the reduced matrix element associated with the generic multipole operator ${\cal T}_{JM}$.
We are now in the position to obtain Eqs. \eqref{eq:ME1}-\ref{eq:ME4} reported in the main text.
We recall that for the problem of interest $J_i=0$ and $J_f=1$ therefore, using selection rules and the identities of Ref.~\cite{Schiavilla2002} we arrive at
\begin{eqnarray}
h_a^0&=&-4\pi i Y_{10}(\hat{\bf q}) \frac{1}{\sqrt{3}}  C_1(q) \label{eq:ME1_pre}\\
\hat{\bf e}^\dagger_{q0}\cdot{\bf h}_a&=&-4\pi i Y_{1,0}(\hat{\bf q})\frac{1}{\sqrt{3}} L_1(q)\label{eq:ME2_pre}\\
\hat{\bf e}^\dagger_{q\lambda}\cdot{\bf h}_a&=&-\sqrt{2\pi}i D^{1\dagger}_{M_f,\lambda}(-\phi,-\theta,\phi) (\lambda M_1(q)+E_1(q))\label{eq:ME3_pre}
\end{eqnarray}
where in the last passage we recall the following definition
\begin{equation}
    D^{1\dagger}_{M_f,\lambda}(-\phi,-\theta,\phi)=\sqrt{\frac{4\pi}{3}}Y^\star_{1,\lambda}(\theta,\phi)\, .
\end{equation}
We can now see that taking $\hat{\bf q}$ along $\hat{\bf z}$, Eqs. ~\eqref{eq:ME1_pre}-\eqref{eq:ME2_pre} lead to Eqs. ~\eqref{eq:ME1}-\eqref{eq:ME2}.  Finally taking in Eq. \eqref{eq:ME3_pre} $\hat{\bf q}$ along $\hat{\bf x}$ we obtain a linear system of two equation for two different values of $\lambda$, whose solution leads to Eqs. \eqref{eq:ME3} and \eqref{eq:ME4}.

We can generalize Eq. \eqref{eq:fhi} to non-standard currents induced by $(\nu)$SMEFT operators. 
The most general Hamiltonian including dimension-six operator in the SMEFT has the form 
\begin{eqnarray}
H_{\textrm{6}} =  - \frac{G_F}{\sqrt{2}}  \int d^3 \textbf{x} \, \left( 
- j^{\textrm{lept}}_S \mathcal J_S - j^{\textrm{lept}}_P \mathcal J_P  + 2 ( j^{\textrm{lept}}_T)^{i0} \mathcal J^{i0}_{T} - (j^{\textrm{lept}}_T)_{ij} \mathcal J^{ij}_{T} \right. \nonumber \\ \left.
\textbf{j}^{\textrm{lept}}_V \cdot \boldsymbol{ \mathcal{J}}_{V} 
+ \textbf{j}^{\textrm{lept}}_A \cdot \boldsymbol{ \mathcal{J}}_{A} 
- \left( j_0^{\textrm{lept}}\right)_V \mathcal J_{V}^0
- \left( j_0^{\textrm{lept}}\right)_A \mathcal J_{A}^0
\right),
\end{eqnarray}
where the leptonic currents are
\begin{eqnarray}
j^{\textrm{lept}}_S &=& 2 \left(C^{(6)}_{\rm SRL} + C^{(6)}_{\rm SLL}\right) \bar e_R  \nu
+ 2 \left(C^{(6)}_{\rm SRR} + C^{(6)}_{\rm SLR}\right)\bar e_L  \nu \\
j^{\textrm{lept}}_P &=& 2 \left(C^{(6)}_{\rm SRL} - C^{(6)}_{\rm SLL}\right) \bar e_R  \nu
+ 2 \left(C^{(6)}_{\rm SRR} - C^{(6)}_{\rm SLR}\right)\bar e_L  \nu \\
(j^{\textrm{lept}}_T )^{\mu \nu}&=& 4 C^{(6)}_{\rm TLL} \bar e_R \sigma^{\mu \nu}  \nu + 
4 C^{(6)}_{\rm TRR} \bar e_L \sigma^{\mu \nu}  \nu,\\
\left(j^{\textrm{lept}}_V\right)^\mu &=& 
2 \left( C^{(6)}_{\rm VRL} + C^{(6)}_{\rm VLL}\right) \bar e_L \gamma^\mu \nu
+ 2 \left(C^{(6)}_{\rm VRR} + C^{(6)}_{\rm VLR}\right)\bar e_R \gamma^\mu  \nu \\
\left(j^{\textrm{lept}}_A\right)^\mu &=& 
2 \left( C^{(6)}_{\rm VLL}  - C^{(6)}_{\rm VRL}  \right) \bar e_L \gamma^\mu \nu
+ 2 \left(C^{(6)}_{\rm VLR} -C^{(6)}_{\rm VRR}  \right)\bar e_R \gamma^\mu  \nu
\end{eqnarray}
and the hadronic currents the nucleon-level realization of
\begin{eqnarray}
 \bar u d \rightarrow \mathcal J_S, \quad
 \bar u \gamma_5 d \rightarrow \mathcal J_P,\quad
 \bar u \sigma^{\mu \nu} d \rightarrow \mathcal J^{\mu\nu}_T, \quad 
 \bar u \gamma^\mu d \rightarrow \mathcal J_V, \quad
 \bar u \gamma^\mu \gamma_5 d \rightarrow -\mathcal J_A.
\end{eqnarray}
Introducing the leptonic matrix elements 
\begin{eqnarray}
\ell_S &=& \langle e \bar\nu | j_S^{\rm lept} | 0 \rangle,\qquad \ell_P = \langle e \bar\nu | j_P^{\rm lept} | 0 \rangle, \qquad 
(\ell_0)_{V,\, A} = \langle e \bar\nu | (j^{\rm lept}_{V,\,A})^0 | 0 \rangle
\end{eqnarray}
which are scalar under rotations, and the vectors 
\begin{eqnarray}
 \boldsymbol{\ell}_T = -\frac{1}{2}\varepsilon^{i j k}\langle e \bar\nu | (j_T^{\rm lept})^{i j} | 0 \rangle, \qquad
\boldsymbol{\ell}^\prime_T = 2 \langle e \bar\nu | (j_T^{\rm lept})^{i 0} | 0 \rangle \qquad \boldsymbol{\ell}_{V,\,A} =  \langle e \bar\nu | (j^{\rm lept}_{V,\,A})^i | 0 \rangle,
\end{eqnarray}
the matrix element of the Hamiltonian becomes  
\begin{eqnarray}
\langle f e \bar \nu| H_{\textrm{6}} | i \rangle  &=&  - \frac{G_F}{\sqrt{2}} \int d^3 \textbf{x} \, \left( 
- \ell_S (\mathcal J_S)_{fi} - \ell_P (\mathcal J_P)_{fi} 
- (\ell_0)_V  \left(\mathcal J^0_{V}\right)_{fi}  
- (\ell_0)_A  \left(\mathcal J^0_{A}\right)_{fi} \right. \nonumber \\ 
& & \left. +  \boldsymbol{\ell}^\prime_T \cdot (\boldsymbol{ \mathcal J}^{\prime}_{T})_{fi} + 
\boldsymbol{\ell}_T \cdot (\boldsymbol{ \mathcal J}^{}_{T})_{fi} +  \boldsymbol{\ell}_V \cdot \left(\boldsymbol{\mathcal J}_{V}\right)_{fi} 
+  \boldsymbol{\ell}_A \cdot \left(\boldsymbol{\mathcal J}_{A}\right)_{fi}
 \right),\label{intro4}
\end{eqnarray}
where 
\begin{eqnarray}
\left(\boldsymbol{ \mathcal J}^{\prime}_{T}\right)^k_{fi} = \langle f | \mathcal J^{k 0}_{T}  | i \rangle \qquad 
\left(\boldsymbol{ \mathcal J}^{}_{T}\right)^k_{fi} =  \varepsilon^{k l m }\langle f | \mathcal J^{l m}_{T}  | i \rangle .
\label{intro5}
\end{eqnarray}

Since at dimension-six in the  
$\nu$SMEFT the leptonic and hadronic currents have at most spin one \footnote{In the SMEFT, spin two charged currents only arise at dimension-eight \cite{Alioli:2020kez}},
for both scalar/tensor and vector/axial operators,  the derivation of the multipole expansion therefore proceeds  as in the SM.
Additional vector and axial operators generate exactly the same multipoles as in the SM.
For scalar and pseudoscalar currents, only the $\mathcal C_{J0}$ multipole is present. Tensor currents generate  electric, magnetic and longitudinal multipoles, but not $\mathcal C_J$. One thus finds
\begin{eqnarray}
\label{fhibsm}
\langle f | H_{6} | i \rangle &=& -\frac{G_F}{\sqrt{2}}  \langle f | \Bigg\{ 
- \sum_{J\ge 1}\sqrt{2\pi (2J+1 )} (-i)^J \sum_{\lambda = \pm 1}\ell^T_\lambda \left[\lambda \mathcal M^T_{J -\lambda}(q)+ \mathcal E^T_{J\, -\lambda}(q)\right] \nonumber \\
& &- \sum_{J\ge 1}\sqrt{2\pi (2J+1 )} (-i)^J \sum_{\lambda = \pm 1}\ell^{\prime T}_\lambda \left[\lambda \mathcal M^{T \prime}_{J -\lambda}(q)+ \mathcal E^{T\prime}_{J\, -\lambda}(q)\right] \nonumber \\
& &
+ \sum_{J \ge 0 } \sqrt{4\pi (2 J +1)} (-i)^J \left(\ell^T_3 \mathcal L^T_{J0}(q)
+ 
\ell^{T \prime}_3 \mathcal L^{T \prime}_{J0}(q)
\right)  \nonumber \\
& & 
- \sum_{J \ge 0 } \sqrt{4\pi (2 J +1)} (-i)^J \
  \left( \ell_S \mathcal C^{S}_{J0}(q) + \ell_P \mathcal C^{P}_{J0}(q)\right)
\Bigg\} | i \rangle,
\end{eqnarray}
with 
\begin{eqnarray}
\{\mathcal C^{S}_{JM}(q) , \mathcal C^{P}_{JM}(q)\} &=& \int d^3 x \,  j_J (q x) Y_{JM}(\Omega_x) \{\mathcal{J}_{S}(x), \mathcal{J}_P(x))\}, \label{multiBSM1}
\end{eqnarray}
and 
\begin{eqnarray}
\mathcal L^{T (\prime)}_{JM}(q) &=& \frac{i}{q}\int d^3 x \, \nabla [  j_J (q x) Y_{JM}(\Omega_x)] \cdot \boldsymbol{\mathcal J}^{(\prime)}_{T}(x), \label{multiBSM2}\\
\mathcal E^{T (\prime)}_{JM}(q) &=& \int d^3 x \, [\nabla \times  j_J (q x) \boldsymbol{\mathcal Y}^{M}_{JJ1}(\Omega_x)  ] \cdot \boldsymbol{\mathcal J}^{(\prime)}_T(x),  \label{multiBSM3}\\
\mathcal M^{T (\prime)}_{JM}(q) &=& \frac{i}{q}\int d^3 x [  j_J (q x) \boldsymbol{\mathcal Y}^{M}_{JJ1}(\Omega_x)  ] \cdot \boldsymbol{ \mathcal J}^{(\prime)}_T(x). \label{multiBSM4}
\end{eqnarray}
Similar results were obtained in Ref. \cite{Glick-Magid:2022erc}.

For the calculation of the  $^6\textrm{He}(0^+) \rightarrow ^6\textrm{Li}(1^+)$ transition, 
only mulitpoles with $J=1$ and positive parity are needed. This leaves
$\mathcal C_{1 0}(q,A)$, $\mathcal L_{1 0}(q,A)$, $\mathcal E_{1 \lambda}(q,A)$ and $\mathcal M_{1\lambda}(q,V)$ for the SM currents and BSM axial and vector currents.
For BSM scalar and tensor currents, the only non-vanishing multipoles are $\mathcal C^P_{1 0}(q)$, $\mathcal L^T_{1 0}(q)$,
$\mathcal E^T_{1 \lambda}(q)$ and $\mathcal M^{T\prime}_{1 \lambda}$.
The steps to express the matrix elements of the operators in Eqs. \eqref{multiBSM1}
-\eqref{multiBSM4} in terms of momentum-space currents are analogous to those discussed for the SM.

\section{Charged currents in chiral EFT}
\label{app:Onebody}

We report in this Appendix, for completeness, the well-known lowest-order expressions of the SM vector and axial currents
\cite{Park:1993jf,Pastore:2009is,Krebs:2019aka,Baroni:2015uza,Krebs:2016rqz}. We also report
the currents induced by SMEFT operators.
We preliminary define for nucleons of incoming momentum ${\bf p}_i$ and outgoing momentum ${\bf p}_i^\prime$ the center of mass and relative momenta ${\bf K}_i$ and ${\bf k}_i$ in the following way
\begin{eqnarray}
{\bf K}_i\equiv \frac{{\bf p}_i^\prime+{\bf p}_i}{2}\, ,\qquad {\bf k}_i={\bf p}_i^\prime-{\bf p}_i\, ,
\end{eqnarray}
and similarly ${\bf q}={\bf p}_1+{\bf p}_2$.
We can express the charge operator up to the order of interest as in the following
\begin{eqnarray}
\rho_{5,a}({\bf q})=\sum_{\nu\in\{-2,-1\}}
\rho_{5,a}^{(\nu)}({\bf q})\, ,
\end{eqnarray}
and similarly for the current we have
\begin{eqnarray}
{\bf  j}_{5,a}({\bf q})=\sum_{\nu\in\{-3,0\}}{\bf j}_{5,a}^{(\nu)}({\bf q})\, ,
\end{eqnarray}
where $\nu$ is the chiral order defined as in Ref.~\cite{Baroni:2016xll}.
 The leading order $\nu=-2$ and next-to-leading order $\nu=-1$ axial charge read as
\begin{eqnarray}
\rho_{a}^{(-2)}({\bf q},A)&=&
\rho^{(-2)}_{\rm recoil}(\mathbf{q}, A) + \rho^{(-2)}_{\rm pseudo}(\mathbf{q}, A) \nonumber \\ 
&=&
-\frac{g_A}{2}\tau_{1,a}
\left( \frac{1}{m_N}
{\bm \sigma}_1\cdot{\bf K}_1
-\frac{q^0}{{\bf q}^2+m_\pi^2}{\bm \sigma_1}\cdot{\bf q}
\right)(2\pi)^3\delta({\bf k_1}-{\bf q})
+1\leftrightarrow 2\, , \label{axial_charge1}\\
\rho_{a}^{(-1)}({\bf q},A)&=&i\frac{g_A}{4f_\pi^2}\left({\bm \tau}_1\times{\bm \tau}_2\right)_a \frac{{\bm\sigma}_2\cdot{\bf k}_2}{\omega_2^2}+1\leftrightarrow 2\,  \label{axialcharge2b},
\end{eqnarray}
where the first term on the second line of Eq. \eqref{axial_charge1}, suppressed by $1/m_N$, contributes to $\rho^{(-2)}_{\rm recoil}(\mathbf{q}, A)$, while the second term, proportional to the electron-neutrino energy $q_0$ is induced by the induced pseuoscalar form factor.
Similarly the leading order $\nu=-3$ and next-to-leading order $\nu=0$ contributions to the axial current read 
\begin{eqnarray}
{\bf j}_{5,a}^{(-3)}({\bf q})&=&-\frac{g_A}{2}\tau_{1,a}\left[{\bm \sigma}_1-\frac{{\bf q}}{q^2+m_\pi^2}{\bm \sigma_1}\cdot{\bf q}\right](2\pi)^3\delta({\bf k}_1-{\bf q})+1\leftrightarrow 2\, \label{eq:axialLO},\\
{\bf j}_{5,a}^{(0)}&=&\widetilde{\bf j}_{5,a}^{(0)}({\bf q})-\frac{{\bf q}}{q^2+m_\pi^2} {\bf q}\cdot \widetilde{\bf j}^{(0)}_{5,a}+\frac{ig_A}{4f_\pi^2 m}\left({\bm \tau}_1\times{\bm \tau}_2\right)_a\frac{\bf q}{q^2+m_\pi^2}\left({\bf K}_1\cdot{\bf k}_1+{\bf K}_2\cdot{\bf k}_2\right)\frac{{\bm \sigma}_2\cdot{\bf k}_2}{\omega_2^2}\nonumber\\ & & +1\leftrightarrow 2\, ,
\label{axial2b}
\end{eqnarray}
where we have defined for convenience
\begin{eqnarray}
\widetilde{\bf j}_{5,a}^{(0)}({\bf q})&=&\frac{g_A^2}{2f_\pi^2}\Bigg\{2c_3\tau_{2,a}{\bf k}_2+\left({\bm \tau}_1\times{\bm \tau}_2\right)_a\bigg[\frac{i}{2m}{\bf K}_1-\frac{c_6+1}{4m}{\bm \sigma}_1\times{\bf q}+\left(c_4+\frac{1}{4m}\right){\bm \sigma}_1\times {\bf k}_2 \bigg]\Bigg\}\nonumber\\ & & \times \frac{{\bm \sigma}_2\cdot{\bf k}_2}{\omega_2^2}\, +1\leftrightarrow 2\, .
\end{eqnarray}
In the code, we replace $g_A$ with the dipole parameterization of the axial form factor 
\begin{eqnarray}
g_A(|{\bf q}|^2) =  g_A \frac{1}{\left(1 + q^2/\Lambda_A^2\right)^2},
\end{eqnarray}
with $\Lambda_A = 1.05$ GeV.
$c_3$, $c_4$ and $c_6$ are NLO low-energy constants. 

Neglecting isospin-breaking effects, the charged vector current is an isospin rotation of the isovector component of the electromagnetic current. The leading term is induced by the isovector magnetic moment and by a recoil correction \cite{Pastore:2009is} 
\begin{eqnarray}\label{eq:magnetic}
{\bf j}^{(-2)}_a = \frac{\tau_{1,\, a}}{2} \frac{1}{2 m_N}
\Big(
2 {\bf K}_1 + 
i (1 +\kappa_V)  {\bm \sigma}_1 \times {\bf q} \Big),
\end{eqnarray}
with $\kappa_V \sim 3.7$.
The NLO contributions originate from the exchange of a pion between nucleon lines, with the  vector current coupling  either to the pion in flight, or to the nucleon. This contribution gives
\begin{eqnarray}
{\bf j}^{(-1)}_a &=& i \frac{g_A^2}{F_\pi^2}
\left({\bm \tau}_1 \times {\bm \tau}_2\right)_a
\left( -     \, {\bm \sigma}_1 \, \frac{{\bm \sigma}_2 \cdot {\bf k}_2}{\omega_2^2}  + \frac{{\bf k}_1 - {\bf k}_2}{2 \omega_1^2 \, \omega_2^2} {\bm \sigma}_1 \cdot {\bf k}_1 {\bm \sigma}_2 \cdot {\bf k}_2 \right) +  1\leftrightarrow 2.
 \end{eqnarray}

For currents induced by SMEFT and $\nu$SMEFT operators, we just retain one-body contributions. 
Recalling that at the quark level,
\begin{equation}
    \mathcal J_S^{a} = \bar q \frac{\tau^a}{2} q, \qquad 
    \mathcal J_P^{a} = \bar q  \gamma_5 \frac{\tau^a}{2} q \qquad 
    \mathcal J_T^{\mu\nu\, a} = \bar q  \sigma^{\mu \nu} \frac{\tau^a}{2} q,
\end{equation}
where $q$ denotes a quark doublet $q = (u, d)^T$, 
the scalar and pseudoscalar densities and the tensor currents are then given by 
\begin{eqnarray}
\mathcal J^{(-3)}_{S,a} &=& g_S \frac{\tau_{1,a}}{2}(2\pi)^3\delta({\bf k}_1-{\bf q})+1\leftrightarrow 2 , \label{S}\\
\mathcal J^{(-4)}_{P,a} &=&  \frac{g_A B}{m_\pi^2 + \textbf{q}^2} \frac{\tau_{1,a}}{2} \boldsymbol{\sigma}_1 \cdot \textbf{q}  (2\pi)^3\delta({\bf k}_1-{\bf q})+1\leftrightarrow 2, \label{T1}\\
\mathcal J^{i j(-3)}_{T,a} &=&  g_T \varepsilon^{i j k} \frac{\tau_{1,a}}{2} \sigma_1^k (2\pi)^3\delta({\bf k}_1-{\bf q})+1\leftrightarrow 2, \label{T2}\\
\mathcal J^{i 0(-2)}_{T, a} &=&  i \frac{g^{\prime}_T}{2 m_N} q^i \,  \frac{\tau_{1,a}}{2} (2\pi)^3\delta({\bf k}_1-{\bf q})+1\leftrightarrow 2\,, \label{T2b}
\end{eqnarray}
and the currents defined in Section \ref{app:multi}
are given by, for example, $\mathcal J^{i j}_{T} = \mathcal J^{i j}_{T,x} + i \mathcal J^{i j}_{T,y}$, where $x$ and $y$ are isospin indices.
The values of the scalar and tensor charges are given in Eq. \eqref{charges}. 
The exact value of $g_T^\prime$ is unknown, but should be a number of order 1.

\section{Fully differential unpolarized decay rate}
\label{app:rate}

In the main text we gave the expression for the decay rate,
differential in the electron energy. We give here more differential expressions, which can be used, for example, to extract corrections to the $\beta$-$\nu$
correlation $a$.
We start from the expression given in Ref.~\cite{Schiavilla2002,Walecka:1995mi} for the specific case of ${}^6{\rm He}$. The starting point is the usual Fermi's golden rule for the unpolarized differential decay rate
\begin{eqnarray}
d\Gamma&=&(2\pi)\delta(E_i-E_f-E_\nu-E_e)\frac{1}{2J_i+1}\sum_{M_i M_f}\sum_{s_e s_\nu}\rvert M\rvert^2\frac{d^3{\bf p}_e}{(2\pi)^3}\frac{d^3{\bf p}_\nu}{(2\pi)^3}\, ,
\end{eqnarray}
with ${\bf P}_f=-{\bf p}_e-{\bf p}_{\nu_e}$. 
Using the multipole expansion of the matrix element of the SM 
weak Hamiltonian, Eq. \eqref{eq:fhi}, one gets
\begin{eqnarray}\label{eq:dGammageneral}
d\Gamma&=&2\pi\delta(M_i-E_f-E_e-E_\nu)G_F^2 V_{ud}^2\frac{4\pi}{2J_i+1}\nonumber\\
&&\bigg[ (1+{\bf v}_e\cdot {\bf v}_{\nu})\sum_{l\geq 0}\rvert C_l(q)\rvert^2+(1-{\bf v}_e\cdot {\bf v}_\nu+2{\bf v}_e\cdot\hat{\bf q}\,{\bf v}_\nu\cdot\hat{\bf q})\sum_{l\geq 0}\rvert L_l(q)\rvert^2\nonumber\\
&&-2\hat{\bf q}\cdot({\bf v}_e+ {\bf v}_\nu)\sum_{l \geq 0}Re\left[C_l(q)L_l^\star(q)\right]+(1-{\bf v}_e\cdot\hat{\bf q}\, {\bf v}_\nu\cdot\hat{\bf q})\sum_{l\geq 1}\left[\rvert M_l(q)\rvert^2+\rvert E_l(q)\rvert^2\right]\nonumber\\
&&-2\hat{\bf q}\cdot({\bf v}_e-{\bf v}_\nu) \sum_{l\geq 1}Re\left[M_l(q)E_l^\star(q)\right]\bigg]
\frac{d^3{\bf p}_e}{(2\pi)^3}\frac{d^3{\bf p}_\nu}{(2\pi)^3}\, ,
\end{eqnarray}
with ${\bf q}={\bf p}_e+{\bf p}_{\nu}$, $\hat{\bf q} = {\bf q}/|\bf q|$ , ${\bf v}_e={\bf p}_e/\sqrt{p_e^2+m_e^2} $ and ${\bf v}_\nu={\bf p}_\nu/E_\nu$. 
We notice that for the transition considered $\rvert J_i-J_f\rvert=\pm 1,0$ and $\pi_i\pi_f=1$, which allow to simplify the expressions to
\begin{eqnarray}
d\Gamma&=&2\pi\delta(M_i-E_f-E_e-E_\nu)G_F^2 V_{ud}^2 \frac{4\pi}{2J_i+1}\nonumber\\
&&\bigg[ (1+{\bf v}_e\cdot {\bf v}_{\nu})\rvert C_1(q;A)\rvert^2+(1-{\bf v}_e\cdot {\bf v}_\nu+2{\bf v}_e\cdot\hat{\bf q}\, {\bf v}_\nu\cdot\hat{\bf q})\rvert L_1(q;A)\rvert^2\nonumber\\
&&-2\hat{\bf q}\cdot({\bf v}_e+ {\bf v}_\nu)Re\left[C_1(q;A)L_1^\star(q;A)\right]+(1-{\bf v}_e\cdot\hat{\bf q}\, {\bf v}_\nu\cdot\hat{\bf q})\left[\rvert M_1(q;V)\rvert^2+\rvert E_1(q;A)\rvert^2\right]\nonumber\\
&&-2\hat{\bf q}\cdot({\bf v}_e- {\bf v}_\nu) Re\left[M_1(q;V)E_1^\star(q;A)\right]\bigg]
\frac{d^3{\bf p}_e}{(2\pi)^3}\frac{d^3{\bf p}_\nu}{(2\pi)^3}\, .
\end{eqnarray}
Retaining terms up to order $q^2$ we have
\begin{eqnarray}
d\Gamma&=&2\pi\delta(E_i-E_f-E_\nu-E_e)G_F^2 V_{ud}^2 \frac{4\pi}{2J_i+1} \frac{1}{9}\nonumber\\
&&\bigg[
(3-{\bf v}_e\cdot {\bf v}_\nu )\rvert L^{(0)}_1(A)\rvert^2
-2\hat{\bf q}\cdot({\bf v}_e+{\bf v}_\nu )  \, q r_\pi \, \textrm{Re}\left[{C}^{(1)}_1(A)L_1^{(0) \star}(A)\right] \nonumber \\
& & 
-2\hat{\bf q}\cdot({\bf v}_e-{\bf v}_\nu) \, q r_\pi \, \textrm{Re}\left[M^{(1)}_1(V)E_1^{(0)\star}(A)\right] +(1+{\bf v}_e\cdot {\bf v}_\nu )(q r_\pi)^2\rvert {C}^{(1)}_1(A)\rvert^2 \nonumber \\
&& +
(q r_\pi)^2 ( 1  -  {\bf v}_e\cdot\hat{\bf q}\, {\bf v}_\nu\cdot\hat{\bf q}) \left[ \lvert {M}^{(1)}_1(V)\rvert^2 -\frac{1}{5}  \textrm{Re}\left( E^{(0)}_{1}(A) E^{(2)}_{1}(A) \right) \right]\nonumber \\
& &- \frac{(q r_\pi)^2}{5}  (1 - {\bf v}_e \cdot {\bf v_\nu} + 2 {\bf v_\nu} \cdot {\bf \hat{q}} {\bf v_e} \cdot {\bf \hat{q}})  {\rm Re} \left( L^{(0)}_{1}(A) L^{(2)}_{1}(A) \right)
\bigg] \frac{d^3{\bf p}_e}{(2\pi)^3}\frac{d^3{\bf p}_\nu}{(2\pi)^3},
\end{eqnarray}
where we used the expansion in Eqs.~\eqref{Fit1}--\eqref{Fit4}. Integrating over the neutrino phase space and the electron angular variables, we obtain Eq. \eqref{rate}. 

With the definitions in Eqs. \eqref{Fit1}--\eqref{Fit4}, the 
coefficients of the momentum expansion of the multipole operators are real. Dropping the labels $A$ and $V$ and using $E_1^{(0)} =  \sqrt{2} L_1^{(0)}$, we can write 
\begin{eqnarray}
& &d\Gamma=2\pi\delta(E_i-E_f-E_\nu-E_e)G_F^2 V_{ud}^2 \frac{4\pi}{2J_i+1} \frac{1}{3}\rvert L^{(0)}_1(A)\rvert^2  \, \frac{d^3{\bf p}_e}{(2\pi)^3}\frac{d^3{\bf p}_\nu}{(2\pi)^3} \nonumber\\
&&\Bigg\{
1 + \frac{2 W_0 r_\pi}{3} \left[   \left( 1 - 2 \varepsilon + \frac{\mu_e^2}{\varepsilon}\right)  \, \frac{\sqrt{2} M^{(1)}_1 }{L_1^{(0)}}
- \left(1 - \frac{\mu_e^2}{\varepsilon} \right)
 \frac{C^{(1)}_1}{L_1^{(0) }} \right] \nonumber\\
 \nonumber \\ & & 
 + \frac{( W_0 r_\pi)^2}{3} 
\Bigg[ 
\left( 1 - \frac{4}{3} \varepsilon (1- \varepsilon) - \frac{\mu_e^2}{\varepsilon} \frac{2+\varepsilon}{3} 
 \right) \left(\frac{ {C}^{(1)}_1 }{L_1^{(0)}}\right)^2 
- \frac{1}{5}  \left(1 
- \frac{\mu^2_e}{\varepsilon}(2-\varepsilon)  \right) \frac{ L^{(2)}_{1}}{ L^{(0)}_{1} } \nonumber \\ &&+ \left( 1 - \frac{10}{3} \varepsilon (1-\varepsilon) + \frac{\mu^2_e}{\varepsilon} \frac{4-7 \varepsilon}{3} 
\right) \frac{\widetilde{M}_1^{(2)}}{L_1^{(0)}}\Bigg] \nonumber \\
& & + {\bf v}_e\cdot {\bf v}_\nu \Bigg[-\frac{1}{3} -\frac{2 W_0 r_\pi}{3}   
\left( 
  (1-2\varepsilon)  \, \frac{\sqrt{2}  M^{(1)}_1 }{L_1^{(0)}}
+  \frac{C^{(1)}_1}{L_1^{(0) }}  \right) \nonumber \\
& & + \frac{( W_0 r_\pi)^2}{3} \Bigg( (1 - \mu_e^2)   
\left( \left(\frac{ {C}^{(1)}_1 }{L_1^{(0)}}\right)^2 
- \frac{1}{5}  \frac{ L^{(2)}_{1}}{ L^{(0)}_{1} }\right)  -  
\left(1 - 4 \varepsilon (1-\varepsilon) - \mu_e^2\right)   \frac{\widetilde{M}_1^{(2)}}{L_1^{(0)}} \Bigg) \Bigg]
 \nonumber \\ & & 
+ \frac{( W_0 r_\pi)^2}{3} \left( ({\bf v}_e\cdot {\bf v}_\nu)^2 - \frac{1}{3} \frac{|{\bf p}_e|^2}{E_e^2} \right)
\  \varepsilon (1 - \varepsilon)  \Bigg(   2  \left(\frac{ {C}^{(1)}_1 }{L_1^{(0)}}\right)^2 
 -  \frac{\widetilde{M}_1^{(2)}}{L_1^{(0)}}
 \Bigg)
\Bigg\}, \label{correlations}
\end{eqnarray}
where we introduced the scaled variables $\varepsilon$ and $\mu_e$ like in Section \ref{sec:spectrum} and we defined
\begin{equation}
\widetilde{M}^{(2)}_1 = L^{(0)}_1 \left( \left(\frac{{M}^{(1)}_1}{L_1^{(0)}}\right) ^2 -\frac{\sqrt{2}}{5} \frac{  E^{(2)}_{1}}{L_1^{(0)}} \right).  
\end{equation}
From Eq. \eqref{correlations}, we can read the corrections to the $\beta$-$\nu$ angular correlation $a$, and to the subleading correlations
$C_{(aa)}$ \cite{Bhattacharya:2011qm}, proportional to 
$({\bf v}_e\cdot {\bf v}_\nu)^2$.
From these expressions, we find that the $\beta$-$\nu$ correlation coefficient, averaged over energy, is
\begin{equation}
    \langle a \rangle = -\frac{1}{3} - 2.3(4) \cdot 10^{-4} - 1.2(2) \cdot 10^{-3}, 
\end{equation}
where we used GFMC matrix elements,
and we did not include radiative corrections.  
The second term is the correction linear in $M_1^{(1)}$ and $C_1^{(1)}$. The term proportional to $M_1^{(1)}$ averages to almost zero, leaving behind a small contribution from $C^{(1)}_1$. The last term appears at second order in the multipole expansion.
Both corrections are still below the experimental sensitivity \cite{Muller:2022jew}.

For tensor and pseudoscalar currents, we find that the interference with the SM is given, at lowest order in the multipole expansion, by
\begin{eqnarray}
d\Gamma&=& 2\pi\delta(E_i-E_f-E_\nu-E_e)G_F^2 V^2_{ud}\frac{4\pi}{2J_i+1} \frac{4 m_e}{E_e}  \frac{d^3{\bf p}_\nu}{(2\pi)^3} \frac{d^3{\bf p}_e}{(2\pi)^3} \nonumber\\
&& 
\bigg[ \epsilon_T \textrm{Re}\left( E_1(q,T) E^*_1(q,A) +  L_1(q,T) L^*_1(q,A)\right)  -  \frac{\epsilon_P}{2}\hat{\bf v}_\nu\cdot\hat{\bf q} \, \textrm{Re} \left(C_1^{}(q,P) L^*_{1}(q,A)\right)
\bigg] ,
\label{BSMspectrum}
\end{eqnarray}
which, using the expansion in Eqs. \eqref{Fit1} -- \eqref{Fit4} becomes
\begin{eqnarray}
& &d\Gamma=2\pi\delta(E_i-E_f-E_\nu-E_e)G_F^2 V^2_{ud}\frac{4\pi}{2J_i+1} \frac{4 m_e}{9 E_e}  \frac{d^3{\bf p}_\nu}{(2\pi)^3} \frac{d^3{\bf p}_e}{(2\pi)^3} \nonumber\\
&&\bigg[
 \epsilon_T \left( E^{(0)}_1(T) E^{(0)}_1(A) +  L^{(0)}_1(T) L^{(0)}_1(A)  \right)
 - \frac{\epsilon_P}{2} 
 W_0 r_\pi \bigg( (1-\varepsilon) + \varepsilon\,   
{\bf v}_e\cdot{\bf v}_\nu \bigg) C_1^{(1)}(P) L^{(0)}_{1}(A)
 \bigg], \nonumber \\
\label{BSMspectrum2}
\end{eqnarray}
so that tensor interactions only induce a Fierz interference term, while pseudoscalar interactons also affect $a$.
The expressions for the decay rates in the presence of sterile neutrinos can be found in a similar way.

\section{Higher-order electroweak and kinematic recoil corrections}
\label{app:radcorr}
\subsection{Coulomb and radiative corrections}
The multipole expansion of the weak Hamiltonian described at the start of this work does not explicitly take into account electromagnetic interactions. The treatment of such effects in nuclear $\beta$ decay has historically been divided into three parts: ($i$) Coulomb corrections via the outgoing $\beta$ particle in the field of the final atomic state; ($ii$) infrared-divergent contributions from various virtual and real emission photon processes up to some $\mathcal{O}(\alpha^nZ^m)$ ($m>n$) not already contained in ($i$); ($iii$) the remainder of electroweak radiative corrections, which are independent of the process kinematics. The latter two are typically described as outer and inner radiative corrections, respectively, and are the topic of a significant body of literature \cite{Sirlin2013, Towner2008} and will only be plugged into the final result. The Coulomb corrections, on the other hand, are of interest and to first order can be understood as the elastic response of the $\gamma W$ box diagrams. Higher-order (i.e. $\mathcal{O}\{[\alpha Z]^n\}$) behaviour can then be absorbed by substituting the electron plane wave by a solution of the Dirac equation in the static, spherical potential of the atomic final state, $\phi_e(\bm{r}, \bm{p}_e)$, and writing the Hamiltonian as
\begin{align}
    \mathcal{M}_{fi} &= \int \mathrm{d}^3 r\, \bar{\phi}_e(\bm{r}, \bm{p}_e)\gamma^\mu(1-\gamma^5)v(\bm{p}_{\bar{\nu}}) \nonumber \\
    &\times \int \frac{\mathrm{d}^3s}{(2\pi)^3}e^{i\bm{s}\cdot \bm{r}}\frac{1}{2}[\langle f(\bm{p}_f+\bm{p}_e-\bm{s})| V_\mu + A_\mu | i(\bm{p}_i) \rangle  + \langle f(\bm{p}_f) | V_\mu + A_\mu | i(\bm{p}_i-\bm{p}_e+\bm{s}) \rangle ].
\end{align}
The multipole expansion then proceeds analogously, and the most immediate modification is the introduction of the Fermi function (Eq. (\ref{eq:Fermi_function})), i.e. the $j=1/2$ large components of the electron wave function for a point charge. The small and $j>1/2$ components show up as small modifications in the radial integrals folded together with the nuclear current and introduce additional small terms to the differential decay rate. These are well-known in the literature and can be found in several places \cite{Behrens1978, Holstein1974}. Additional correction terms are well-known for moving past a point-charge model of the nucleus and introducing additional subdominant electromagnetic corrections such as screening by atomic electrons and atomic exchange processes. We use the results of Ref. \cite{Hayen2018} by translating our results into the Behrens-B\"uhring formalism \cite{Behrens1982,Hayen:2020nej}
\begin{subequations}
\begin{align}
    L_1^0 &= \mathcal{C}\sqrt{3}{}^AF_{101}^{(0)}; \qquad E_1^0 = \mathcal{C}\sqrt{6}{}^AF_{101}^{(0)} \\
    C_1^1 &= -\mathcal{C}\frac{R}{r_\pi} {}^AF_{110}^{(0)}; \qquad M_1^1 = -\mathcal{C}\frac{R}{r_\pi} {}^VF_{111}^{(0)} \\
    L_1^2 &= \mathcal{C}\frac{1}{\sqrt{3}}\left(\frac{R}{r_\pi}\right)^2(5{}^AF_{101}^{(1)}-2\sqrt{2}{}^AF_{121}^{(0)}) \\
    E_1^2 &= \mathcal{C}\sqrt{\frac{2}{3}}\left(\frac{R}{r_\pi} \right)^2(5{}^AF_{101}^{(1)}+\sqrt{2}{}^AF_{121}^{(0)})
\end{align}
\end{subequations}
where $R$ is the nuclear radius of the uniformly charged sphere, i.e. $R=\sqrt{5/3}\langle r^2\rangle_\mathrm{exp}$ and $\mathcal{C}=\sqrt{2J_i+1/4\pi}$. This results in the following additional terms
\begin{align}
    C_\text{Coulomb}(Z, \varepsilon) = \alpha ZW_0R &\left[|L_1^0|^2\left(\frac{6}{35}-\frac{233}{210}\frac{\alpha Z}{W_0R}-\frac{3}{70}\frac{\mu_e^2}{\varepsilon}-\frac{12}{7}\varepsilon\right)\right. \nonumber \\
    &\left.+\left(\frac{r_\pi}{R}\right)^2\left(\frac{E_1^0E_1^2}{2}-L_1^0L_1^2\right)\left(-\frac{20}{35}+\frac{4}{7}\varepsilon\right)\right]
\end{align}
where the fractional prefactors were calculated assuming a uniformly charged sphere as described in Ref. \cite{Hayen2018}.

Putting everything together, the inclusion of electroweak corrections modifies the $\beta$ spectrum by
\begin{equation}
\frac{d\Gamma}{d\varepsilon} \propto (1+\Delta_R^V)(1+\delta_R(\varepsilon))F_0(Z, \varepsilon)L_0(Z, \varepsilon)S(Z, \varepsilon)[C_0(\varepsilon)+C_\text{Coulomb}(Z, \varepsilon)]
\end{equation}
where $\Delta_R^V$ is the inner radiative corrections to vector transitions (the difference induced by axial transitions is small and absorbed into an experimentally determined $g_A$ value), $\delta_R(\varepsilon)$ are outer radiative corrections to $\mathcal{O}(\alpha^3Z^2)$. Further, $F_0L_0$ is the Fermi function for a uniformly charged sphere, $S$ describes the shielding of the nuclear charge by atomic electrons and $C_\text{Coulomb}(Z, \varepsilon)$ are modifications due to higher-order Coulomb corrections folded with the nuclear current discussed above.

\subsection{Kinematic recoil corrections}
In App. \ref{app:multi} we defined the multipole expansion of the weak Hamiltonian as introduced by Donnelly and Walecka \cite{Walecka:1995mi}. This expansion, however, is not Lorentz covariant but implicitly performed in the Breit (brick wall) system, i.e. where $\bm{p}_i = -\bm{p}_f$. In the approximation of an infinitely massive nucleus the Breit and lab frames agree. For a consistent description, however, results must be Lorentz boosted back into the lab frame, leading to additional $\mathcal{O}(q/M)$ results. This is discussed in more detail in Ref. \cite{Hayen:2020nej} and will not be repeated here.

The kinematic recoil corrections originating from the phase space integral can also be easily written as a multiplicative factor
\begin{eqnarray}
d\Gamma \propto |\mathcal{M}|^2\left( 1 + \frac{3E_e - W_0 - 3\bm{p}_e\cdot \hat{\bm{v}}_\nu}{M} \right)
\end{eqnarray}
where the final term gives a finite contribution in the neutrino angular integral when combined with the $\beta$-$\nu$ correlation.

Combining both leads to well-known expressions for the total kinematic recoil corrections \cite{Wilkinson1982, Hayen2018}
\begin{align}
    R_N &\approx 1 - \frac{2W_0}{3M} + \frac{10E_e}{3M} - \frac{2m_e^2}{3E_eM}
\end{align}
where we kept only terms to first order in $q/M$, thereby neglecting terms of $\mathcal{O}(10^{-6})$ at most.

\bibliography{biblio}

\end{document}